\documentclass[acmsmall,screen,natbib,numbers]{acmart}

\setcitestyle{numbers,square}

\usepackage[normalem]{ulem}
\usepackage[utf8]{inputenc} 
\usepackage{amsmath}
\usepackage{thmtools}
\usepackage[english]{babel}
\usepackage{pdfpages}
\usepackage{graphicx}
\usepackage{setspace}
\usepackage{longtable}
\usepackage{booktabs}
\usepackage{algpseudocode}
\usepackage{tabu}
\usepackage[referable]{threeparttablex}
\usepackage{varwidth}
\usepackage{multirow}
\usepackage{wrapfig}
\usepackage{amsfonts}
\usepackage{multicol}
\usepackage{bbm}

\usepackage[inline]{enumitem}

\usepackage[ruled,vlined,resetcount,linesnumbered]{algorithm2e}

\usepackage[]{hyperref}

\usepackage{comment}

\usepackage{todonotes}
\usepackage{tikz}
\usetikzlibrary{arrows,automata,shapes,decorations,decorations.markings,calc, matrix,decorations.pathmorphing, patterns,backgrounds,shapes.misc,arrows.meta}

\input{macros}

\begin{document}

\title{Efficient Timestamping for Sampling-based Race Detection}

\author{Minjian Zhang}
\affiliation{%
  \institution{University of Illinois at Urbana-Champaign}
  \country{USA}}
\email{minjian2@illinois.edu}

\author{Daniel Wee Soong Lim}
\affiliation{%
  \institution{National University of Singapore}
  \country{Singapore}}
\email{dws.lim@nus.edu.sg}

\author{Mosaad Al Thokair}
\affiliation{%
  \institution{University of Illinois at Urbana-Champaign}
  \country{USA}}
\email{mosaada2@illinois.edu}

\author{Umang Mathur}
\affiliation{%
  \institution{National University of Singapore}
  \country{Singapore}}
\email{umathur@comp.nus.edu.sg}

\author{Mahesh Viswanathan}
\affiliation{%
  \institution{University of Illinois at Urbana-Champaign}
  \country{USA}}
\email{vmahesh@illinois.edu}


\begin{abstract}
Dynamic race detection based on the happens before ({\acrhb}) partial order has now become the \textcolor{black}{de facto} approach to quickly identify data races in multi-threaded software. Most practical implementations for detecting these races use timestamps to infer causality between events and detect races based on these timestamps. Such an algorithm updates timestamps (stored in vector clocks) at every event in the execution, and is known to induce excessive overhead. Random sampling has emerged as a promising algorithmic paradigm to offset this overhead. It offers the promise of making sound race detection scalable. In this work we consider the task of designing an efficient sampling based race detector with low overhead for timestamping when the number of sampled events is much smaller than the total events in an execution. To solve this problem, we propose (1) a new notion of \emph{freshness timestamp}, (2) a new data structure to store timestamps, and (3) an algorithm that uses a combination of them to reduce the cost of timestamping in sampling based race detection. Further, we prove that our algorithm is close to optimal --- the number of vector clock traversals is bounded by the number of sampled events and number of threads, and further, on any given dynamic execution, the cost of timestamping due to our algorithm is close to the amount of work \emph{any} timestamping-based algorithm must perform on that execution, that is it is instance optimal. Our evaluation on real world benchmarks demonstrates the effectiveness of our  proposed algorithm over prior timestamping algorithms that are agnostic to sampling.
\rmv{
Recently, a property tester for {\acrhb}-race detection --- that is, one that never erroneously reports a race and at the same time guarantees detecting a race with high probability on traces that are ``very far'' from any race-free trace --- was proposed. Though this property tester processes only constantly many events in the observed trace, it works ``offline'' on traces that are recorded and stored. In this paper, we present an ``online'' property tester for {\acrhb}-race detection, namely, one that can run concurrently as the program is executing and the trace is being observed. Our online property tester also processes only constantly many events during its analysis. Key challenges overcome by our algorithm are finding mechanisms for individual threads to locally turn on and off sampling, and for these decisions to be communicated to other threads with little additional synchronization. Our experiments demonstrate the promise of the new algorithm.
Dynamic data race detectors have emerged as indespensible tool in the
toolkit of developers of concurrent software.
Yet, the runtime overhead introduced by them limits the frequency
of their use.
Sampling has emerged as a promising general-purpose algorithmic
solution to lower the overhead of dynamic race detectors.
Existing sampling algorithms as well as tools, however, impose additional
synchronization in the underlying software 
to determine which events to sample during execution of the concurrent code under test.
This synchronization not only adds additional overhead to the otherwise lightweight approach,
but can also potentially limit the effectiveness of the race detector
in exposing subtle concurrency errors.
In this paper, we propose a sampling algorithm that can be implemented without the use of such
additional synchronizations \ucomment{well, we should be fixing this.}.
We implemented our algorithm as an extension of ThreadSanitizer
and our evaluation demonstrates the effectiveness of our algorithm \ucomment{\ldots}.
}
\end{abstract}

\keywords{data race, concurrency, sampling, timestamping}

\maketitle

\section{Introduction}
\seclabel{intro}

Writing correct concurrent code is a particularly challenging task
even for experienced programmers, since simple reasoning paradigms do not
naturally translate to programs where multiple threads may non-deterministically interleave.
{\color{black}It is no surprise that concurrency bugs routinely make their way into production-grade code, degrading code quality, impacting user experience, and often compromising key properties such as security and crash freedom.}
A first line of defense against concurrency bugs such as data races
are dynamic analysis tools such as \tsan~\cite{threadsanitizer}
and Helgrind~\cite{Helgrind} that
detect races as the underlying program executes.
{\color{black}These tools are widely adopted and have been integrated into actively maintained tools such as
the LLVM compiler framework~\cite{lattner2004llvm} or the Valgrind debugger~\cite{Valgrind}.} 

Despite their popularity and efficiency, 
their use is often limited to in-house testing~\cite{TSanChromium,TSanFirefox}, 
limiting their bug detection capability to small scale workloads, which are often not
sufficient to expose heisenbugs~\cite{heisenbugs,evil2012}.
The primary bottleneck in advocating runtime tools
like \tsan for production use is the performance overhead they induce;
\tsan, for example, reportedly induces
upto $20\times$ overhead~\cite{threadsanitizer}.
A popular emerging paradigm for reducing the overhead of sanitizers
is the use of \emph{sampling}~\cite{RPT2023,pacer-tool,marino2009literace,racemob}, 
where one limits data race detection or bug detection to a small ``sample set'' of events that is tiny fraction of the full execution. The sample set may either be identified through random sampling~\cite{marino2009literace,pacer-tool,RPT2023} or through static analysis~\cite{racemob}. 
While a challenging endeavor in general, sampling has shown
promise for in-production use in the context of
sanitizers for memory-safety~\cite{GWPASan2024}, where
the problem of large overhead could be \textcolor{black}{mitigated} by 
optimizing the cost of instrumentation (call backs inserted at every event of interest).
Unfortunately, simply addressing the cost of instrumentation
will likely not be enough for making sampling-based race detection
amenable for production use.
This is because the analysis itself can be
expensive when data race detectors employ 
vector clock based timestamping. In vector clock based algorithms, each event costs
$O(\numthr)$ arithmetic operations, giving a total
of $O(\trsz \cdot \numthr)$ operations for executions with $\trsz$ events and $\numthr$ threads.
This can be prohibitive 
even for moderately large applications.

In this paper, we ask --- \emph{can we reduce
overhead due to timestamping operations in the context of sampling-based data race detection?}
While prior works on sampling-based race detection~\cite{pacer-tool,marino2009literace,racechaser,RPT2023}
propose new algorithms, they do not directly address the
cost of analysis \emph{after a sample set has been identified}. Their focus is on the composite problem of sampling and analysis. By focusing our attention on the cost of timestamping during analysis, we hope to improve the performance of all sampling-based approaches, no matter how the sample set is identified. 
Towards this we propose a clean formulation of 
the algorithmic problem underlying our question, namely the
\emph{analysis problem} --- given an execution and a sample
set of events $S$, how much cost should
any timestamping algorithm pay if it were to detect races only on events in $S$?
In our setting, the set $S$ has size typically much smaller than the size of the execution. However, the set $S$ is often identified on the fly while the execution is being analyzed. Therefore, our analysis methods must work even when membership of an event $e$ in the sample set $S$ is only known after $e$ is actually observed. 

How do we solve the \emph{analysis problem} optimally? 
The na\"{i}ve solution --- simply process all events --- takes $O(\trsz \cdot \numthr)$
operations and is not ideal.
In particular, can we reduce the multiplicative factor on $\trsz$
and instead only pay the timestamping cost proportional to $|S|$?
Observe that the other na\"{i}ve solution --- simply perform analysis when 
the observed event is known to be in the set $S$, and skip
otherwise --- is unsound, as it may miss synchronization events that are otherwise necessary to rule out false positives.
In other words, synchronization events
such as lock acquire and release events cannot be skipped
if soundness is paramount.
At the same time, paying full timestamping cost for all synchronization
events (whether or not they are in $S$) will \textcolor{black}{in the worst case} take $O(\trsz \cdot \numthr)$ time.
The central contribution of this work is a timestamping algorithm
for sampling-based race detection that spends $O(|S| \cdot \numthr^2)$ time
in timestamping operations, which can be significantly smaller than
the vanilla $O(\trsz \cdot \numthr)$ algorithm.
Our algorithm stems from several interesting technical insights
which we discuss next.

First, we observe that local increments in a timestamping algorithm
primarily serve the purpose of distinguishing different events
across different \emph{epochs} (i.e., region in a thread
between two synchronization events in the thread).
When limiting race detection to a small set $S$,
it suffices to only increment \emph{local}
clocks of threads, at most $|S|$ times since there are only so many
events that potentially need to be distinguished. {
\color{black}More specifically, unlike the case without sampling, where an increment occurs on every release, only the first release after any sampled event is relevant for incrementing.
}
As a consequence, vector timestamps get updated less frequently.
In turn, this means that 
a vast majority of synchronization events often communicate redundant
information.
This brings us to our second insight --- if we can capture when
timestamp communication is redundant, we can \textcolor{black}{proactively} avoid
sending and/or receiving timestamps when unnecessary.
Towards this, we propose a \emph{freshness timestamp},
that tracks metadata about when the timestamp of a thread
is fresh and not yet communicated via a given synchronization channel (such as a lock).
We show that this freshness timestamp can be additionally tracked accurately
and efficiently, giving us an algorithm that
spends $O(|S| \cdot \numthr^3)$ time in timestamping.
Third, we show that a fine-grained data structure
and optimistic sharing (i.e. shallow copying) of clocks
can further reduce the complexity to $O(|S| \cdot \numthr^2)$.
Finally, the time spent in updating and accessing \emph{access histories} for reporting data races at read and write events
is $O(|S| \cdot \numthr)$, giving
us an algorithm that spends a total of $O(|S| \cdot \numthr^2)$
time in the overall race detection algorithm.

We evaluated our proposed algorithms by implementing
it in \tsan and in the offline dynamic analysis
framework \rapid~\cite{rapid}.
{\color{black}We evaluated our \tsan implementation on popular database workloads. Our experiments show that the algorithmic overhead introduced by timestamp computation constitutes a major portion of the overall cost of dynamic race detection. Our innovations reduces this component by a non-trivial fraction, with the reduction particularly noticeable in scenarios where vanilla timestamping exacerbates existing lock contention in the application.}

Our evaluation on \rapid explains the performance of our
algorithm by pointing out that through the usage of the freshness timestamp with
the data structure and object sharing that we propose,
the number of operations for timestamping can be significantly reduced.


\section{A Gentle Introduction to Dynamic Race Detection}
\seclabel{preli}

Here, we recall basic background on data races and 
algorithmic details underlying dynamic data race detectors.

\myparagraph{Events and programs executions}{
Data race detectors such as \tsan~\cite{threadsanitizer} work by instrumenting instructions
of a concurrent program under test and insert callbacks to observe events.
An event $e$ of a concurrent program is of the form 
$\ev{t}{o}$, where $t = \ThreadOf{e}$ is the identifier of the thread that 
performs the event and $o = \OpOf{e}$ is the operation of $e$. 
For the purpose of our work, it suffices to consider an
operation $o$  to be one of the following:
(a) read/write access (i.e., $o = \rd(x)$ or $o = \wt(x)$ for some memory location $x$), or 
(b) acquire/release (i.e., $o = \acq(\lk)$ or $o = \rel(\lk)$ for some lock $\lk$).
An execution of a concurrent program can then be viewed
as a sequence of events $\tr = e_1 e_2 \ldots, e_n$.
We will use $\events{\tr}$ to denote the set of events of $\tr$.
The set of threads, locks and memory locations of $\tr$
will be denoted $\threads{\tr}$, $\locks{\tr}$ and $\vars{\tr}$ respectively.
For each lock $\lk \in \locks{\tr}$, the semantics of locking operations ensure
that the sub-sequence of $\tr$ corresponding to events that access $\lk$
is a prefix of some word that matches the regular expression
$(\ev{t_1}{\acq(\lk)}\ev{t_1}{\rel(\lk)} + \cdots + \ev{t_k}{\acq(\lk)}\ev{t_k}{\rel(\lk)})^*$,
where $\set{t_1, \ldots, t_k} = \threads{\tr}$;
in other words, \textcolor{black}{at most one thread can hold $\lk$ at a given time.}
For two distinct events $e_1, e_2 \in \events{\tr}$, we will use
$e_1 \trord{\tr} e_2$ to denote that 
$e_1$ appears before $e_2$ in $\tr$.
Likewise, we use $e_1 \tho{\tr} e_2$ to denote that
$e_1 \trord{\tr} e_2$ and further $\ThreadOf{e_1} = \ThreadOf{e_2}$.
The corresponding irreflexive orders are denoted
as $\stricttrord{\tr}$\textcolor{black} {(trace order)} and $\stricttho{\tr}$ \textcolor{black}{(thread order)} respectively.
}

\myparagraph{Happens-before data races}{
    While several definitions of data races have been
    introduced in the literature \cite{SyncP2021,Shi2024,shb2018,wcp2017,Mathur20,rvpredict,cp2012,PavlogiannisPOPL20}, the one based on the \emph{happens-before} partial order is a popular choice.
    It is also the one that widely used tools like
    \tsan build upon.
    The happens before partial order $\hb{\tr}$ of an execution
    $\tr$ is the smallest partial order on the set of events
    $\events{\tr}$ of $\tr$ such that for any two events $e_1, e_2 \in \events{\tr}$,
    we have:
    \begin{enumerate}
        \item $e_1 \tho{\tr} e_2$, then $e_1 \hb{\tr} e_2$, and
        \item if $e_1 \stricttrord{\tr} e_2$ and there is a lock $\lk$ such that $\OpOf{e_1} = \rel(\lk)$ and $\OpOf{e_2} = \acq(\lk)$, then $e_1 \hb{\tr} e_2$.
    \end{enumerate}
    A pair of events $(e_1, e_2)$ in an execution $\tr$ with $e_1 \stricttrord{\tr} e_2$
    is said to be \emph{conflicting} if they do not share a thread
    (i.e. $\ThreadOf{e_1} \neq \ThreadOf{e_2}$) and further,
    there is a common memory location $x \in \vars{\tr}$ such that
    they both access $x$ and not both are read accesses, i.e.,
    $\set{\wt(x)} \subseteq \set{\OpOf{e_1}, \OpOf{e_2}} \subseteq \set{\wt(x), \rd(x)}$. 
    A pair of events $(e_1, e_2)$
    is said to be a happens-before
    data race, or \acrhb-race or simply data race, if
    it is a conflicting pair and 
    $e_1 \not\hb{\tr} e_2$. Note that in such a case, $e_2 \not\hb{\tr} e_1$ as well since we are assuming that in this pair $e_1 \stricttrord{\tr} e_2$.
    An execution $\tr$ is said to have a data race if there is a pair $(e_1, e_2)$
    of events in $\tr$ that is a data race.
}

\subsection{Vector clock algorithm for data race detection}

Dynamic race detectors such as \tsan are based on the 
\fasttrack~\cite{fasttrack} optimization on top of the 
\djitp~\cite{Pozniansky:2003:EOD:966049.781529} algorithm,
which uses \emph{vector clocks} to infer causality.
Since the epoch optimization of \fasttrack is independent of our innovations, we stick to \djitp for simplicity of discussion. Given that our presentation will share some key 
ideas underlying these algorithms, we share some of their details.  At a high level, the \djitp algorithm (\algoref{djitp}) processes events in a streaming fashion, calling the appropriate {\bf handler}
based on the type of the event.
The handlers are designed to achieve two key 
tasks --- (a) compute timestamps of each event
in the execution as a proxy for the \acrhb partial order, and
(b) use these timestamps to check for the presence of data races.
Let us first recall the notion of timestamps used by \djitp.

\myparagraph{\djitp Timestamps}{
    Instead of explicitly constructing the partial order
    (say, by constructing a graph whose vertices are events 
    and whose edges reflect the  $\acrhb$ partial order), 
    the \djitp algorithm implicitly infers the partial order 
    between events in a streaming manner by associating
    each event with a \emph{timestamp}.
    We present here the precise declarative definition of the timestamp
    that underlies this algorithm.
    We fix an execution $\tr$ in the following.
    For an event $e$ of $\tr$, the \emph{local time} of $e$ 
    represents the number of release events that have been performed
    in $\tr$ before $e$ in the same thread as $e$:
    \begin{align}
        \ltimeft^\tr(e) = |\setpred{f}{\exists \lk \cdot \OpOf{f} = \rel(\lk), f \stricttho{\tr} e}| + 1
    \end{align}
    Using this, we can associate with each event the \emph{causal time}
    of an event $e$ as follows; we use the convention that $\max \emptyset = 0$:
    \begin{align}
        \ctimeft^\tr(e) = \lambda t \cdot \max \setpred{\ltimeft^\tr(f)}{\ThreadOf{f} = t, f \hb{\tr} e}
    \end{align}
    That is, $\ctimeft^\tr(e) : \threads{\tr} \to \nats$
    captures the \emph{knowledge} of $e$ about
    other threads, via the \acrhb partial order.
    Indeed, the above notion of timestamps 
    is sufficient to check when a pair of events is in a data race 
    (\propref{ft-timestamps-capture-hb}).
    In the following,
    we use $\cle$ to denote the \emph{pointwise}
    comparison operator defined 
    for two timestamps $T_1, T_2: \threads{\tr} \to \nats$; it is defined as follows:
    \begin{align}
    T_1 \cle T_2 \equiv \forall t.\ T_1(t) \leq T_2(t)
    \end{align}
    \begin{proposition}
    \proplabel{ft-timestamps-capture-hb}
    For an execution $\tr$ events $e_1, e_2 \in \events{\tr}$
    with $\ThreadOf{e_1} \neq \ThreadOf{e_2}$, we have:
    \begin{align*}
        \ctimeft^\tr(e_1)(\ThreadOf{e_1}) \leq \ctimeft^\tr(e_2)(\ThreadOf{e_1}) \quad \text{iff} \quad \ctimeft^\tr(e_1) \cle \ctimeft^\tr(e_2) \quad \text{iff} \quad e_1 \hb{\tr} e_2
    \end{align*}
    \end{proposition}
}


{
\begin{algorithm}[t]
\small
\vspace{-0.2in}
\begin{multicols}{2}
\myfun{\init}{
	\ForEach{$t \in \threads{}$}{
		$\CFT_t \gets \bot[t \mapsto 1]$
	}
	\ForEach{$\lk \in \locks{}$}{
		$\CFT_\lk \gets \bot$
	}
	\ForEach{$x \in \vars{}$}{
		$\WFT_x \gets \bot$ ;
		$\RFT_x \gets \bot$
	}
}

\myhandler{\rdhandler{$t$, $x$}}{
	\lIf{$\WFT_x \not\cle \CFT_t$}{\declare race \linelabel{djitp-check-race-read}} 
	$\RFT_x\gets \RFT_x[t \mapsto \CFT_t(t)]$ \linelabel{djitp-update-read-history}
}

\myhandler{\wthandler{$t$, $x$}}{
	\lIf{$\RFT_x \not\cle \CFT_t$ or $\WFT_x \not\cle \CFT_t$}{\declare race \linelabel{djitp-check-race-write}}
	$\WFT_x\gets \CFT_t$ \linelabel{djitp-update-write-history}
}

\myhandler{\acqhandler{$t$, $\lk$}}{
	$\CFT_t \gets \CFT_t \mx \CFT_\lk$ \linelabel{djitp-join-acq} \;
}

\myhandler{\relhandler{$t$, $\lk$}}{
	$\CFT_\lk \gets \CFT_t$ \linelabel{djitp-copy-rel} \; 
	$\CFT_t \gets \CFT_t[t \mapsto \CFT_t(t)+1]$ \linelabel{djitp-inc-rel}
}

\end{multicols}
\normalsize
\caption{\small Vector clock algorithm for detecting \acrhb-races}
\algolabel{djitp}
\end{algorithm}
}

\myparagraph{Computing Timestamps}{
    Instead of storing the timestamps of all events,
    the algorithm maintains the timestamps of a small,
    dynamically changing, set of events
    and computes the timestamp of each new event using these small set of stored timestamps.
    More precisely, after having processed the prefix $\pi$ of $\tr$,
    it maintains the timestamp of 
    (1) the last event $e_{\pi, t}$ of thread $t$, for each thread $t \in \threads{\tr}$,
    (2) the last release event $e_{\pi, \lk}$ of lock $\lk$, for each lock $\lk \in \locks{\tr}$,
    {\color{black} (3) the last write event $w_{\pi,x}$ on every memory location $x\in\vars{\tr}$, and
    (4) the last read event $r_{\pi,x,t}$ of thread $t$ on $x$, for each thread $t\in \threads{\tr}$ and every memory location $x\in\vars{\tr}$.}.

    For this purpose, the algorithm uses \emph{vector clocks}, 
    which are variables that take values over the space of timestamps.
    Precisely, for each thread $t \in \threads{\tr}$, it maintains
    the vector clock $\CFT_t$ to track the last event $e_{\pi, t}$,
    and for each lock $\lk$, it maintains the vector clock
    $\CFT_\lk$ to track the timestamp of the
    last event $e_{\pi, \lk}$, where $\pi$ is the prefix of the execution seen so far.
    At each release event of lock $\lk$ by thread $t$,
    the algorithm \emph{sends} the timestamp of $e_{\pi, t}$
    to the next event that acquires $\lk$ by copying $\CFT_t$ to $\CFT_\lk$ (\lineref{djitp-copy-rel}),
    and also increments the local component of $\CFT_t$ (\lineref{djitp-inc-rel}).
    At an acquire event of lock $\lk$ by thread $t$, the algorithm
    updates the clock $\CFT_t$ by performing a \emph{join} operation
    with the timestamp of the last release of $\lk$ stored in the clock $\CFT_\lk$ (\lineref{djitp-join-acq}).
    Here, the join ($\mx$) operation computes the pointwise maximum;
    for timestamps $T_1, T_2: \threads{\tr} \to \nats$, the join
    of $T_1$ and $T_2$, we have:
    \begin{align} 
    T_1 \mx T_2 = \lambda t \cdot \max \set{T_1(t), T_2(t)}
    \end{align}
}

\myparagraph{Checking for races}{
    Let us now see how \djitp performs data race detection.
    As before, it stores the timestamps of only a few events.
    In particular, for each memory location $x \in \vars{\tr}$,
    the algorithm maintains a write access history vector clock  $\WFT_x$ that stores 
    the timestamp of the last event $e_{\pi, \wt(x)}$ that writes to 
    $x$ in the prefix $\pi$ seen so far. 
    Likewise, it also maintains the read access history vector clock
    $\RFT_x$ that satisfies $\RFT_x(t) = \ltimeft(e_{\pi, t, \rd(x)})$,
    where $e_{\pi, t, \rd(x)}$ is the last read event of $x$ in thread $t$.
    Observe that the updates in \lineref{djitp-update-read-history} and
    \lineref{djitp-update-write-history} accurately maintain these invariants.
    With access to these clocks, the race check at a read or a write event $e$
    can be performed by checking if an earlier conflicting read or write event
    is unordered with respect to $e$, by comparing the $\WFT_x$ or the $\RFT_x$ 
    clock to the timestamp of $e$ (stored in $\CFT_t$), 
    as in \lineref{djitp-check-race-read} and \lineref{djitp-check-race-write}.
    The correctness guarantee of this algorithm, formalized in \lemref{djitp-soundness},
    states that this algorithm solves the \acrhb-race detection problem. 
    \begin{lemma}
    \lemlabel{djitp-soundness}
    For an execution $\tr$, \algoref{djitp} declares a race iff $\tr$ has an \acrhb-race and runs in time $O(\trsz\numthr)$.
    \end{lemma}
}

\section{The Analysis Problem for Sampling-Based Data Race Detection}
\seclabel{problem}
\noindent
Although dynamic race detection is the go-to technique for automatically finding data races in practice, it still adds a significant overhead to space and running time due to expensive vector clock operations performed at each event. Since large software often induce executions exceeding billions of events, dynamic race detection is typically limited to in house-testing of moderate size software. 

One popular approach to address this limitation is ``sampling''. Roughly, instead of trying to check if any pair of data access events is in a race, in sampling a small subset, say $S$, of events is identified, and race detection is limited to searching for a race involving events in $S$. The hope in sampling is that by limiting the set $S$ in which races are searched, the overhead of dynamic race detection can be reduced. 

Existing sampling-based approaches vary in how the set $S$ --- the events that race detection is limited to --- is identified. Events in $S$ could be randomly sampled from an appropriate distribution; examples of this approach include \literace~\cite{marino2009literace}, \pacer~\cite{pacer-tool}, and \RPT~\cite{RPT2023}. Or events in $S$ can be identified through static analysis, like in \racemob~\cite{racemob}. The different approaches use sophisticated techniques to identify the set $S$ in order to establish the mathematical guarantees that each algorithm provides. Even though prior works on sampling-based dynamic race detection do not abstractly decompose the task as we outline here, they solve two basic problems: (a) the \emph{\samprb} which identifies the sample set $S$, and (b) the \emph{\analprb} that analyzes the trace to check for the existence of a race involving events in $S$. Decoupling sampling-based race detection into these two problems allows one to isolate the challenges of each phase, enabling one to overcome them effectively. In this paper, we will focus on finding efficient algorithmic solutions for the {\analprb}. Success in tackling the {\analprb} will help improve the efficiency of all the sampling-based race detection approaches.

Abstractly, the {\analprb} can be stated as follows: Given a program execution $\tr$ and a subset $S\subseteq \events{\tr}$, determine if there are events $e,e' \in S$ such that  $(e,e')$ is a race in $\tr$. However, this formulation obfuscates a subtle issue --- how is $S$ given? Is it known before $\tr$ is presented? In sampling-based race detection, the {\samprb} and the {\analprb} are not necessarily solved sequentially in stages, but may be solved simultaneously and adaptively --- identification of the set $S$ and its analysis happen together as the program execution is observed. Thus, we change the way we define the {\analprb} subtly to make explicit the fact that the set $S$ can be revealed to the analyzer as the execution is observed and is not known at the very beginning. We will consider program executions where some of the events are ``marked''; these marked events indicate events that belong to the set $S$. For an execution $\tr$ with marked events, the subset of marked events will be denoted as $\mevents{\tr}$.
\begin{problem}[\analprb]
\problabel{race-detection-set}
Given a program execution $\tr$ with marked events, determine if there is a pair of events $e,e' \in \mevents{\tr}$ such that $(e,e')$ is a race in $\tr$.
\end{problem}

The problem formulation we propose encompasses a wide range of sampling-based techniques for reducing the overhead of data race detection proposed in the literature. The set $S$ can represent accesses to specific memory locations (coming from, say specific shared data structures, critical sections, or memory hotspots)~\cite{racemob}. Alternatively, when the focus is not on specific memory locations, the set 
$S$ can be constructed according to a chosen distribution~\cite{pacer-tool,RPT2023,marino2009literace}.
{\color{black}It is also worth noting that while the analysis problem is particularly relevant in sampling, it is not limited to this scenario. For instance, programs that are synchronization-heavy, as considered in prior work~\cite{TSVD}, naturally have a relatively small set of read/write events. In such cases, an efficient solution to the analysis problem can provide significant benefits.}

So what is an efficient algorithm for the {\analprb}? It is clear that since the set of marked events $S = \mevents{\tr}$ of an execution $\tr$ of size $\trsz$, is only known as it is observed, every event in $\tr$ must be processed leading to a running time of $\Omega(\trsz)$. But then what are we trying to optimize? It is useful to compare against the running time of {\djitp} (which is the same as \fasttrack). {\djitp}, in the worst case, performs an expensive vector clock operation for each event, and this is the cost that the sampling-based approaches try to ameliorate by focusing on a small set $S$. An ideal algorithm for the {\analprb} is one that can achieve a running time of $O(\trsz)$ plus $O(|S|)$ vector clock \textcolor{black}{traversals}. This goal seems beyond our reach right now. Taking the cost of a single vector clock \textcolor{black}{traversal} to be $O(\numthr)$, the number of threads, in this paper we present algorithms that solve the {\analprb} in time $O(\trsz) + \widetilde{O}(|S|)O(\numthr)$, where the notation $\widetilde{O}(\cdot)$ hides some factors that depend on parameters of the trace like the number of locks and the number of threads. We will also show that such an algorithm is close to optimal not just in its worst case behavior, but that on any execution, its running time is close to the number of updates any vector clock based algorithm needs to make; we will make this notion precise. Thus, we provide strong theoretical evidence for the effectiveness of our algorithm.

\section{Freshness Timestamp For Solving the {\analprb}}
\seclabel{timestamps} 

In this section we will present an efficient algorithm to solve the {\analprb}. For an execution $\tr$ with marked events $S = \mevents{\tr}$ of length $\trsz$, the algorithm we present in this section will run in $O(\trsz) + \widetilde{O}(|S|)O(\numthr)$, where $O(\numthr)$ is the time taken to perform vector clock operations. However, it will not be our fastest algorithm which will be presented in \secref{orderedlist}. But the algorithm we present here will use one of the key innovations we need, which is a vector timestamp that counts the changes to vector timestamps that track the {\acrhb}-partial order. In \secref{orderedlist}, further improvements to the running time will be achieved through the use of a new data structure for storing vector timestamps.

Before presenting the main algorithm in \secref{aug-clk}, we first begin by modifying {\djitp} to obtain an algorithm that computes the {\acrhb} partial order only among events involving the restricted sample set $S$. Even though the {\djitp} modification presented in \secref{hb-subset} solves the {\analprb}, its asymptotic complexity is the same as {\djitp}. However, it has a few key features. One can show that the vector clocks maintained by each thread in this algorithm only change as many times as the size of the sample set $S$. By tracking changes to these clocks, and using that to decide whether to perform a vector clock operation improves the asymptotic running time leading to the main algorithm of this section. 

\subsection{Tracking {\acrhb}-partial order for a subset of events}
\seclabel{hb-subset}

\rmv{
\begin{figure}[t]
\includegraphics[width=0.7\textwidth, height=0.4\textwidth]{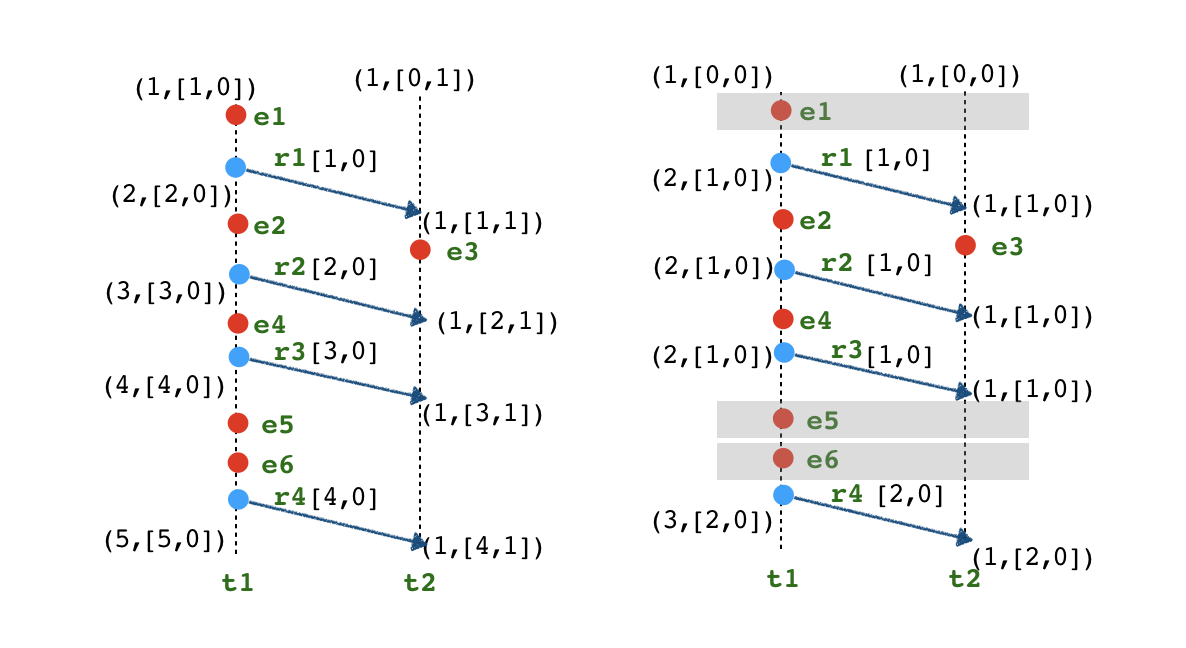}
 \vspace{-0.1in}
\caption{examples of computing time stamp on an execution of two threads. Shaded events are events sampled.
The left figure simulates how \djitp computes the \djitp time stamp and the right figure computes a sampled time stamp. Edges are labelled with the vector time being passed to other thread. The pairs (a,[b,c]) represents the local time and the vector time.}
\figlabel{moti2}
\end{figure}
}

\begin{figure}[h]
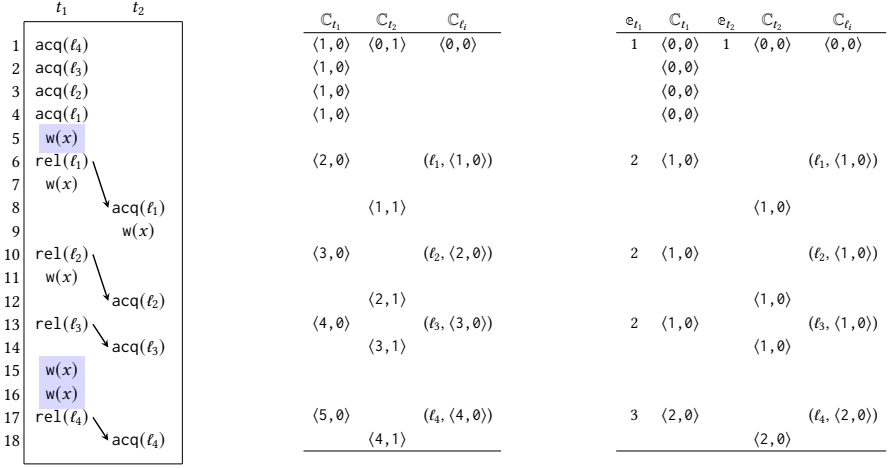

\begin{minipage}{0.25\textwidth}
\scalebox{0.8}{
\execution{2}{
  \figev{1}{\acq(\lk_4)}
  \figev{1}{\acq(\lk_3)}
  \figev{1}{\acq(\lk_2)}
  \figev{1}{\acq(\lk_1)}
  \figmev{1}{\wt(x)}
  \figev{1}{\rel(\lk_1)}
  \figev{1}{\wt(x)}
  \figev{2}{\acq(\lk_1)}
  \figev{2}{\wt(x)}
  \figev{1}{\rel(\lk_2)}
  \figev{1}{\wt(x)}
  \figev{2}{\acq(\lk_2)}
  \figev{1}{\rel(\lk_3)}
  \figev{2}{\acq(\lk_3)}
  \figmev{1}{\wt(x)}
  \figmev{1}{\wt(x)}
  \figev{1}{\rel(\lk_4)}
  \figev{2}{\acq(\lk_4)}
  \orderedge{1}{6}{0.6}{2}{8}{-0.6}
  \orderedge{1}{10}{0.6}{2}{12}{-0.6}
  \orderedge{1}{13}{0.6}{2}{14}{-0.6}
  \orderedge{1}{17}{0.6}{2}{18}{-0.6}
}
}
\end{minipage}
\hspace*{0.2in}
\begin{minipage}{0.25\textwidth}
\scalebox{0.64}{
\renewcommand{\arraystretch}{1.13}
\begin{tabular}{ccc}
$\CFT_{t_1}$	& $\CFT_{t_2}$	& $\CFT_{\lk_i}$	\\ \hline
\vc{1,0}	&	\vc{0,1}			&	\vc{0,0}			    \\ 
\vc{1,0}    &                   &       \\
\vc{1,0}    &                   &       \\
\vc{1,0}    &                   &       \\
\\
\vc{2,0}    &                   &   ($\lk_1$, \vc{1,0})\\
\\
            &   \vc{1,1}        &   \\
\\
\vc{3,0}    &                   &   ($\lk_2$, \vc{2,0})\\
\\
            &   \vc{2,1}        &   \\
\vc{4,0}    &                   &   ($\lk_3$, \vc{3,0})\\
            &   \vc{3,1}        &   \\
\\
\\
\vc{5,0}    &                   &   ($\lk_4$, \vc{4,0})\\
            &   \vc{4,1}        & \\
\hline
\end{tabular}
}
\end{minipage}
\hspace*{0.2in}
\begin{minipage}{0.35\textwidth}
\scalebox{0.64}{
\renewcommand{\arraystretch}{1.13}
\begin{tabular}{ccccc}
$\epch_{t_1}$   & $\CFT_{t_1}$      & $\epch_{t_2}$     &   $\CFT_{t_2}$	& $\CFT_{\lk_i}$	\\ \hline
1           &   \vc{0,0}	    &	1           &     \vc{0,0}   	&	\vc{0,0}	  \\ 
            &       \vc{0,0}    &               &                   &  \\
            &       \vc{0,0}    &               &                   &     \\
            &       \vc{0,0}    &               &                   &   \\
\\
2           &       \vc{1,0}    &               &                   & ($\lk_1$, \vc{1,0})\\
\\
            &                   &               &   \vc{1,0}        &   \\
\\
2           &       \vc{1,0}    &               &                   & ($\lk_2$, \vc{1,0})\\
\\
            &                   &               &   \vc{1,0}        &   \\
2           &       \vc{1,0}    &               &                   &   ($\lk_3$, \vc{1,0})\\
            &                   &               &   \vc{1,0}        &   \\
\\
\\
3           &       \vc{2,0}    &               &                   &   ($\lk_4$, \vc{2,0})\\
            &                   &               &   \vc{2,0}        & \\
\hline
\end{tabular}
}
\end{minipage}
\caption{Example execution of a two threaded program shown on the left. Marked events that form the set $S$ are shown shaded in \textcolor{blue!45!white}{light blue}. Arrows indicate that information is communicated from a release to the next acquire. Vector clocks maintained by  {\djitp}( \fasttrack) are shown in the table in the middle. Columns 1 and 2 of the table show the clocks of threads $t_1$ and $t_2$, respectively. Column 3 of the table shows the clock of locks $\lk_i$; to save space they are combined into one column. An entry ($\lk$, \vc{a,b}) in this column means that $\CFT_{\lk}$ has value \vc{a,b} at that step. The table on the right shows the values of vector clocks maintained by \algoref{algo2}. Column 1 now records the local time of $t_1$, column 2 the vector clock of $t_1$, column 3 the local time of $t_2$, column 4 the vector clock of $t_2$, and column 5 shows the clock of lock $\lk_i$.}
\figlabel{moti2}
\end{figure}

The {\analprb} requires races to be detected only among a subset of events. Therefore, the full {\acrhb} partial order between all pairs of events does not need to be computed in order to solve it. For example, consider the example (partial) execution shown on the left in \figref{moti2}. We will use $e_i$ to denote the $i$th event listed in the execution. So for example, as per this notation, $e_7 = \ev{t_1}{\wt(x)}$ and $e_9 = \ev{t_2}{\wt(x)}$. The marked events whose races we wish to track are shown shaded and so in this example, our sample set $S$ (within this partial execution) contains three events $\{e_5,e_{15},e_{16}\}$ and there may be a potentially conflicting event in $S$ that occurs outside the partial execution. Since we wish to only find out if $e_5$, $e_{15}$, and $e_{16}$ are in race with something appearing later in the execution, we do not need to track the {\acrhb}-order between $e_7$ and $e_9$ for example; the race between $e_7$ and $e_9$ does not matter when solving the {\analprb} since neither $e_7$ nor $e_9$ are in the set $S$. In this section, we present a modification of {\djitp} that accomplishes this goal. We begin by introducing some notation for the relation we need to track to solve the {\analprb}.

\begin{definition}[Sampling Partial Order]
\deflabel{samp-po}
    For an execution $\tr$ and a set of marked events $S = \mevents{\tr}$, define $\hb{(\tr,S)}=\{(e_1,e_2)|e_1\in S, e_1\hb{\tr}e_2\}$.
\end{definition}
The name ``sampling partial order'' is really a misnomer. The relation $\hb{(\tr,S)}$ is not a partial order --- it is not even reflexive.
But we will use that name and hope the reader is not too bothered by it. 
To illustrate the definition, in the example execution from \figref{moti2}, $\{(e_5,e_5),(e_5,e_9),(e_5,e_7)\} \subseteq \hb{(\tr,S)}$ but $\{(e_7,e_7),(e_7,e_{11}),(e_{11},e_{14}) \subseteq \hb{\tr} \setminus \hb{(\tr,S)}$ since neither $e_7$ nor $e_{11}$ are in $S$. 
Observe that the definition of the partial order is stronger than necessary for solving the Analysis problem — it not only orders events within $S$,  but also specifies whether every event is \acrhb-ordered after some event in $S$. We note that in a single-pass algorithm, where future events are unknown, computing this set is almost certainly required. In the following sections, we will also demonstrate that the number of vector clock operations required to compute this set is bounded by the size of $S$, which is also the best one can hope for when solving the the Analysis problem. 

\myparagraph{Timestamps to track the sampling partial order}{
Let us develop the notion of a timestamp that will allow us to track the sampling partial order $\hb{(\sigma,S)}$. {\djitp} tracks the {\acrhb}-partial order by assigning to each event a \emph{local time} $\ltimeft^\tr(e)$ that records the number of releases that have been performed by the thread of the event $e$. This is because two events $e_1$ and $e_2$ performed by the same thread (say) $t$ that occur between consecutive releases of thread $t$ are ``equivalent'' with respect to {\acrhb} from the viewpoint of other threads. In other words, for any event $e$ with $\ThreadOf{e} \neq t$, $e_1 \hb{\tr} e$ iff $e_2 \hb{\tr} e$. In addition, by sending messages through locks at releases and receiving messages through locks at acquires, each thread $t$ maintains a vector timestamp that tracks the local time of events in other threads that are $\hb{}$ ordered before the latest event of $t$. 

The table in the middle in \figref{moti2} shows the run of {\djitp} on the execution shown on the left in \figref{moti2}. Column 1 shows the vector clock time of thread $t_1$, and Column 2 shows the vector clock time of thread $t_2$. Times in these columns are only shown at steps when they are updated due to an acquire or after a release. Column 3 shows the vector clock time associated with the locks $\lk_1,\lk_2,\lk_3$, and $\lk_4$ in a common column. Entries of the form ($\lk$, \vc{a,b}) in column 3 indicate that the clock associated with $\lk$, i.e. $\CFT_{\lk}$, was updated to value \vc{a,b} at that step.  The clocks of threads $t_1$ and $t_2$ start at values \vc{1,0} and \vc{0,1} respectively, while the clocks of all the locks are \vc{0,0}. The local clock of $t_1$ is incremented after every release. In particular, this ensures that the timestamp of event $e_7$ is \vc{2,0}, while that of event $e_{11}$ is \vc{3,0}. These two events must get different timestamps because they are not ``equivalent'' from the standpoint of {\acrhb}-partial order --- $e_7 \hb{\tr} e_{12}$ but $e_{11} \not\hb{\tr} e_{12}$. On the other hand, $e_{15}$ and $e_{16}$ get the same time because they are equivalent with respect to $\hb{}$. 

When the goal is to track $\hb{(\tr,S)}$, notice that $e_7$ and $e_{11}$, which received different timestamps in {\djitp}, can now in fact be treated ``equivalent'' with respect to $\hb{(\tr,S)}$. This is because neither $e_7$ nor $e_{11}$ are in $S$, and so $(e_7,e_{12}) \not\in \hb{(\tr,S)}$ and $(e_{11},e_{12}) \not\in \hb{(\tr,S)}$. Thus, $e_7$ and $e_{11}$ need not be distinguished when solving the {\analprb}. This allows us to unlock a new optimization --- the new local time at threads only need to be incremented at certain releases and not \emph{all} releases as done in {\djitp}. The releases that increment local time are those that are the first one after some event in $S$. In addition, at a release performed by thread $t$, the algorithm will send the time of the \emph{last event in $S$} of thread $t$ as opposed to the time of the \emph{last event} of thread $t$; when tracking races between all events this distinction does not exist. 

Let us see how this works on the example in \figref{moti2}. The rightmost table tracks the various timestamps that the new algorithm will keep. The local time at threads $t_1$ and $t_2$ are listed explicitly in columns 1 and 3. Threads also maintain a vector clock $\CFT_{t_1}$ and $\CFT_{t_2}$ (columns 2 and 4). The local component of $\CFT_{t_1}$ (and $\CFT_{t_2}$) does not store the local time but rather the local time of the last event of the thread that is also in $S$. This distinction is important to maintain because at a release the algorithm sends this time rather than the current local time as there may have been no new events in $S$. Finally, column 5 records the clocks of the locks $\lk_1,\lk_2,\lk_3$, and $\lk_4$; again, a particular entry in this column will be a pair where the first component of the pair indicates which lock's clock is being updated. The local times start at 1, and the vector clocks at \vc{0,0}. At the first release event $e_6$, we send the current vector clock with the time of last event in $S$ (which is $e_5$) to lock $\lk_1$. So the clock of $\lk_1$ gets updated to \vc{1,0}. We update the local clock of $t_1$ as well since $e_6$ is the first release after an event in $S$. In contrast, we will not update the local time at release event $e_{10}$ because the events $e_7$ and $e_{12}$ are not distinguishable with respect to $\hb{(\tr,S)}$. Also note that at event $e_{10}$, the time sent to $\lk_2$ is \vc{1,0} whose $t_1$-th component is the local time of $e_5$ ($1$), the last event in $S$, and not the local time of $e_7$, which is $2$. Similarly, the release event $e_{13}$ will not change the local time, but the local time will be changed after $e_{17}$ because it comes after events $e_{15}$ and $e_{16}$ which belong to $S$. \algoref{algo2} shows the full algorithm which will be discussed in more detail after we introduce some new definitions.

We now give a definition of the new timestamp used in our algorithm for computing $\hb{(\tr,S)}$. For the new timestamp, only certain releases will update the local time of threads. Let us define which ones those are.
\newcommand{\SRel}{\mathsf{RelAfter}_S\xspace}
\begin{align}
\SRel = \setpred{f}{\exists e \in S.\ f \text{ is the first release event after } e \text{ with } \ThreadOf{e} = \ThreadOf{f}}
\end{align}
The local time of an event then counts the number of such releases that have been performed by the thread. Formally,
\begin{align}
\ltimesampling^{(\tr,S)}(e) = |\setpred{f}{f\in \SRel, f \stricttho{\tr} e}|+1
\end{align}
As in the case of {\djitp}, we can use the local time to define a vector timestamp for each event. 
\begin{align}
\ctimesampling^{(\tr,S)}(e) = \lambda t \cdot \max \setpred{\ltimesampling^{(\tr,S)}(f)}{f\in S, \ThreadOf{f} = t, f \hb{\tr} e}
\end{align}
Tracking the \textit{sampling timestamp} $\ctimesampling^{(\tr,S)}$ allows one to compute the relation $\hb{(\tr,S)}$ as shown by the following proposition.
\begin{proposition}
    \proplabel{ft-samplestamps-capture-hb}
    For an execution $\tr$, a set of sampled events $S$,  events $e_1\in S$ and $e_2\in \events{\tr}$
    with $\ThreadOf{e_1} \neq \ThreadOf{e_2}$, we have:
    \begin{align*}
        \ctimesampling^{(\tr,S)}(e_1)(\ThreadOf{e_1}) \leq \ctimesampling^{(\tr,S)}(e_2)(\ThreadOf{e_1}) \quad \text{iff} \quad \ctimesampling^{(\tr,S)}(e_1) \cle \ctimesampling^{(\tr,S)}(e_2) \quad \text{iff} \quad e_1 \hb{\tr} e_2
    \end{align*}
    \end{proposition}

Before presenting an algorithm to compute $\ctimesampling^{(\tr,S)}$, we present an important property about it. Observe that $|\SRel| \leq |S|$. Therefore, it follows that
$\sum_{t\in\threads{\tr}} \ctimesampling^{(\tr,S)}(e)(t) \leq |S|
$
for every event $e \in \events{\tr}$. This will be exploited in \secref{aug-clk} to present an improved algorithm for solving the {\analprb}. 


{
\begin{algorithm}[t]
\small
\vspace{-0.2in}
\begin{multicols}{2}
\myfun{\init}{
	\ForEach{$t \in \threads{}$}{
		$\CFT_t \gets \bot$;
            $\epch_t \gets 1$ \;
	}
	\ForEach{$\lk \in \locks{}$}{
		$\CFT_\lk \gets \bot$
	}
	\ForEach{$x \in \vars{}$}{
		$\WFT_x \gets \bot$ ;
		$\RFT_x \gets \bot$
	}
}

\myhandler{\rdhandler{$t$, $x$}}{
        \lIf{the event is not sampled}{
        skip;
        }
	\lIf{$\WFT_x \not\cle \CFT_t$}{\declare race \linelabel{algo2-check-race-read}} 
	$\RFT_x\gets \RFT_x[t \mapsto \epch_t]$ \linelabel{algo2-update-read-history}
}

\myhandler{\wthandler{$t$, $x$}}{
        \lIf{the event is not sampled}{
        skip;
        }
	\lIf{$\RFT_x \not\cle \CFT_t$ or $\WFT_x \not\cle \CFT_t$}{\declare race \linelabel{algo2-check-race-write}}
	$\WFT_x\gets \CFT_t[t \mapsto \epch_t]$ \linelabel{algo2-update-write-history}
}

\myhandler{\acqhandler{$t$, $\lk$}}{
	$\CFT_t \gets \CFT_t \mx \CFT_\lk$ \linelabel{uclock-join-acq} \;
}

\myhandler{\relhandler{$t$, $\lk$}}{
 
 		\If{$\exists e \in S$, with $\ThreadOf{e} = t$, since last release in $t$}{
			$\CFT_t \gets \CFT_t[t \mapsto \epch_t]$ \;

                $\epch_t\gets \epch_t+1$
		}
 
        $\CFT_\lk \gets \CFT_t$ \linelabel{uclock-copy-rel} \;
        
	\linelabel{uclock-inc-rel-sample}
}


\end{multicols}
\caption{\small Vector clock algorithm for computing the sampling timestamp}
\normalsize
\algolabel{algo2}
\end{algorithm}
}
\myparagraph{Algorithm to compute the sampling partial order}{
The algorithm that computes $\hb{(\tr,S)}$ and solves the {\analprb} using the timestamp $\ctimesampling^{(\tr,S)}$ is presented in \algoref{algo2}. The timestamp $\ctimesampling^{(\tr,S)}$ can be computed in a manner similar to how {\djitp} computes $\ctimeft^\tr$. Roughly, the two main differences between \algoref{djitp} and \algoref{algo2} are that in \algoref{algo2} (a) the local time is only updated at a release that is in $\SRel$, and (b) the race checks are only done on events in $S$. This requires making only a few modifications to \algoref{djitp} to get \algoref{algo2}.

In addition to the vector clock $\CSmp_t$, each thread $t$ maintains its local clock in an epoch $\epch_t$. This is because $\CSmp_t(t)$ by definition only stores the local time of the last event of $t$ that also belongs to the set $S$. Thus, the local time of the current event needs to be stored separately as $\epch_t$. The most significant change to the code is in the handler for release. If $t$ has performed an event in $S$ since the last release, then we first update $\CSmp_t(t)$ with $\epch_t$ and then increment the local epoch $\epch_t$. 
The modifications to the read/write handlers are more straightforward. If an event is not in S, it can be entirely disregarded. Consequently, the total number of vector clock operations across all read/write handlers is at most |S|. In the remainder of this paper, the proposed algorithms will incorporate the same read/write handlers, and for brevity, their detailed presentation will be omitted.
}

\begin{lemma}
\lemlabel{algo2-correctness}
For an execution $\tr$, a set of sampled events $S$,  \algoref{algo2} runs in time $O(\trsz\numthr)$ and declares a race on event $e$ if and only if $e\in S$ and 
there exists $e'\in S$ that such $(e',e)$ is an \acrhb-race in $\tr$. 
\end{lemma}

\algoref{algo2} has the same running time as {\djitp} because it still performs a vector clock operation for every release and acquire event. However, as $\sum_{t\in\threads{\tr}} \ctimesampling^{(\tr,S)}(e)(t) \leq |S|$ for any event $e$, the vector clocks $\CFT_\lk$ and $\CFT_t$ are updated at most $|S|$ times. This is because the vector clocks increase monotonically in this algorithm. This will be exploited to get further improvements.

\subsection{An Efficient Algorithm for the {\analprb}}
\seclabel{aug-clk}

Let us look at the example in \figref{moti2}. The right table shows the states of \algoref{algo2} for the execution trace shown on the left. Thread $t_2$ receives thread $t_1$'s vector timestamp four times through acquire events, namely $e_8$, $e_{12}$, $e_{14}$, and $e_{18}$. But it receives \textit{new information} at only two of these events --- $e_8$ and $e_{18}$. This is because the timestamps that $t_1$ sends through its release events $e_6$, $e_{10}$, and $e_{13}$, which are read by $t_2$ at $e_8$, $e_{10}$ and $e_{14}$, respectively, are the same. In \algoref{algo2}, $t_2$ performs a vector clock join at each of the events $e_8$, $e_{10}$, and $e_{14}$, even though no new information is obtained from two of them. Can we somehow avoid performing a vector clock operation when no new information is going to be received? Before answering the question of how to improve the algorithm, it is worth asking whether this is even worth the effort. Will a mechanism to avoid performing vector clock operations at these acquires lead to an improvement in the asymptotic running time? For this let us recall our observation at the end of the previous section which says that throughout a run of \algoref{algo2}, none of the vector clocks change more than $|S|$ times. But an execution $\tr$ of length $\trsz$ can have $O(\trsz)$ many releases and acquires, and if $|S| \ll \trsz$ then many of these releases are sending the same information. 

Thus, a mechanism that allows one to be aware of the ``freshness'' of a message is beneficial for avoiding redundant vector clock operations on acquires, which will lead to an improvement in the asymptotic running time. To accomplish this goal, our new algorithm will track the freshness of information by counting the number of updates on a vector clock. Let us define a new timestamp that accomplishes this goal.

\rmv{
For the execution on the left of \figref{moti2}, the freshness of $t_1$'s vector clock at the end is $4$ and similarly, for the execution on the right, the freshness of $t_1$'s vector clock at the end is $1$ while that of $t_2$'s vector clock is $2$. 
At the high level, when $t_2$ learns about $t_1$'s information via $r_1$, it also remembers that the freshness of what it knows of $t_1$ is $0$, then when in $r_2$ and $r_3$, it will realize that the freshness of the information sent by $t_1$ is still 0, which means that no change have been applied so it can safely ignore the operation. Let's now summarize this intuition with a mathematical expression that defines another time stamp.
}

\myparagraph{The Freshness Timestamp}{
Let us first formalize the quantity
$\s{diff}(e_i,e_j)$ that captures the number of entries where the vector timestamps of events $e_i$ and $e_j$ differ:
 \begin{align}
 \s{diff}(e_i,e_j)=|\{t|\ctimesampling^{(\tr,S)}(e_i)
 (t)\neq\ctimesampling^{(\tr,S)}(e_j)(t)\}|
 \end{align}
We can now define $\text{VT}(e)$ to capture how much the timestamp of $e$ has evolved in its history:
\begin{align}
 \s{VT}(e)=\sum_{e'\stricttho{\tr}e}\s{diff}(e',\text{next}(e'))
 \end{align}
where $\text{next}(e')$ denotes the next event after $e'$ in the same thread. Thus, $\s{VT}(e)$ counts the number of updates to components of the clock $\CFT_{\ThreadOf{e}}$ throughout the computation until event $e$. We will use $\s{U}$ to define a new vector timestamp that measures an event's knowledge of how many times the $\CFT_t$ clock changed for each thread $t$:
\begin{align}
        \s{U}(e) = \lambda t \cdot \max \setpred{\s{VT}(f)}{f\in S,\ThreadOf{f} = t, f \hb{\tr} e}
\end{align}

}
\begin{proposition}
    \proplabel{uclock-fresh}
    For an execution $\tr$, a set of sampled events $S$, events $e_1, e_2\in \events{\tr}$
    with $t_1 = \ThreadOf{e_1} \neq \ThreadOf{e_2}$, we have:
    \begin{align*}
     \text{if }\quad \s{U}(e_1)(t_1)\leq \s{U}(e_2)(t_1), \text{ then }\ \ctimesampling^{(\tr,S)}(e_1) \cle \ctimesampling^{(\tr,S)}(e_2) 
    \end{align*}
\end{proposition}

The ``freshness'' timestamp $\s{U}$ of an event $e$ can not only help determine the ordering of the $\ctimesampling^{(\tr,S)}$ timestamp of events, but also indicates the degree to which one is ahead of the other.
\begin{proposition}
    \proplabel{uclock-numentry}
    For an execution $\tr$, a set of sampled events $S$, events $e_1, e_2\in \events{\tr}$, let  $k=\s{U}(e_1)(\ThreadOf{e_1})-\s{U}(e_2)((\ThreadOf{e_1}))$. The number of threads $t$ such that $\ctimesampling^{(\tr,S)}(e_1)(t)>\ctimesampling^{(\tr,S)}(e_2)(t)$ is at most $\min(\numthr, \max(k,0))$.

\end{proposition}
}

We here remark that \propref{uclock-fresh} and \propref{uclock-numentry} can be extended to the cases where $\s{U}(e_2)$ is replaced
with any $\s{U'}$ such that $\s{U'}\cle \s{U}(e_2)$. The correctness of the key results presented in the paper hinges on this observation.


{
\begin{algorithm}[t]
\small
\vspace{-0.2in}
\begin{multicols}{2}
\myfun{\init}{
	\ForEach{$t \in \threads{}$}{
		$\CSmp_t \gets \bot$ ;
		$\CUpd_t \gets \bot$ ;
            $\epch_t\gets 1$
	}
	\ForEach{$\lk \in \locks{}$}{
		$\CSmp_\lk \gets \bot$ ;
		$\CUpd_\lk \gets \bot$ ;
		$\lastrelthr_\lk \gets \nil$ \;
	}
}

\myhandler{\acqhandler{$t$, $\lk$}}{
	\If{$\CUpd_\lk(\lastrelthr_\lk) > \CUpd_t(\lastrelthr_\lk)$}{\linelabel{acq-guard-chk}
            
		$\CUpd_t \gets \CUpd_t \mx \CUpd_\lk$ \;\linelabel{u-join-acq}
            \ForEach{$t^*\in \threads{\tr}$}{\linelabel{c-join-acq-loop}
                \If{$\CSmp_\lk[t^*] > \CSmp_t[t^*]$}{
                $\CSmp_t[t^*]\gets\CSmp_\lk[t^*]$\;\linelabel{c-join-acq}
                $\CUpd_t \gets \CUpd_t[t \mapsto \CUpd_t(t)+1]$\linelabel{c-join-u-upt}
                }
            }
	
	}
}

\myhandler{\relhandler{$t$, $\lk$}}{
     $\lastrelthr_\lk \gets t$ \;
		\If{$\exists e \in S$, with $\ThreadOf{e} = t$, since last release in $t$}{\linelabel{u-acq-skip}
			$\CSmp_t \gets \CSmp_t[t \mapsto \epch_t]$ \;
			$\CUpd_t \gets \CUpd_t[t \mapsto             \CUpd_t(t)+1]$\;
                $\epch_t\gets \epch_t+1$
		}	

	\If{$\CUpd_t(t) \neq \CUpd_\lk(t)$}{\linelabel{rel-skip}
		$\CSmp_\lk \gets \CSmp_t$ \;
		$\CUpd_\lk \gets \CUpd_t$ \;

	}

}

\end{multicols}
\caption{\small Vector clock algorithm partially computing the VT timestamp}
\normalsize
\algolabel{smp-upd}
\end{algorithm}
}

\myparagraph{Using Freshness Timestamps to solve the {\analprb}}{
\propref{uclock-fresh} indicates that the freshness of the sampling timestamps of two events can be compared  by checking two scalars if the freshness timestamp is also properly maintained. In \algoref{algo2}, a lock carries the sampling timestamp of the latest sampled event from the thread that last released it.  When a thread $t$ acquires the lock, it reads this information and updates its own timestamp for future events. On an acquire event, if the timestamp of the previous event is fresher than the timestamp carried by the lock, then the acquire event can be "omitted". Similarly a lock's timestamp need not be updated by a thread on a release event if the thread's timestamp has not changed since the lock was acquired.

\algoref{smp-upd} is the algorithm that results from modifying \algoref{algo2} using the ideas outlined. Each thread and lock now has a ``$\CUpd$'' vector clock storing $\s{U}$ timestamps, in addition to a ``$\CSmp$'' vector clock storing $\ctimesampling^{(\tr,S)}$ timestamps. Further, at each lock, we also store the thread ID of the thread that performed the last release of the lock --- this is the variable $\lastrelthr_\lk$. Let us start by examining the acquire handler. If $t'$ is the last thread that released the lock and $\CUpd_t(t') \geq \CUpd_{\lk}(t')$ then based on \propref{uclock-fresh}, we can conclude that $\CSmp_\lk$ does not contain any new information. In this case the acquire handler needs to do nothing. On the other hand, if $\CUpd_t(t') < \CUpd_{\lk}(t')$ then the acquire handler performs a join to update both $\CUpd_t$ (\lineref{u-join-acq}) and $\CSmp_t$ (for loop on \lineref{c-join-acq-loop}). The update of $\CSmp_t$ requires tracking the number of components that changed so that $\CUpd_t(t)$ can be updated correctly (\lineref{c-join-u-upt}). The release handler is very similar to the release handler of \algoref{algo2} except that it needs to (a) update $\lastrelthr_\lk$ to reflect the thread ID of the releasing thread, (b) if this is the first release after an event in $S$, increment $\CUpd_t(t)$ in addition to updating the clock $\CSmp_t$ and incrementing the local epoch $e_t$, and (c) update both $\CSmp_\lk$ and $\CUpd_\lk$, if the thread has new information (if check on \lineref{rel-skip}).
\algoref{algo2}'s run on the trace from \figref{moti2} is shown in \figref{moti3}. The table in  \figref{moti3} extends the right-hand table from  \figref{moti2} by adding the $\CUpd$ vector clocks, which stores the freshness timestamps (columns 3, 6, and 9), and the $\lastrelthr$ scalar, which records the last thread to release each lock (column 7). Note that in event $e_8$, a join operation is performed because $\CUpd_{t_2}(1)<\CUpd_{\lk_1}(1)$ holds prior to $e_8$. However, $e_{12}$ and $e_{14}$ are successfully skipped because $\CUpd_{t_2}(t_1)=\CUpd_{\lk_2}(t_1)$ and $\CUpd_{t_2}(t_1)=\CUpd_{\lk_3}(t_1)$ prior to these events respectively. 
\begin{figure}[h]
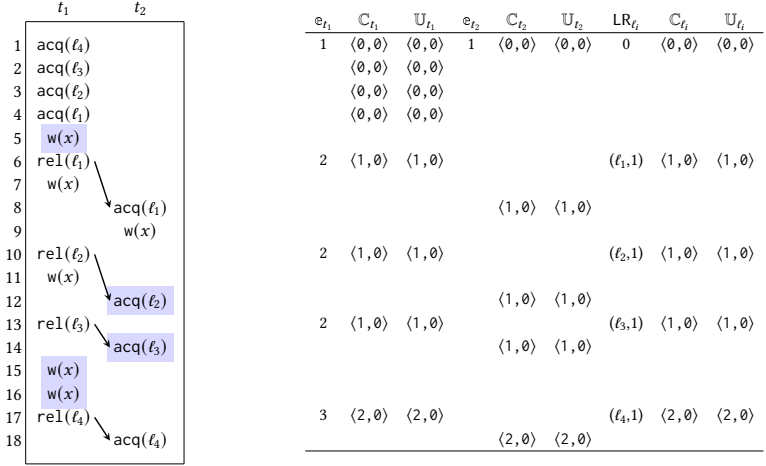

\begin{minipage}{0.25\textwidth}
\scalebox{0.8}{
\execution{2}{
  \figev{1}{\acq(\lk_4)}
  \figev{1}{\acq(\lk_3)}
  \figev{1}{\acq(\lk_2)}
  \figev{1}{\acq(\lk_1)}
  \figmev{1}{\wt(x)}
  \figev{1}{\rel(\lk_1)}
  \figev{1}{\wt(x)}
  \figev{2}{\acq(\lk_1)}
  \figev{2}{\wt(x)}
  \figev{1}{\rel(\lk_2)}
  \figev{1}{\wt(x)}
  \figmev{2}{\acq(\lk_2)}
  \figev{1}{\rel(\lk_3)}
  \figmev{2}{\acq(\lk_3)}
  \figmev{1}{\wt(x)}
  \figmev{1}{\wt(x)}
  \figev{1}{\rel(\lk_4)}
  \figev{2}{\acq(\lk_4)}
  \orderedge{1}{6}{0.6}{2}{8}{-0.6}
  \orderedge{1}{10}{0.6}{2}{12}{-0.6}
  \orderedge{1}{13}{0.6}{2}{14}{-0.6}
  \orderedge{1}{17}{0.6}{2}{18}{-0.6}
}
}
\end{minipage}
\hspace*{0.2in}
\begin{minipage}{0.6\textwidth}
\scalebox{0.64}{
\renewcommand{\arraystretch}{1.13}
\begin{tabular}{ccccccccc}
$\epch_{t_1}$   & $\CFT_{t_1}$   &$\CUpd_{t_1}$    & $\epch_{t_2}$     &   $\CFT_{t_2}$	&  $\CUpd_{t_2}$   & $\lastrelthr_{\lk_i}$ & $\CFT_{\lk_i}$ & $\CUpd_{\lk_i}$  	\\ \hline

1           &  \vc{0,0}	     & \vc{0,0}        &	1          &   \vc{0,0}     &   \vc{0,0}       & 0           &	 \vc{0,0}     & \vc{0,0}  \\ 
            &  \vc{0,0}      & \vc{0,0}        &               &                &                  &             &                &            \\
            &  \vc{0,0}      & \vc{0,0}        &               &                &                  &             &                &            \\
            &  \vc{0,0}      & \vc{0,0}        &               &                &                  &             &                &            \\
\\

2           &  \vc{1,0}      & \vc{1,0}        &               &                &                  &  ($\lk_1$,1)&   \vc{1,0}    &  \vc{1,0}   \\
\\
            &                &                 &               &    \vc{1,0}    &     \vc{1,0}     &             &                &           
\\
\\
2           &    \vc{1,0}    &  \vc{1,0}       &               &                &                  &  ($\lk_2$,1)&    \vc{1,0}    &  \vc{1,0}   \\
\\
            &                &                 &               &    \vc{1,0}    &     \vc{1,0}     &             &                &        \\    

2           &    \vc{1,0}    &  \vc{1,0}       &               &                &                  &  ($\lk_3$,1)&    \vc{1,0}    &  \vc{1,0}   \\
            &                &                 &               &    \vc{1,0}    &     \vc{1,0}     &             &                &        \\
\\
\\
3           &    \vc{2,0}    &   \vc{2,0}      &               &                &                  &  ($\lk_4$,1)&    \vc{2,0}    &  \vc{2,0}   \\
            &                &                 &               &    \vc{2,0}    &    \vc{2,0}      &             &                &        \\

\hline
\end{tabular} 
}
\end{minipage}
\caption{The same execution as \figref{moti2} The table on the right shows the values of vector clocks maintained by \algoref{smp-upd}. Acquires which can be skipped are shown shaded in \textcolor{blue!45!white}{light blue}.}
\figlabel{moti3}
\end{figure}

\begin{lemma}
\lemlabel{algo3-correctness}
For an execution $\tr$, a set of sampled events $S$, \algoref{smp-upd} declares a race on an event $e$ if and only if \algoref{algo2} declares a race on the same event and \algoref{smp-upd} runs in time $O(\trsz) + O(|S|\numthr(\numthr+\numlk))O(\numthr)$ and performs $O(|S|\numthr(\numthr+\numlk))$ many vector clock operations.
\end{lemma}

Due to space limitations, the proof of \lemref{algo3-correctness} has been moved to the appendix. We note that the proof relies on two key observations: (a) the sampling timestamp is bounded by $|S|$ which also implies that the freshness timestamp is bounded by $|S|\numthr$, and (b) the vector clock held by threads and locks grow monotonically because every release always follows an acquire by the same thread. This guarantees that, for each lock, the number of attempts to update any timestamp entry remains bounded.
Although \algoref{smp-upd} is asymptotically faster than \djitp when the sampled set of events $S$ is small, it still faces certain limitations. First, extending the algorithm to handle generic acquire and release operations—where releases do not necessarily follow acquires—would cause the time complexity to revert to that of \djitp, as the vector clocks held by locks would no longer grow monotonically. Second, readers may notice that the definition of the $\s{VT}$ timestamp is unnecessarily complex for the theoretical results achieved. A scalar distinguishing releases that transmit different information would suffice. This is because \algoref{smp-upd} does not yet fully exploit the power of the timestamp, a point we will illustrate with a simple example in the next section, showing how the running time can be further optimized.
Finally, the number of vector clock operations scales linearly with the number of synchronization objects. This becomes particularly significant when considering other synchronization mechanisms such as volatile and atomic variables, barriers, and wait operations, where the number of such synchronization objects can far exceed the number of threads. 
In the next section, we will tackle each of these challenges with a surprisingly simple solution.


\section{A nearly optimal algorithm for solving the analysis problem}
\seclabel{orderedlist}
\begin{figure}[h]
\begin{minipage}{0.25\textwidth}
\scalebox{1.0}{
\executionnonumber{2}{
  \figevnonumber{1}{\ldots}
  \figevnonumber{1}{\rel(\lk)}
  \figevnonumber{1.9}{\acq(\lk)}
  \figevnonumber{1}{\ldots}
}
}
\end{minipage}
\hspace*{0.2in}
\begin{minipage}{0.6\textwidth}
\scalebox{0.8}{
\renewcommand{\arraystretch}{1.13}
\begin{tabular}{cccc}
 $\CFT_{t_1}$   &$\CUpd_{t_1}$       &   $\CFT_{t_2}$	&  $\CUpd_{t_2}$  	\\ \hline

  \vc{9,6,3,0,1,0}	     & \vc{15,12,4,0,1,0}               &   \vc{8,18,3,0,1,0}     &   \vc{14,22,3,0,1,0}     \\ 

  \vc{9,6,3,0,1,0}	     & \vc{15,12,4,0,1,0}               &   \vc{9,18,3,0,1,0}     &   \vc{15,22,3,0,1,0}     \\ 

\hline
\end{tabular} 
}
\end{minipage}
\caption{The figure on the left shows a pair of release and acquire of the same lock done by two threads in an execution of a program with 6 threads. The right table shows the the vector clocks \algoref{smp-upd} maintains for the two threads. }
\figlabel{moti4}
\end{figure}

Let us begin the discussion by examining the simple example illustrated in \figref{moti4}. The figure on the left depicts an acquire operation on a lock by thread $t_2$, which follows a preceding release of the lock by thread $t_1$. The chart on the right presents the vector clocks that \algoref{smp-upd} would maintain for both threads prior to these respective events.
The acquire operation cannot be omitted, as indicated by the freshness timestamp, where $\CUpd_{t_1}(t_1) > \CUpd_{t_2}(t_1)$. It is important to recall that by \propref{uclock-numentry}, the freshness timestamp not only identifies which vector clock is more up-to-date but also indicates the degree to which one is ahead of the other. Given that $\CUpd_{t_1}(t_1) - \CUpd_{t_2}(t_1)=1$, we can infer that there is at most one entry in $\CFT_{t_1}$ that $\CFT_{t_2}$ is unaware of. If we could efficiently determine which entry this is, we could avoid the $O(\numthr)$ vector clock join operation. In this example, the relevant entry would be the first one in $\CFT_{t_1}$. Our proposed solution is to extend the vector clock into a new data structure that also tracks the order in which each entry is updated.

\myparagraph{Ordered lists}{
Let us fix an execution $\tr$ with $\numthr$ threads. An \emph{ordered list} is a data structure that stores a vector timestamp. It consists of the following.
\begin{enumerate}
    \item A doubly linked list $l$ of length $\numthr$ with nodes of the form $(\text{tid},\text{time})\in\threads{\tr} \times \nats$. Every node $u$ in $l$ other than the tail and the head, has a unique predecessor and successor. For $u=(\text{tid},\text{time})$, let $u.\text{tid}=\text{tid}$ and $u.\text{time}=\text{time}$. For every thread $t$, there is a unique node $u$ with $u.\text{tid}=t$.    
    \item A thread map $\s{ThrMap}: \threads{\tr} \to \text{nodes}(l)$ that maps every thread to the unique node in $l$ with the same thread id. That is, for every $t$, $\s{ThrMap}(t).\text{tid} = t$
\end{enumerate}
Intuitively, the order in which nodes appear in $l$ indicate the order in which the entries were updated. 

We now list some operations that ordered lists support. For an ordered list $O$, we use $O[0:k]$ to denote the first k elements of $O.l$. If $k > \numthr$, then $O[0:k]$ denotes all of elements. Additionally we have the following operations.
\begin{enumerate}
    \item $O.\text{get(tid)}$ returns $u.\text{time}$ for $u.\text{tid}=\text{tid}$. 
    \item $O.\text{set(tid, time)}$ sets $u.\text{time} = \text{time}$ for the unique node $u$ with $u.\text{tid}=\text{tid}$. The operation also moves node $u$ to the head of $l$. 
    \item $O.\text{increment(tid, $k$)}$ increments $u.\text{time}$ by $k$ for the unique node $u$ with $u.\text{tid}=\text{tid}$. The operation also moves node $u$ to head of $l$. 
\end{enumerate}
Each of the above operations can be implement in $O(1)$ time. Finally, given two ordered lists $O,O'$ (or an ordered list $O$ and a vector clock $C$), $O \cle O'$ (or $O \cle C$) if for every thread $t$, $O.\text{get}(t) \leq O'.\text{get}(t)$ ($O.\text{get}(t) \leq C(t)$). 
}

\begin{figure}[t]
\includegraphics[width=0.9\textwidth]{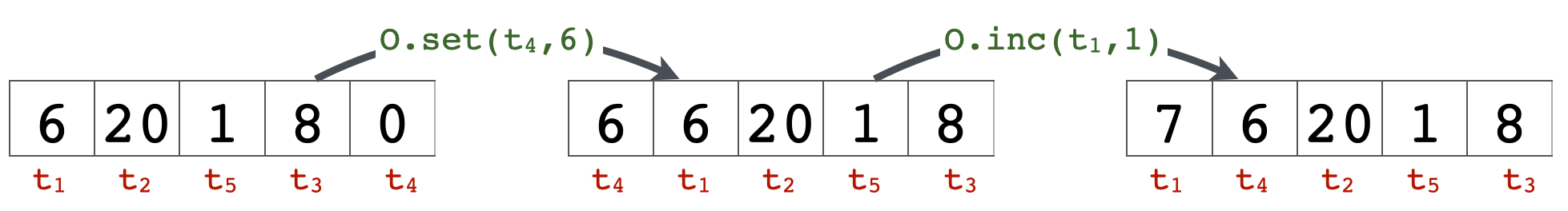}
 \vspace{-0.1in}
\caption{Example ordered list $O$ (left). Result of operation $O.\text{set}(t_4,6)$ (middle), and followed by $O.\text{inc}(t_1,1)$ (right).}
\figlabel{list}
\end{figure}

\begin{example}
    Let us look at an example to understand some of the operations on ordered lists. An ordered list $O$ is shown pictorially on the left in \figref{list}. It represents a vector timestamp, involving $5$ threads $t_1, t_2, t_3, t_4$ and $t_5$.
    The timestamp in the first vector is given by the map $t_1 \mapsto 6, t_2 \mapsto 20, t_3 \mapsto 8, t_4 \mapsto 0, t_5 \mapsto 1$. The list order is given by: $t_1 < t_2 < t_5 < t_3 < t_4$. Thus, $O.\text{get}(t_3)$ would return $8$. The result of $O.\text{set}(t_4,6)$ is shown as the structure in the middle of \figref{list}. Notice that the value assigned to the node corresponding to $t_4$ has been changed to $6$, and it has been moved to the head of the list. $O.\text{inc}(t_1,1)$ applied to the ordered list in the middle results in the list shown on the right. The entry for $t_1$ is incremented by $1$ to $7$ and moved to the head of the list.
\end{example}

\myparagraph{Exploiting Ordered Lists}{
The ordered list data structure has been designed to reduce the overhead during acquires. The idea is to replace the vector clocks $\CSmp_t$ and $\CSmp_\lk$ by ordered lists $\COmp_t$ and $\COmp_\lk$, respectively. Then in the acquire handler, instead of going through all entries to perform the join in \lineref{c-join-acq-loop} (of \algoref{smp-upd}), we only need to go through the first $\CUpd_\lk(\lastrelthr_\lk) - \CUpd_t(\lastrelthr_\lk)$ entries of the ordered list $\COmp_\lk$. 

However, the use of ordered lists introduces a subtle issue when handling releases.
First, when updating the timestamp of the lock, we will require $\COmp_\lk$ to be a copy of $\COmp_t$, which means that not only the values of the timestamp need to be changed, the structural order of $\COmp_\lk$ must also be correctly updated to align with that of $\COmp_t$. Although the freshness timestamp reduces the number of entries that need to be traversed, it is independent of the structure's order, indicating that an $O(\numthr)$ operation cannot be omitted. This seems to doom us to the same complexity as \djitp. However, there is a way out. We can use the idea of ``shallow copies'', which has been previously applied  in~\cite{pacer-tool,TSVD}}. 

\myparagraph{A holistic solution--lazy copy}{
In ~\cite{pacer-tool,TSVD}, lazy copy was introduced as an optimization. However, we remark that in our work, it serves as a holistic solution that addresses all the challenges, including those highlighted at the end of the previous section.
The rough idea of the lazy copy is that instead of copying the timestamp entry-by-entry, let the lock and the thread share the same ordered list object. 
When the thread needs to update its vector clock (in the sampling case, the sampling timestamp) later, it then creates a deep copy. At a high level, the optimization shifts the $O(\numthr)$ operation in each release event to the acquire events, and that is only performed when a deep copy is necessary. In the normal case without sampling, the shallow copy makes no asymptotic difference. However, with sampling, the scenario where a thread needs to create a deep copy is infrequent—it is limited by the number of changes to the sampling timestamp, which is $|S|$. This suggests that, for a thread $t$, the total cost of processing all release events performed by $t$ can be consolidated into creating at most  $|S|$ deep copies. 
In \algoref{smp-upd},  locks are treated as objects that maintain individual vector clocks, updated in the same manner as threads. However, by employing lazy copy, a lock object merely passes a reference to the ordered list of the thread that last released it. This design allows for an algorithm whose complexity is determined exclusively by the number of threads, thereby eliminating the dependency on the number of locks.
Furthermore, the algorithm can be extended to accommodate generic acquire and release operations, while maintaining the optimized running time, as monotonicity is now only required for the vector clocks maintained by threads and no longer so for locks/synchronization objects. 
In the appendix, we provide more details on how the algorithm can be extended to handle non-mutex synchronizations.} 


{
\begin{algorithm}[t]
\small
\vspace{-0.2in}
\begin{multicols}{2}
\myfun{\init}{
	\ForEach{$t \in \threads{}$}{
		$\CUpd_t \gets \bot$ ;
		$\COmp_t \gets \bot$ ;
            $\shared_t\gets\false$ ;
            $\epch_t\gets 1$
	}
	\ForEach{$\lk \in \locks{}$}{
		$\COmp_\lk \gets \bot$ ;
		$\lastrelthr_\lk \gets \nil$;
            $\lastrelu_l \gets \nil$\;
	}
}

\myhandler{\acqhandler{$t$, $\lk$}}{
	\If{$\lastrelu_\lk > \CUpd_t(\lastrelthr_\lk)$}{
            $d\gets\lastrelu_\lk - \CUpd_t(\lastrelthr_\lk)$\;
             $\CUpd_t \gets \CUpd_t[\lastrelthr_\lk \mapsto \lastrelu_\lk]$
            
            \ForEach{$(t^*,n)\in \COmp_\lk[0:d]$}{\linelabel{orderlist-forloop}
                \If{$n > \COmp_t.\textbf{get}(t^*)$}{
                \If{$\shared_t$}{
                    $\COmp_t = \textbf{deepcopy}(\COmp_t)$\;
                    $\shared_t=\false$
                }
                 $\COmp_t.\textbf{set}(t^*,n)$\;
                $\CUpd_t \gets \CUpd_t[t \mapsto \CUpd_t(t)+1]$\;
                }
            }
	
	}
}

\myhandler{\relhandler{$t$, $\lk$}}{

\If{$\exists e \in S$, with $\ThreadOf{e} = t$, since last release in $t$}{\linelabel{o-acq-skip}
        \If{$\shared_t
        $}{
         $\COmp_t = \textbf{deepcopy}(\COmp_t)$\;
        }
                
			$\COmp_t\gets\COmp_t[t\mapsto \epch_t]$ \;
                $\epch_t\gets \epch_t+1$\;
			$\CUpd_t \gets \CUpd_t[t \mapsto \CUpd_t(t)+1]$
		}	
	$\COmp_\lk=\textbf{shallowcopy}(\COmp_t)$\;
        $\shared_t=\true$\;
 $\lastrelthr_\lk \gets t$ \;
 $\lastrelu_\lk=\CUpd_t.\textbf{get}(t)$\;
		
}

\end{multicols}
\caption{\small Ordered list algorithm partially computing the VT timestamp}
\normalsize
\algolabel{algo4}
\end{algorithm}
}

\myparagraph{Final algorithm}{
\algoref{algo4} is the final algorithm. As we described earlier, vanilla vector clocks $\CSmp_t$ and $\CSmp_\lk$ are replaced by ordered lists $\COmp_t$ and $\COmp_\lk$. Further, locks no longer have a vector clock $\CUpd_\lk$. This is because the only way to reduce the overhead when doing the join of $\CUpd$-clocks in \lineref{u-join-acq} of \algoref{smp-upd}, is by having another timestamp that measures the number of entries of $\CUpd$ that have changed! Therefore, we instead only store the $\CUpd_t(t)$ component of the thread $t$ performing the last release at the lock, which is just a scalar. Next, in the release handler, we always perform only a shallow copy. In the acquire handler, when the thread learns new information from the lock, it only traverses  $\lastrelu_\lk - \CUpd_t(\lastrelthr_\lk)$ many entries.
}
\begin{lemma}
\lemlabel{algo4-correctness}
For an execution $\tr$, a set of sampled events $S$, \algoref{algo4} declares a race on an event $e$ if and only if \algoref{algo2} declares a race on $e$. \algoref{algo4} runs in time $O(\trsz) + O(|S|\numthr)O(\numthr)$ and performs $O(|S|\numthr)$ many vector clock operations and $O(|S|\numthr)$ many deep copies .
\end{lemma}

\begin{lemma}
\lemlabel{algo4-optimality}
For an execution $\tr$, a set of sampled events $S$, let $\s{VTWORK}(\tr)$ be the number of times any of the vector clocks maintained by \algoref{algo2} changes when run on $\tr$. 
\algoref{algo4} runs in time $O(\trsz) + O(\s{VTWORK}(\tr)\numthr)$.
\end{lemma}

\myparagraph{Optimality of running time}{
The lemma above indicates that the runtime of \algoref{algo4} is nearly optimal. It is important to note that $O(\trsz)+\s{VTWORK}(\tr)$ represents a lower bound for any algorithm computing the relation $\hb{(\tr,S)}$. \algoref{algo4} operates in time $O(\trsz) + O(\s{VTWORK}(\tr)\numthr)$, which is close to the best achievable performance. However, it is open if the algorithm can be improved to meet the lower bound.

\rmv{
Although the running time of \algoref{algo4} has a quadratic dependence on the number of threads $\numthr$, we now argue that its performance is close to optimal. Consider \algoref{algo2} which is the first most natural vector clock algorithm to compute the relation $\hb{(\tr,S)}$. Let $\s{VTWORK}(\tr)$ be the number of times any of the vector clocks maintained by \algoref{algo2} changes when run on execution $\tr$ --- note these are the times when some entry in some vector clock is actually changed to a \emph{different} value. This is a lower bound on the amount of time any algorithm must take to compute $\hb{(\tr,S)}$. Using a counting argument similar to the one used to analyze the running time of \algoref{algo4}, we can show that the running time of \algoref{algo4} on any execution $\tr$ is at most $\s{VTWORK}(\tr)*\numthr$, which is close to the best one can hope. Is there an algorithm whose running time if just $\s{VTWORK}(\tr)$? We have unfortunately not been able to resolve this question and we leave it for future investigation.}}

\rmv{
\myparagraph{Treeclocks}{It is worth mentioning that for computing the full Happens before order, the tree clock data structure \cite{MathurTreeClocks2022}has been proved to optimal. However, for the sampling time stamp, its optimality is not preserved. The high level image is that tree clocks' optimality relies on the observation that to compute the full happens before,  every local time instance kept by any thread there has a unique sender(namely the first release
that send the message) and unique receiver (the first remote acquire after the release). Such an observation does not hold for the sampling time stamp. }

\myparagraph{On-demand race detection}{
As we discussed before, the RaceMob tool applied on-demand race detection for dynamic analysis. The high level is that the approach globally controls the processing of synchronization events. It turns on the processing once an event is sampled and turns it off once all sampled events so far have established happens-before relation to all threads. 
We note that such an approach is an heuristic that does not improve the running time complexity. One can easily construct a trace where even when only one event is sampled, the on-demand race detection still processes all synchronization events. Also, it is not addressed in \cite{racemob} whether such an mechanism can be implemented without introducing additional synchronizations while our approach introduces no additional synchronizations. At the end, we would also love to emphasize that our approach nevertheless is orthogonal to the principle of on-demand race detection and can improve it as it can improve all other sampling approaches. 
}
}





\section{Evaluation}
\seclabel{experiment}

We implemented our proposed data structures and algorithms in \tsan (TSan) v3~\cite{threadsanitizer} to evaluate their effectiveness.
\tsan is a state-of-the-art data race detector
that performs online race detection on a running process. Our evaluation on \tsan is catered to gauge our algorithms' effect on the performance of real-world systems running large workloads. We also implemented our algorithms in \rapid~\cite{rapid} for offline experiments, enabling us to fully eliminate non-determinism caused by thread interleavings and gain an unbiased understanding of each algorithm's performance. Due to space limitations, our experimental results using {\rapid} are presented in the appendix.
\subsection{Modifications to \tsan}
\seclabel{methodology}

We modified TSan v3 (in LLVM's \texttt{compiler-rt}) to use our proposed clocks in place of the existing vector clock for handling synchronization operations, and modified the memory access handlers to perform sampling. \textcolor{black}{We disabled {\tsan}’s slots' preemption mechanism --- which is used to enable data race detection on any number of threads with a fixed vector clock size --- to simplify our implementation and focus solely on the core race detection logic.}
Below, we briefly discuss some noteworthy design choices and optimizations in our implementation.

\myparagraph{Sampling Strategy and Race Detection}{
The algorithms we propose in previous sections
are agnostic to how the events $S$ were chosen.
For our evaluation, we stick to the standard choice of
random sampling~\cite{marino2009literace} where each read or write
access event is sampled independently with a fixed probability.
Upon encountering an access event, we generate a random number and skip the event if the number is above a fixed threshold.
The choice of random sampling allows us to evaluate the effectiveness of our solution on a broad and general distribution of sampled events, ensuring robust analysis. 

We do not compare against other sampling-based race detection techniques~\cite{RPT2023, pacer-tool, racemob, prorace}, as our work addresses only the \analprb, making it a complementary enhancement rather than a competing approach. \textcolor{black}{Moreover, existing techniques typically rely on system-level innovations to reduce overhead: for example, controlling garbage collection~\cite{pacer-tool}, crowdsourced dynamic validation~\cite{racemob}, and hardware-assisted sampling with offline reconstruction~\cite{prorace}. While these approaches have demonstrated effective overhead reduction, they may not be generally applicable across settings. In contrast, our approach is purely algorithmic and does not depend on any system-level or hardware support.}

}



\myparagraph{Local Epoch Optimization}{
Our implementations closely follow the algorithms presented in the technical sections. We applied an optimization to potentially improve the performance of \algoref{algo4}. The high level idea is to disentangle the `local time epoch' from the entire vector clock when communicating them over \acrhb edges. This saves individual threads from creating deep copies when incrementing their local epoch. A similar optimization was also applied in TSan v2.
}
\subsection{Evaluation on \tsan}

\subsubsection{Evaluation Setup} 
We first describe our experiment and benchmark setup.

\myparagraph{Benchmarks}{
Since our improvements only pertain to the synchronization handlers, we wanted to evaluate them on executions with heavy lock usage. Instead of testing on small benchmarks, we chose MySQL Server 8.0.39 as our evaluation subject because a database server runs with many threads and uses locks very frequently.
{\color{black}Our evaluation suite consists of 15 benchmarks from the BenchBase framework (commit 82af61)~\cite{DifallahPCC13}, excluding two benchmarks for documented and reproducible reasons. Specifically, we omit CH-benCHmark due to its long runtime idle periods, as reported in GitHub issue 318, and TPC-DS due to missing configuration files, also noted in the repository.
Each benchmark includes both a schema and a workload (i.e., a sequence of queries), collectively offering a broad range of execution characteristics, including varying levels of lock contention.

During experimentation, we identified three benchmarks as outliers. The noop benchmark performs no operations; resourcesstresser focuses solely on I/O operations; and OT-Metrics exhibits highly inconsistent performance across runs under identical configurations. Further analysis suggests that MySQL's execution of OT-Metrics queries may rely on randomized heuristics, leading to non-deterministic behavior. Given these issues, we exclude these three benchmarks from our reported results.}
}

\begin{figure}[t]
\vspace{-0.2in}
\includegraphics[height=4cm]{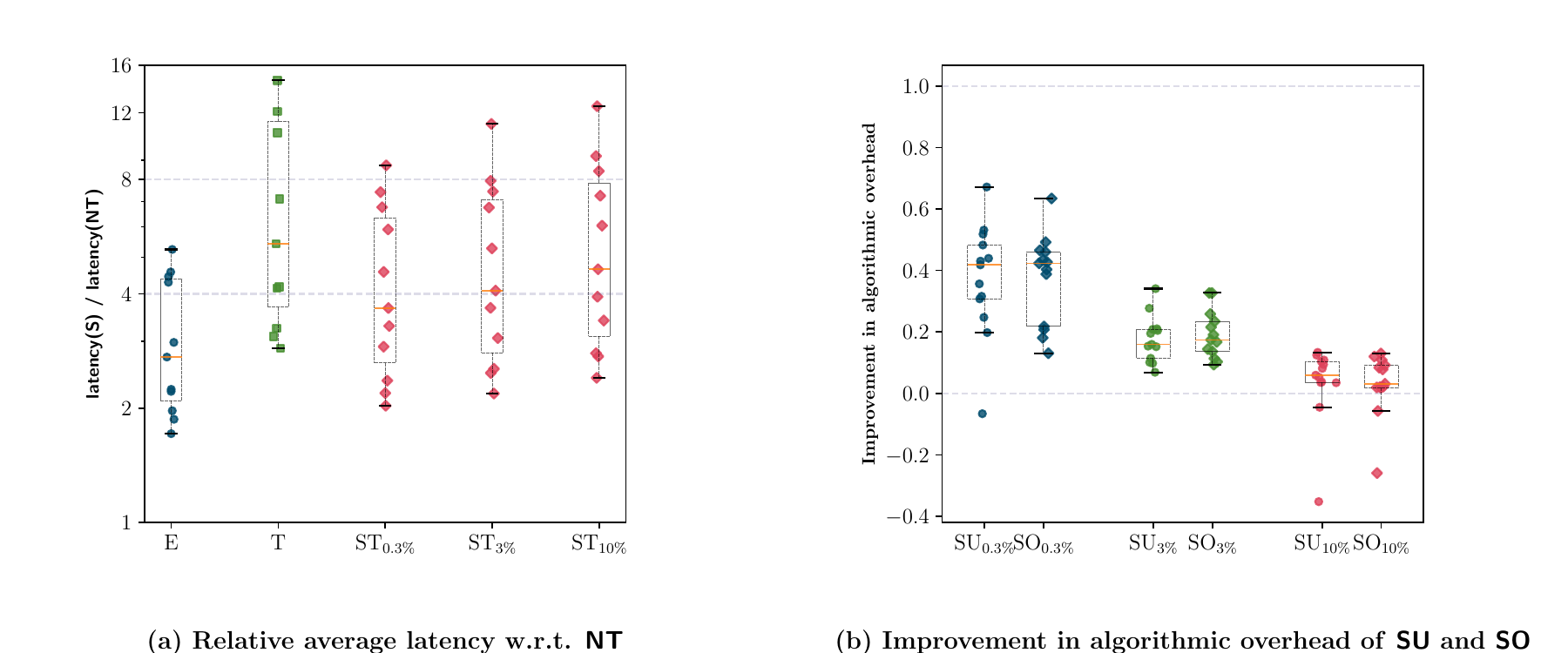}
\caption{Latency Relative to \confNT and Algorithmic Overhead Improvement}
\figlabel{mysql-tsan}
\end{figure}

\begin{figure}[ht]
  \centering

  \begin{minipage}[t]{0.33\textwidth}
    \centering
    \includegraphics[height=4cm,width=\textwidth]{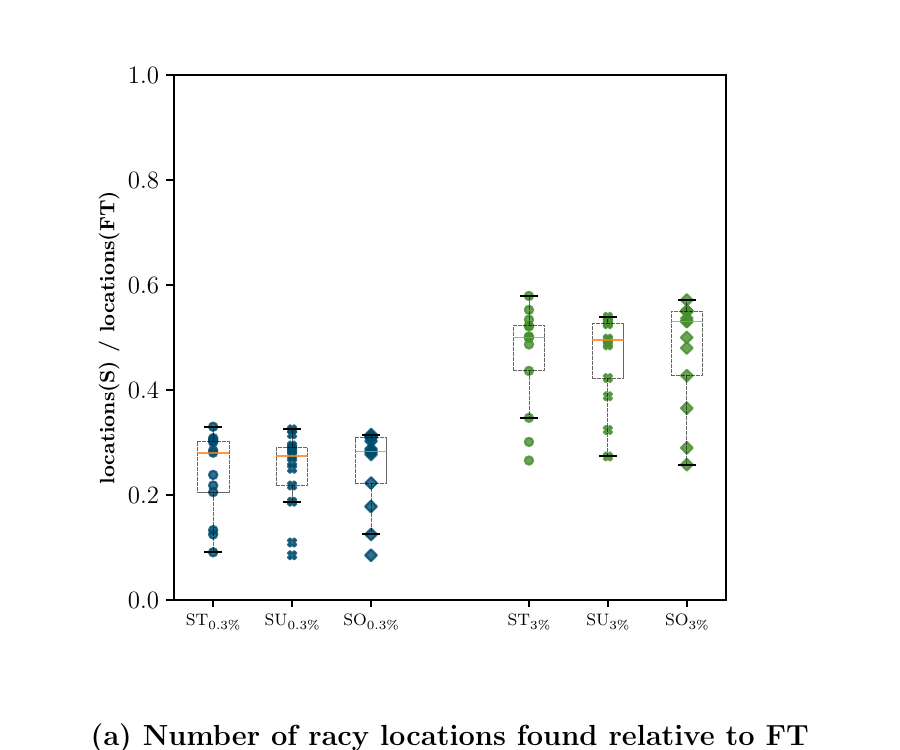}
  \end{minipage}
  \hfill
  \begin{minipage}[t]{0.66\textwidth}
    \centering
    \includegraphics[height=4cm,width=\textwidth]{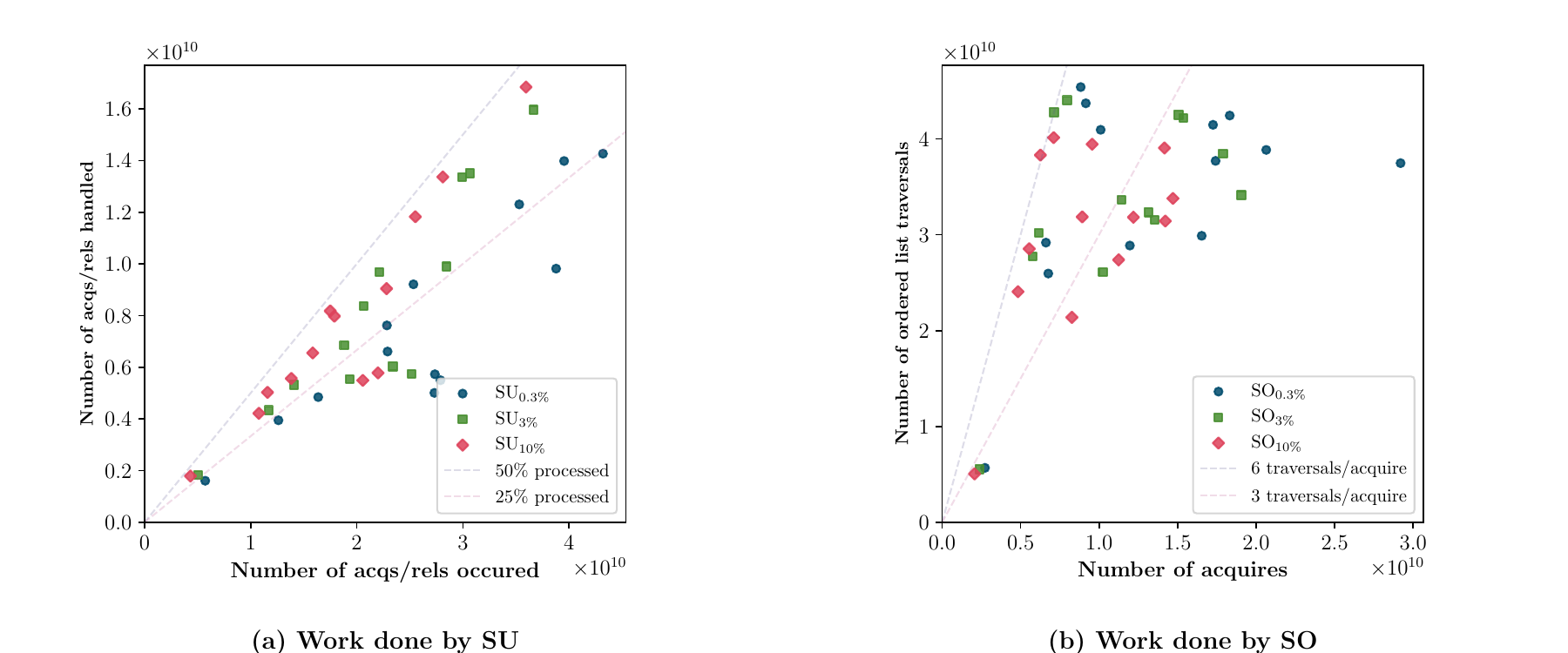}
  \end{minipage}

  \caption{ratios of exposed racy locations(left), work done by SU(center), and work done by SO(right).}
  \label{fig:race-comparison}
\end{figure}

{\color{black}
\subsubsection{Baselines, Configurations and Evaluation Metric}

In this experiment, we considered three fundamentally different baselines: (1) \underline{No-TSan}(\confNT) --- running benchmarks without any instrumentation, (2) \underline{Empty-TSan}(\confET) --- running benchmarks instrumented with TSan but without performing any race detection, and (3) \underline{Full-TSan}(\confFT) --- running benchmarks instrumented with TSan and perform race detection for all events. 

Intuitively, \confNT represents the zero overhead baseline, while \confET captures the pure instrumentation overhead, unavoidable by algorithmic solutions. \confFT, on the other hand, reflects the total overhead introduced by the framework for dynamic data race detection. The performance difference between \confET and \confFT represents the overhead induced by running the dynamic analysis algorithm \fasttrack and can potentially be optimized using approaches like dynamic sampling. 

To accurately measure the improvement of our innovations which are specific to the synchronization handlers, we compiled MySQL in different configurations:
\begin{enumerate*}
    \item \underline{Sampling-TSan (\confST)} --- the naive sampling algorithm without optimizations on synchronization handlers, compiled with TSan v3.
    \item \underline{Sampling-UClock (\confSU)} --- compiled with TSan v3 with modified timestamping as in \algoref{smp-upd}.
    \item \underline{Sampling-OrderedList (\confSO)} --- compiled with TSan v3 with modified timestamping as in \algoref{algo4}.
\end{enumerate*}
\confST serves as a more accurate baseline for our algorithms (\confSU and \confSO) as all three differ only in how they handle synchronizations, with \confST using the most naive approach. 
We have three variants for each build above that samples events with 0.3\%, 3\% and 10\% sampling rate. The sampling rate is indicated with a subscript, e.g. \confSTsr{0.3\%}. The selection of the 0.3\% and 3\% sampling rates is guided by prior work~\cite{pacer-tool,prorace}, while 10\%, though rarely considered in the sampling literature, is included to evaluate the performance of our algorithms on larger sets of events. All baselines, including \confNT, as well as the evaluated configurations, are compiled with optimization level O1, which is the default for MySQL when built with TSan.

Since our experiment is conducted in a stress-testing manner, we consider average latency as an indicator of efficiency, where latency is the time taken to finish a request. This metric serves as a strong indicator of potential improvements in the speed of completing bug-testing tasks in a non-stress testing scenario.

Because our solutions are purely algorithmic, we introduce the notion of \textbf{algorithmic overhead} (AO) to better quantify the improvements:
\[
	\text{AO}(\mathsf{S})= \mathsf{latency(}\mathsf{S})-\mathsf{latency(}\mathsf{ET}). 
\]
where $\mathsf{S}$ is one of \confST,\confSO,\confSU and \confFT.

} 

\myparagraph{Setup}{
We ran all experiments on an Intel(R) Xeon(R) w9-3495X 1.9GHz system with $64$ CPUs and $64$GB memory running Ubuntu 24.04.
We configured the BenchBase suite to run for 1 hour, using the 
\texttt{SERIALIZABLE} isolation level, with $12$ client terminals, $1$ minute warm-up,
a rate limit of $10$ million requests, and a 
fixed seed to ensure that all runs, irrespective of their configurations,
processed the same distribution of requests. 
{\color{black}We first disabled race reporting in TSan v3 to eliminate the associated I/O overhead and ensure precise latency measurements for all configurations. We then repeated the experiments with race reporting enabled to record the number of data races detected under specific configurations. Finally, we conducted an additional set of runs in profiling mode to measure the amount of work performed by \confSU and \confST.}

}
{\color{black}\subsubsection{Baseline Overheads}
In \figref{mysql-tsan}(a), we present the relative average latency of \confET, \confFT, and \confST across three sampling rates, all measured with respect to the unintrumented baseline \confNT. 

Notably, \confET introduces an average latency of $3.1\times$ compared to  \confNT; this overhead is inherent to Tsan v3's instrumentation mechanism and cannot be eliminated by algorithmic improvements at the analysis level. While optimizing instrumentation overhead for the sampling setting is beyond the scope of this work and not required to evaluate our algorithmic improvements, prior works~\cite{racemob,prorace,Choi02,Wilcox2015,bigfoot,GMrace} have shown that this overhead can be significantly reduced through system level engineering, static analysis techniques, or hardware support.
On top of the instrumentation overhead, \confFT incurs a significantly higher average relative latency of $9\times$, primarily due to algorithmic overhead. This suggests that the algorithmic component is the dominant contributor to the overall slowdown in dynamic race analysis. The figure also shows that naive sampling can reduce this overhead, though not ideally: the three sampling configurations of \confST (0.3\%, 3\%, and 10\%) yield relative latencies of $4.5\times$, $5.1\times$, and $5.8\times$, respectively. The algorithmic overheads of these configurations relative to \confNT, calculated as $\mathsf{AO}(\confST) / \mathsf{latency}(\confNT)$, are $1.4\times$, $2.1\times$, and $2.7\times$, respectively. This indicates that naive sampling still introduces substantial algorithmic overhead, even when only a small fraction of memory access events are analyzed. 

Lastly, we note that three benchmarks (TATP, Wikipedia, and YCSB) are omitted from this graph because their uninstrumented versions exhibit very low latency and quickly reach the saturation point—i.e., the maximum throughput of the database system—during execution. As a result, they remain at this upper bound, or even fall below it due to system overload, making the \confNT measurement unreliable. Other configurations and baselines for these benchmarks are unaffected, and the results involving them are included in other sections.

\subsubsection{Improvements In Algorithmic Overhead}
To gauge the efficiency of our innovations, we evaluate the improvement in 
algorithmic overhead introduced by our algorithms \confSO and \confSU with respect to \confST, the naïve sampling algorithm. Precisely, the improvement of a configuration $\mathsf{S}$ is calculated by:
$1-\frac{\text{AO(S)}}{\text{AO(\confST)}}$.  In \figref{mysql-tsan}(b), we show this relative improvement achieved by our algorithms \confSO and \confSU compared to \confST at each sampling rate. Overall, we observe encouraging improvements for most executions, up to over 60\% under both \confSUsr{0.3\%} and \confSOsr{0.3\%}. 
The improvement tends to decrease as the sampling rate increases, with an average improvement of 37\% for both \confSUsr{0.3\%} and \confSOsr{0.3\%}, 17\% and 19\% for \confSUsr{3\%} and \confSOsr{3\%}, respectively, and 3\% for both \confSUsr{10\%} and \confSOsr{10\%}.

We believe this trend is due to two main factors: (a) the algorithmic overhead becomes increasingly dominated by the cost of analyzing memory access events, and (b) the number of synchronization operations that can be skipped decreases as more memory access events are sampled.

In a few rare cases, our algorithms resulted in higher algorithmic overhead. Upon investigation, we found that these benchmarks perform very few synchronizations relative to memory accesses, leaving limited opportunity to reduce overhead by optimizing synchronization handling. 
\subsubsection{Race Detection Rate}
The previous section shows that at lower sampling rates (0.3\% and 3\%), our algorithms yield encouraging improvements in algorithmic overhead. In this section, we investigate whether this reduction translates to stronger predictive power by comparing the number of racy locations exposed by \confST, \confSU, and \confSO in our experiments.

As shown in \figref{race-comparison}(a), we report the number of racy locations relative to those exposed by \confFT, the full \tsan . The results suggest that there is no strong correlation between reduced algorithmic overhead and the number of racy locations detected as races are inherently rare under sampling, and lower latency from reduced overhead alone does not necessarily translate into consistently higher race detection rates.

Nonetheless, we observe that lower sampling rates still uncover a substantial portion of the racy locations found by \confFT. This surprisingly strong result may be partly due to the fact that lower sampling rates reduce latency effectively, allowing sufficiently more events to be processed within the runtime budget. Even so, we believe this observation demonstrates that small sampling rates can be practically beneficial.

\subsubsection{Work done} {
In this section, we investigate how our algorithms \confSU and \confSO achieve their performance improvements. Recall that the savings in \confSU are mostly binary: it either skips a synchronization operation entirely or performs a full vector clock traversal. \confSO, on the other hand, can partially skip the traversal by leveraging the ordered list of clock entries.

In \figref{race-comparison}(b), the x-axis shows the total number of acquire and release events during execution, while the y-axis shows how many of those events triggered an $O(n)$ vector clock traversal under \confSU. In most runs, \confSU skipped more than 50\% of acquire and release operations combined.

\figref{race-comparison}(c) shows the average number of ordered list entries processed in each acquire operation by \confSO per benchmark. Notably, in most runs, \confSO performed an average of six or fewer traversals of $\COmp_\lk$ per acquire—significantly lower than 64, the number of concurrently runnable threads (i.e., number of CPUs), and much lower than 256, the fixed vector clock size used by \tsan.

}
}

\subsection{Summary and Offline Experiment}{
\textcolor{black}{Our evaluation on \tsan demonstrates that the two algorithms we propose yield meaningful improvements in algorithmic overhead compared to the naive sampling algorithm. Profiling results further indicate that the timestamps and the data structure introduced in this paper enable the reduction of workload for most vector clock operations, corroborating our theoretical analysis.} Additionally, in the appendix, we present an offline experiment conducted on \rapid, where all analyses were run with identical execution traces and seeds (for random number generation) for consistency. The experiment focused on two specific sampling rates: $3\%$, which achieves an effective balance between high recall and low overhead, as shown in \cite{pacer-tool}, and $100\%$, which allows us to investigate the potential uses of timestamps beyond sampling.
The results of the offline experiment are consistent with those in this section, further supporting the effectiveness of the innovations proposed in this paper.
}

\section{Related Work}
\seclabel{related}
\myparagraph{Data Race Detection, Runtime Predictive analysis and Concurrency Testing}{
Data race detection techniques can primarily be classified
as static or dynamic analyses.
Static analysis techniques primarily rely on type systems~\cite{Abadi:2006:TSL:1119479.1119480,racerx}
and often report false positives.
Recently though, RacerD~\cite{racerd2018} and its successor~\cite{racerdx2019}
have emerged as promising static analyzers with reduced false alarms.
Nevertheless, dynamic data race detectors remain the tool of choice.
Lockset-based race detectors, popularized by Eraser~\cite{savage1997eraser}
look for violations of the locking discipline, are lightweight but unsound.
Sound dynamic race detectors are instead primarily based on the
happens-before (\acrhb) partial order~\cite{lamport1978time},
use either lock-set like algorithm~\cite{elmas2007goldilocks}
or faster vector clock~\cite{Mattern1988,Fidge:1991:LTD:112827.112860} based algorithms,
popularized by~\cite{Pozniansky:2003:EOD:966049.781529}, and later
improved by~\cite{fasttrack}.
Delay injection based race detectors~\cite{datacollider2010} insert active delays and sidestep timestamping.
Data race prediction techniques aim to enhance coverage
by reasoning about alternate interleavings~\cite{Said2011,rv2014,cp2012,wcp2017,shb2018,SyncP2021,Shi2024,Roemer20,Roemer18,PavlogiannisPOPL20}.
Runtime predictive analysis has been extended to other properties such as deadlocks, atomicity violations as well as more general properties~\cite{syncpdeadlocks2023,MathurAtomicity20,AngPatternLanguages2024,AngTracePrefixes2024}, but is known to be intractable in general~\cite{Mathur20,kulkarniCONCUR2021,FarzanMathur2024,kulkarniCONCUR2021} and \acrhb-based race detection, based on Mazurkiewicz-style trace-based reasoning has remained popular because of the performance benefits it offers.
Concurrency testing approaches, on the other hand, aim to explore bugs by executing the underlying program systematically multiple times using randomization~\cite{SURW2025,PCT2010,pos}, together with feedback-guidance~\cite{rff2024} or in a strictly enumerative manner~\cite{kokologiannakis2022truly,abdulla2014optimal,Abdulla14,agarwal2021stateless}. 
}

\myparagraph{Sampling-based techniques}{
LiteRace~\cite{marino2009literace} performs sampling to reduce overhead due
	to instrumentation, switching back and forth between instrumented
	and uninstrumented code, based on a cold-region hypothesis.
	Our work is orthogonal and can improve the cost of timestamping here.
	The \pacer~\cite{pacer-tool} algorithm splits program executions into 
	alternating sampling and non-sampling periods and observes the read/write events in all sampling periods.
	 Optimizations incorporated by \pacer include selective clock increments and use
	 of version clocks to avoid redundant vector clock computations in non-sampling period. 
	 While similar in spirit, our proposed freshness timestamp is more 
	 fine-grained and allows us to exploit ordered lists to 
	 further omit redundant communication.
	Further, the use of sampling phases is particularly catered towards
	a language with managed runtime, such as Java, that allows control over when
	to start and stop these phases.
	Implementing a similar strategy in a language like C requires additional
	global synchronization, degrading the performance of the underlying application-under-test.
  The recently proposed \RPT~\cite{RPT2023} algorithm uses ideas from property testing, and provides an probabilistic guarantee for detecting data races, assuming the execution is sufficiently racy.
  \RPT is designed to sample constantly many events, and performs only constantly many operations. 
  In such a setting, our algorithm also performs constantly many vector clock operations, and can potentially further enhance the timestamping cost incurred by \RPT. 
  {\color{black} \ProRace focuses on low-overhead race detection through hardware-assisted sampling and offline reconstruction. \ProRace demonstrated the instrumentation overhead can be significantly reduced with low sampling rates (0.1\%, 0.01\% and 0.001\%).  
  Its contributions are primarily systems-level, combining PEBS and Intel PT with a custom lightweight tracing stack. In contrast, our innovation is purely algorithmic and does not rely on any hardware support. }
}

\myparagraph{Other techniques for reducing the overhead of race detection}{
The epoch optimization due to \fasttrack~\cite{fasttrack} is perhaps the most popular work on reducing the overhead due to vector clock operations involved in race checks.
Approaches such as \cite{redcard,bigfoot} often perform static analysis to optimize the placement of vector clock checks, and thus reduce timestamping overhead.
Optimistic concurrency control~\cite{Wood2017,Bond2013} offer an orthogonal approach to reduce the overhead due to shared vector clocks.
Multiple works~\cite{racemob,prorace,Choi02,Wilcox2015,bigfoot,GMrace} have demonstrated that combining static analysis with system-level engineering can effectively reduce instrumentation overhead and eliminate redundant checks in dynamic analysis.
Our proposed algorithms can naturally enhance the timestamping cost of these approaches.
The tree clock data structure \cite{MathurTreeClocks2022,treeClocksCPP2024} is an optimal data structure for computing the full happens-before relation. However, this data structure ceases to be optimal in the context of computation of the sampling partial order. On the other hand, as we showed in \secref{orderedlist}, the ordered list structure is indeed more suitable for the sampling partial order as it reduces the running time complexity of the vector clock algorithm presented in \secref{timestamps} by an order of $\numthr$. This follows because the hierarchical structure of tree clocks does not exploit the redundant operations introduced by the sampling timestamp.
}
{\color{black}{

\myparagraph{Sync-dominated programs}{
The \TSVD work presented in \cite{TSVD} focuses on a special class of synchronization-dominated programs, where the number of read and write events is relatively small even without sampling. The analysis problem we formulate naturally subsumes the race detection problem in such settings. Although \TSVD targets thread-safety violations rather than traditional data races and leverages structured parallel constructs, some of the optimizations proposed in the implementation share similarities with ours; for instance, they use mutable timestamp objects (akin to shallow copies) and reduce redundant communication. However, their timestamping scheme increments on every memory access, whereas ours does so only for the first release after each sampled event. Moreover, while their techniques improve practical efficiency within a language-specific runtime, they do not offer the same asymptotic complexity improvements as our algorithmic solution.
}
}}

\section{Conclusion and Future Work}
\seclabel{conclusions}

We consider the {\analprb} that arises naturally in the context 
of sampling-based dynamic race detection --- given a set $S$ of marked events, determine if there
is a data race which involves an event from $S$.
We show that, for an execution with $\trsz$ events
performed by $\numthr$ threads,
this problem can be solved while spending
only $O(|S|\numthr^2)$ time for vector clock traversals
and $O(\trsz) + O(|S|\numthr^2)$ total time; strictly speaking the number of vector clock operations is bounded by $\trsz$  and for each operation at most $O(\numthr)$ work will be done so the running time is $O(\trsz) + O(\min(\trsz\numthr, |S|\numthr^2))$, which reduces to the same complexity as {\fasttrack} when $|S| = O(\trsz)$. 
As part of our approach, we proposed two new timestamp notions and 
a data structure to exploit redundancy in vector clock operations. 
Our proposed timestamping notions may be of independent interest outside
of sampling based race detection.
Our algorithms are implemented in \tsan and in the offline analysis framework 
\rapid and our evaluation shows promising results and indicates 
our solution can be a significant step towards in-production sampling-based race detection. 

While there are many possible avenues for future work, 
we list the most relevant ones.
We believe that optimizing the data structure we propose here, 
can further improve performance and is likely goint to be important for practicability, but is also a challenging task.
Another interesting avenue is to
further improve the dependence on the parameter $\numthr$, and possibly design an algorithm which can be proved to have optimal running time for solving the {\analprb}.
\newpage

\begin{acks}
Umang Mathur is partially supported by a Google SE 
\& SEA research award, and by the National Research Foundation, Singapore, and Cyber Security Agency of Singapore under its National Cybersecurity R\&D Programme (Fuzz Testing <NRF-NCR25-Fuzz-0001>). Any opinions, findings and conclusions, or recommendations expressed in this material are those of the author(s) and do not reflect the views of National Research Foundation, Singapore, and Cyber Security Agency of Singapore. Minjian Zhang and Mahesh Viswanathan are partially supported by NSF SHF 1901069 and NSF CCF 2007428. 
\end{acks}
\appendix

\section{Additional Evaluation Details}


\subsection{Offline Evaluation using \rapid}
\seclabel{eval-rapid}
Here, we present our evaluation using \rapid, an offline 
dynamic analysis framework detector that analyzes execution trace logs
with various race detection algorithms including \fasttrack. 
Unlike our evaluation using TSan, which suffers from inevitable non-determinism because of
uncontrolled thread interleaving, \rapid enables a controlled study and
allows us to understand fine-grained metrics to evaluate our algorithms.

\subsubsection{Evaluation Setup} 
We first describe our experimental and benchmark setup.

\myparagraph{Implementation}{
We implemented four analysis algorithms in \rapid: \SUThree, \SOThree,
\SUFull, \SOFull.
Here, algorithm {\sf A}-($p$\%)
denotes that it samples $p$\% of access events
(according to the strategy described in \secref{methodology}),
and the core algorithm is either \algoref{smp-upd} (if {\sf A} is \confSU)
or \algoref{algo4} (if {\sf A} is \confSO).
We remark that our sampling algorithms do not converge to 
\fasttrack even when $p=100\%$! 
Although our algorithms are designed to solve the {\analprb} of sampling race detection, 
the optimizations also apply to the case when all access events are being observed.  
}

\myparagraph{Benchmarks}{
We conducted our experiments on execution traces from~\cite{Shi2024} 
which include 30 Java programs from the IBM Contest benchmark 
suite~\cite{Farchi2003}, DaCapo~\cite{DaCapo2006}, SIR~\cite{SIR2005}, the Java Grande forum benchmark suite~\cite{sen2005detecting}, and some other standalone benchmarks. 
The traces only contain accesses to shared variables and synchronizations 
via acquiring or releasing lock objects.
We omit 
}

\myparagraph{Setup}{
We analysed each benchmark trace $30$ times with each engine. 
Across different analysis engines, the same sequence of seeds is used
to ensure apples-to-apples comparison.
We count different fine-grained metrics such as
the number of times the algorithm determines that it can skip
processing certain events, or number of entries in the vector clocks that the algorithm traverses 
throughout its execution.
Our experiments are conducted on an AMD EPYC Milan 7713 supercomputer cluster with 64GB memory.
}

\subsubsection{Results} 
To evaluate the effectiveness
of our algorithms in the \rapid framework, we measure how many vector clock
operations do our algorithms skip, as compared to vanilla \fasttrack.
 
\begin{figure}[t]
\includegraphics[width=1\textwidth]{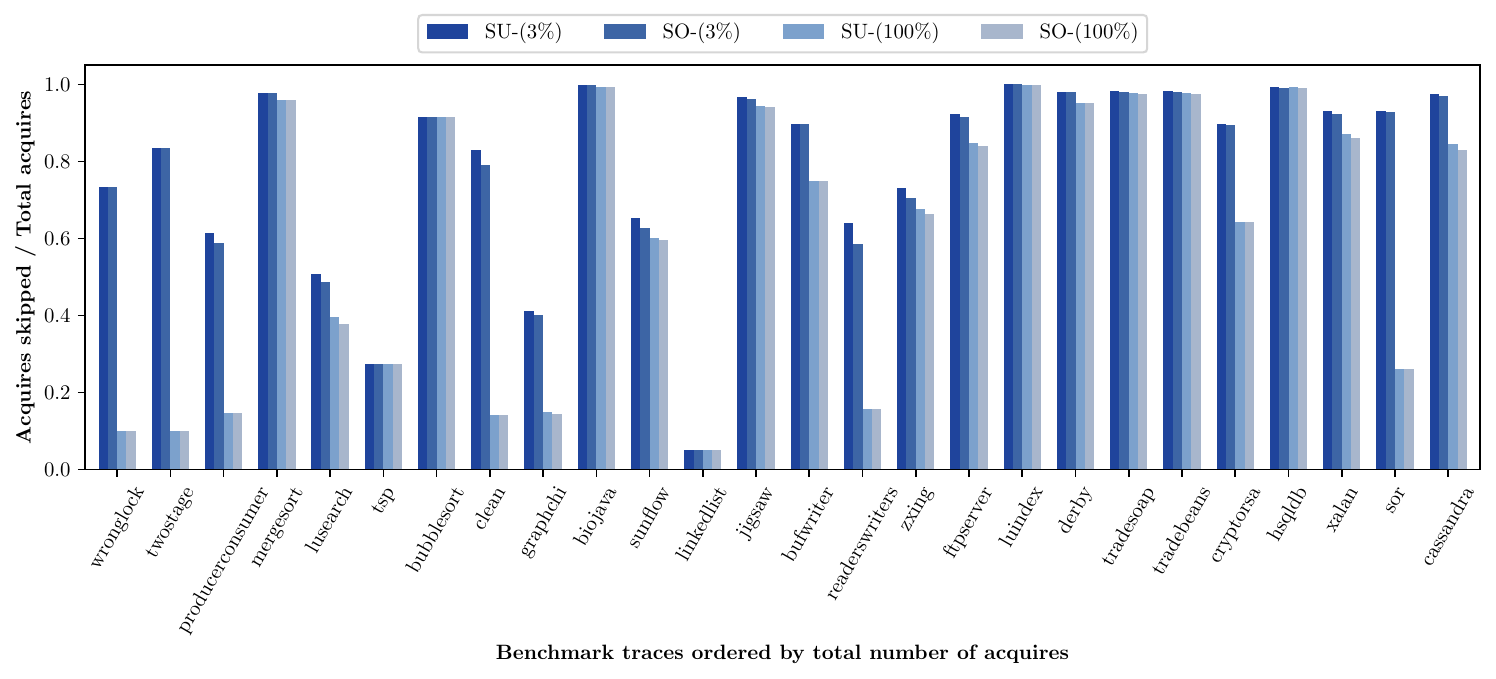}
\vspace{-0.1 in}
\caption{Ratio of acquires skipped over total number of acquires for four engines. 
For each benchmark, bars are ordered (from left to right, marked with darkest to lightest shades of blue) \SUThree, \SOThree, \SUFull, and \SOFull.}
\figlabel{skipped-acq}
\end{figure}

\myparagraph{Acquire events skipped}{
The key idea of \algoref{smp-upd} and \algoref{algo4} is to
detect and avoid redundant vector clock operations. 
In this context, for each benchmark, we recorded the number of 
acquire events that are skipped in each algorithm (respectively \lineref{u-acq-skip} in \algoref{smp-upd} and \lineref{o-acq-skip} of \algoref{algo4}) and averaged them over 30 runs.
\figref{skipped-acq} shows the ratio of acquire events skipped over 
the total number of acquire events in the execution trace, for each
of the four engines.
We can make the following observations:
\begin{enumerate}
    \item The two sampling engines \SUThree and \SOThree skipped more than $50\%$ acquires for $23/26$ benchmarks and skipped more than $80\%$ for $16/26$ benchmarks.
    \item \SUThree always skips more acquires than \SOThree and similarly \SUFull always skips more than \SOFull but the difference is always small. This implies that computing the freshness timestamp does not lead to visible improvement in reducing redundant vector clock operations. 
    \item The two non-sampling engines also skipped a significant amount of acquires in the majority of benchmarks. Such skipping of the algorithms are due to  (a) threads frequently acquire locks released by themselves, (b) threads frequently acquire locks in order reverse to the order of how the locks got released. 
\end{enumerate}
}

\begin{figure}[t]
\includegraphics[width=1\textwidth]{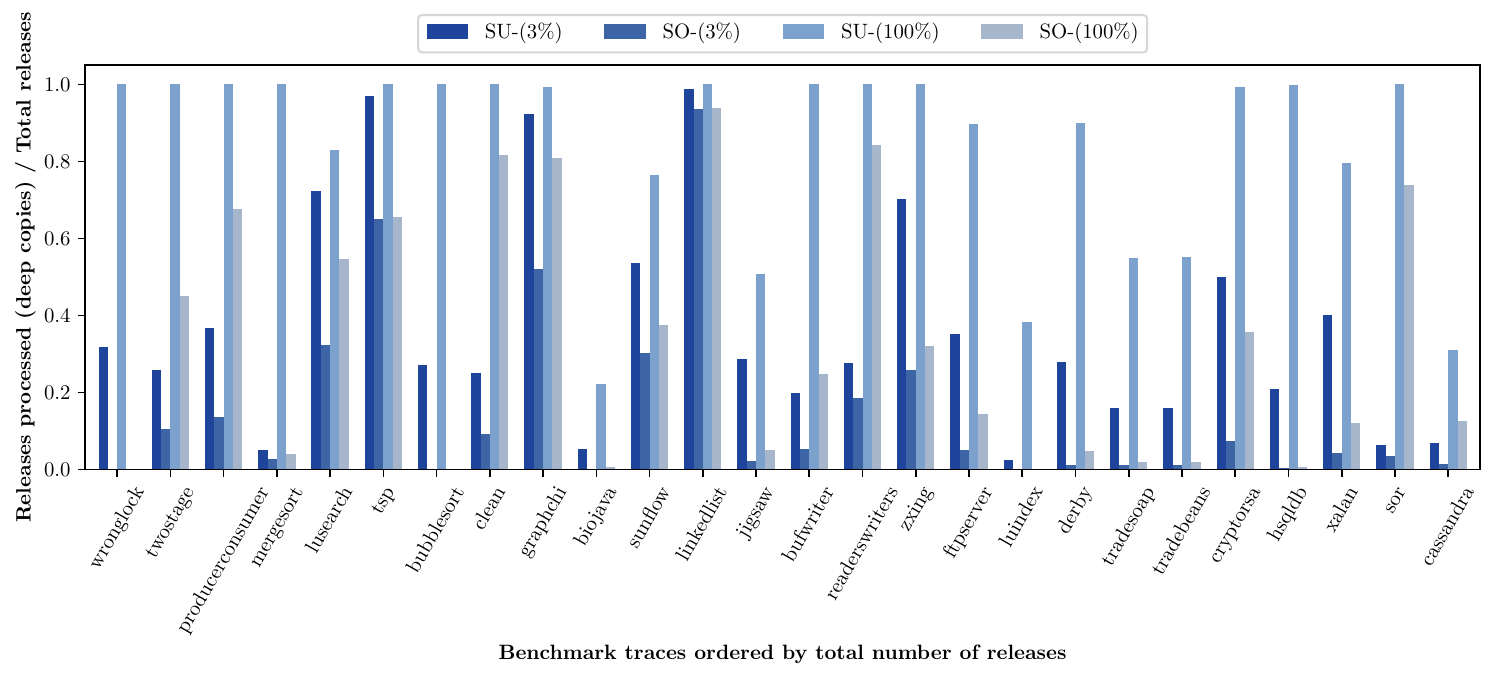}
 \vspace{-0.1in}
\caption{Ratio of releases processed or 
deep copies performed over total number of releases for four engines. 
For each benchmark, bars are ordered (from left to right, marked with darkest to lightest shades of blue) 
\SUThree, \SOThree, \SUFull, and \SOFull.}
\figlabel{releases}
\end{figure}

\myparagraph{Release events processed and deep copies created}{
Another analogous metric is the number of $O(\numthr)$ vector clock
operations performed when processing release events. 
We remark that, this case differs subtly from the case of
acquire events, since \algoref{smp-upd} and \algoref{algo4} perform
different operations at release events. 
Recall that \algoref{smp-upd} skips release events 
based on the freshness timestamp
associated with locks and threads, 
whereas \algoref{algo4} creates a shallow copy for every release
event, and shifts the $O(\numthr)$ cost of join operations onto
the deep copy operations that take place only when timestamps
are actually updated. 
Next, both \SOThree and \SOFull employ the dirty epoch optimization,
which further reduces the number of deep copies.

\figref{releases} presents the ratio of number of release events
processed and deep copies created, aggregated over
all the release events (for each algorithm). 
In contrast to~\figref{skipped-acq}, 
we can see in~\figref{releases} that the number of deep copies created by 
\SOThree is generally much smaller than the release events processed by \SUThree.
This is in line with our theoretical analysis that 
shallow copy reduces the running time complexity by a factor of $\numlk$ (i.e., number of locks).
Another interesting observation is --- the non-sampling algorithm \SUFull 
did not process all release events in some benchmarks. 
These cases arise when execution traces contain critical sections
that do not contain any shared memory access.
}

\begin{figure}[t]
\includegraphics[width=1\textwidth]{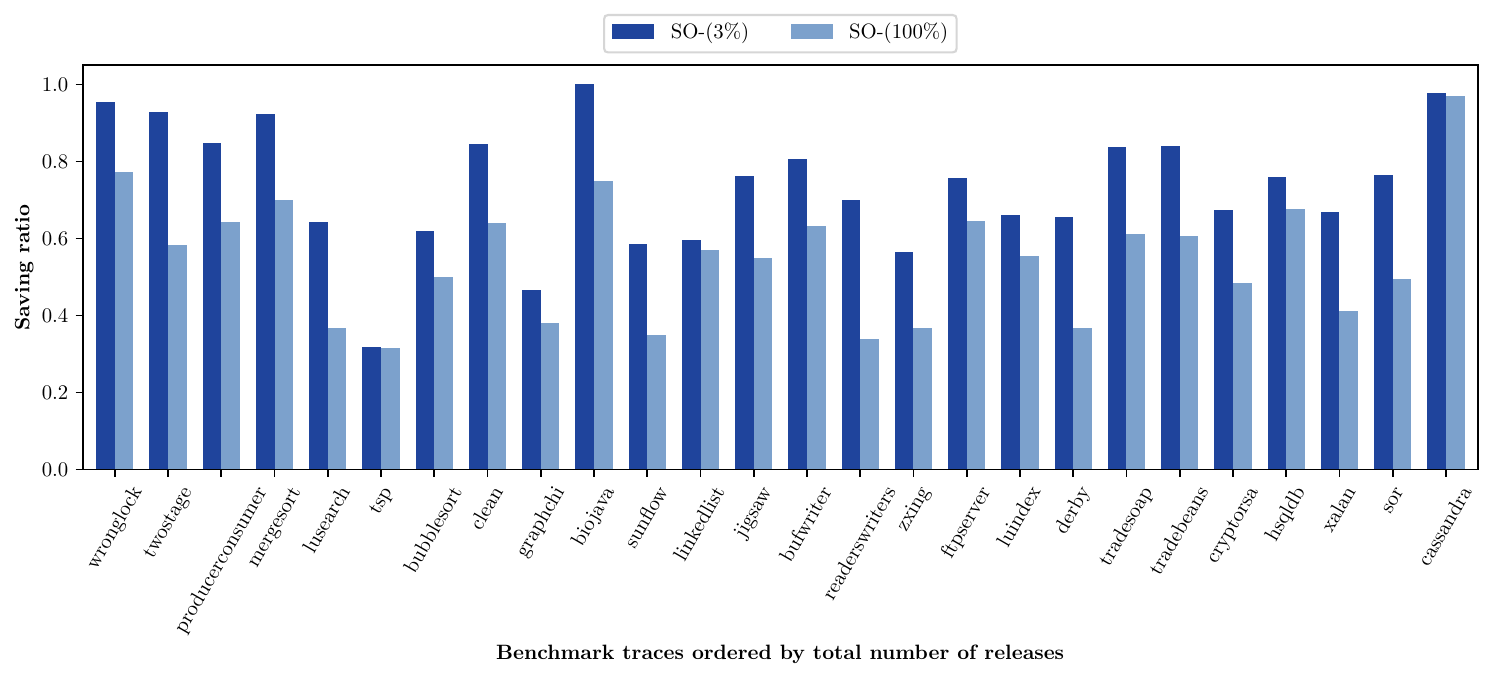}
 \vspace{-0.1in}
\caption{Saving ratio of the ordered list structure of \SOThree (dark blue) and \SOFull (light blue). }
\figlabel{saving_ratio}
\end{figure}

\myparagraph{Improvements due to the ordered list data structure}{
Next, we investigate the quantitative impact of 
the ordered list data structure in \confSO. 
To measure this, we count the number of vector clock entries that
we could afford to skip, thanks to our data structure.
More concretely, for each acquire event $e$ that was \underline{not skipped}
in \SOThree and \SOFull (this way we only measure the impact exactly due to the data structure),
we count $s_e$, the total number of entries in the vector clock
that were not traversed (i.e, the difference of $\numthr$ and the number of entries actually traversed);
see \lineref{orderlist-forloop} of \algoref{algo4}.
We then count the sum $\textsf{SavedTraversals} = \sum_{e} s_e$ over all acquire events that were not skipped.
Likewise, the total number of entries that would have been traversed
in absence of the data structure
is $\textsf{AllTraversals} = \sum_{e} \numthr$.
In \figref{saving_ratio} we report the ratio $\textsf{SavedTraversals}/\textsf{AllTraversals}$
for each benchmark, both for \SOThree and \SOFull.
We remark that this ratio is considerably high for both these algorithms.  
Further, the saving ratio of \SOThree is always higher than \SOFull,
as expected, confirming our hypothesis that the ordered-list data structure
is particularly well-suited in the context of sampling based race detection.
}

\myparagraph{Summary}{
In this section, we evaluated the effect of the key components of \algoref{smp-upd} and \algoref{algo4}. In particular, the experiment shows that the new timestamps, shallow copy and ordered list structure indeed lead to significant savings as suggested by our theoretical analysis. 
}

\subsection{Non-mutex Synchronizations of TSan}


\algoref{smp-upd} and \algoref{algo4} were discussed under the context where the only synchronization mechanism is locking, in which every release follows an acquire by the same thread. In TSan, there are various acquire and release handlers that support other synchronization mechanisms with varying semantics. Below, we outline the vector clock operations that TSan implemented for these handlers, along with example use cases and adaptations of our innovations(the time stamps and the ordered list structure) to support them.

\texttt{ReleaseStore}.
This handler is responsible for operations such as mutex unlocking, atomic release-store, or thread forking. TSan performs a vector clock copy, forcing the sync to carry the information of the thread. The innovation of \algoref{smp-upd} can't be adopted if the release is performed without the same thread having acquired the same sync beforehand (as done with mutexes).
This is because such a release may result in a non-monotonic update of vector clocks, which happens in the case of message passing using atomic release-stores for example, in which a thread might only release and never acquire. 
This handler can be optimized as per the release handler in \algoref{smp-upd} when otherwise. The innovations of  \algoref{algo4} can always be adopted.

\texttt{Release}.
 This handler is responsible for operations such as unlocking of shared locks, barrier entries, or atomic read-modify-write (RMW) and compare-and-swap (CAS) operations within a release sequence, for which a sync does not receive its timestamp from a unique thread. TSan performs a vector clock join, updating the sync's vector clock with information from the thread's vector clock. 
 We did not adopt our innovation for this case as the sync has to carry information from multiple threads simultaneously, which is not the focus of this work.

\texttt{Acquire}.
 This handler is responsible for every acquire operation such as locking, atomic acquire-load or thread join. TSan performs a vector clock join, updating the thread with the sync. No innovations can be adopted if the last release on the synchronization object was done by \texttt{Release}. Otherwise, the handler can be optimized with our innovations.
 
While it may appear that our innovations are not applicable to certain non-mutex optimizations, it is important to note that such cases are generally rare, as indicated in the experiment.



\section{Proofs}

\myparagraph{Proof of correctness of \fasttrack}{
For correctness of \propref{ft-timestamps-capture-hb} and
\lemref{djitp-soundness}, we refer readers to \cite{fasttrack} for further details.

}

\myparagraph{Proof of \propref{ft-samplestamps-capture-hb}}{
We first demonstrate that $\ctimesampling^{(\tr,S)}(e_1)(\ThreadOf{e_1}) \leq \ctimesampling^{(\tr,S)}(e_2)(\ThreadOf{e_1})$ $\text{iff}$ $e_1 \hb{\tr} e_2$. 
Now assume that $\ctimesampling^{(\tr,S)}(e_1)(\ThreadOf{e_1}) \leq \ctimesampling^{(\tr,S)}(e_2)(\ThreadOf{e_1})$.
By the definition of $\ctimesampling^{(\tr,S)}$, there must exists an event $e'$ from $\ThreadOf{e_1}$ such that  $\ltimesampling^{(\tr,S)}(e')\geq \ltimesampling^{(\tr,S)}(e_1)$ and $e' \hb{\tr} e_2$. 
Note that the local time $\ltimesampling^{(\tr,S)}$ grows  monotonically for events in the same thread so if $\ltimesampling^{(\tr,S)}(e')>\ltimesampling^{(\tr,S)}(e_1)$, we have $e_1\tho{\tr} e'$ which implies that $e_1 \hb{\tr} e_2$. 
When $\ltimesampling^{(\tr,S)}(e')=\ltimesampling^{(\tr,S)}(e_1)$, it must be the case that $e'$ and $e_1$ are from the same critical section, which also implies $e_1 \hb{\tr} e_2$. 
For the reverse direction: if $e_1 \hb{\tr} e_2$, then by definition of $\ctimesampling^{(\tr,S)}$ we must have $\ctimesampling^{(\tr,S)}(e_1)(\ThreadOf{e_1}) \leq \ctimesampling^{(\tr,S)}(e_2)(\ThreadOf{e_1})$ which completes the proof.

Then let's argue that $\ctimesampling^{(\tr,S)}(e_1) \cle \ctimesampling^{(\tr,S)}(e_2) \quad \text{iff} \quad e_1 \hb{\tr} e_2$. Similarly let's assume $\ctimesampling^{(\tr,S)}(e_1) \cle \ctimesampling^{(\tr,S)}(e_2)$ holds, which directly implies that $\ctimesampling^{(\tr,S)}(e_1)(\ThreadOf{e_1}) \leq \ctimesampling^{(\tr,S)}(e_2)(\ThreadOf{e_1})$ and the previous proof indicates that $e_1 \hb{\tr} e_2$. For the reverse direction: Assume that $\ctimesampling^{(\tr,S)}(e_1) \cle \ctimesampling^{(\tr,S)}(e_2)$ does not hold and let $i$ be a thread id such that $\ctimesampling^{(\tr,S)}(e_1)(i)>\ctimesampling^{(\tr,S)}(e_2)(i)$. By definition of $\ctimesampling^{(\tr,S)}$,  there must be a event $e'$ with $\ltimesampling^{(\tr,S)}(e')=\ctimesampling^{(\tr,S)}(e_1)(i)$ and $e'\hb{\tr} e_1$. However, since $\ctimesampling^{(\tr,S)}(e_2)(i)<\ltimesampling^{(\tr,S)}(e')$, $e'\not\hb{\tr} e_2$ which means $e_1\not\hb{\tr} e_2$. 

}
\myparagraph{Proof of \lemref{algo2-correctness}}{
It is straightforward to see the update of each $\epch_t$ and $\CFT_t$ variable aligns precisely with the definition of $\ltimesampling^{(\tr,S)}(e)$ and $\ctimesampling^{(\tr,S)}(e)$.
Therefore, the correctness follows directly from \propref{ft-samplestamps-capture-hb}.  
The $O(\trsz\numthr)$ running time comes from the fact that the algorithm does a vector clock operation for each of the acquire and release event. 

}

\myparagraph{Proof of \propref{uclock-fresh}}{
 Observe that for events $e$ and $f$ performed by the same thread $t$,  $\s{U}(e)(t)=\s{U}(f)(t)$ implies $\ctimesampling^{(\tr,S)}(e) = \ctimesampling^{(\tr,S)}(f) $; this follows from the definition of the freshness timestamp. Next, if for events $e_1$(performed by $t_1$) and $e_2$(performed by $t_2$) , $\s{U}(e_1)(t_1)\leq\s{U}(e_2)(t_1)$ then there must exists an event $e'$ performed by $t_1$ such that $\s{U}(e_1)=\s{U}(e')$ and $e'\hb{\tr} e_2$. Since $e'\hb{\tr} e_2$, by \propref{ft-samplestamps-capture-hb}, we have  $\ctimesampling^{(\tr,S)}(e_1)=
\ctimesampling^{(\tr,S)}(e')\cle \ctimesampling^{(\tr,S)}(e_2)$, which completes the proof.
}

\myparagraph{Proof of \propref{uclock-numentry}}{
When $k<=0$, the case is covered by \propref{uclock-fresh}. 
Since it is also trivial the number of threads $t$ such that $\ctimesampling^{(\tr,S)}(e_1)(t)>\ctimesampling^{(\tr,S)}(e_2)(t)$ is upper bounded by $\numthr$,
we can assume that $0<k<\numthr$. If $\s{U}(e_2)((\ThreadOf{e_1})=0$, we simply have $k=\s{U}(e_1)(\ThreadOf{e_1})$.  By definition, $\ctimesampling^{(\tr,S)}(e')$ have only updated $k$ times across all $e'\tho{tr} e_1$. So there are at most $k$ none-zero entries of $\ctimesampling^{(\tr,S)}(e_1)$. When $\s{U}(e_2)((\ThreadOf{e_1})>0$, let $e'$ be the event from $\ThreadOf{e_1}$ such that $e'\hb{\tr} e_2$ and $\s{U}(e')(\ThreadOf{e_1})=\s{U}(e_2)(\ThreadOf{e_1})$. Following \propref{uclock-fresh} we have $\ctimesampling^{(\tr,S)}(e')\cle \ctimesampling^{(\tr,S)}(e_2)$. By definition, $\ctimesampling^{(\tr,S)}(e'')$ have only updated $k$ times across all events $e''$ such that $e'\tho{\tr}e''\tho{\tr} e_1$, which completes the proof.
}

\myparagraph{Proof of \lemref{algo3-correctness}}{First observe that the $\CUpd_t$ variable kept by \algoref{smp-upd}
stores a timestamp $\s{U'}\cle \s{U}(e)$ for every event $e$ with  $\s{U'}(e)(t)=\s{U}(e)(t)$ and therefore \propref{uclock-fresh} can be applied. 
Then correctness follows from showing that the $\ctimesampling^{(\tr,S)}$ timestamp for each event computed by \algoref{smp-upd} are the same as those computed by \algoref{algo2}. This is established by induction on the number of events processed to date. Base case follows from the fact that sampling clocks are initialized to the same value. The inductive case follows when the processed event is not an acquire or release. When the new event is an acquire or release and conditions in line 11 or 23 are satisfied, \algoref{smp-upd} updates clocks in the same way as \algoref{algo2}. When conditions in line 11 and 23 are not satisfied, \propref{uclock-fresh} ensures that the copy/join operations performed by \algoref{algo2} do not alter state, ensuring correctness.

Running time:
Let us fix the execution length to be $\trsz$, the number of threads to be $\numthr$, and the number of locks to be $\numlk$. To determine the running time, we need to count the number of times we perform vector clock operations, each of which take $O(\numthr)$ time. At an acquire, we perform a vector clock operation when $\CUpd_\lk(\lastrelthr_\lk) > \CUpd_t(\lastrelthr_\lk)$. Note that $\CUpd_t(t')$ is at most the number of times $\CSmp_{t'}$ changes, which we argued is at most $|S|$. Since vector clocks increase monotonically, for a fixed thread $t$, the number of acquires that perform a vector clock operation is at most $|S|\numthr$. As there are $\numthr$ threads, the number of vector clock operations in all the acquires is at most $O(|S|\numthr^2)$. Next, let us count the number of vector clock copies that take place in releases. In a release, \algoref{smp-upd} does a vector clock copy when $\CUpd_t(t) \neq \CUpd_\lk(t)$. For a fixed lock $\lk$, by an argument similar to the case of acquires, this can happen at most $|S|\numthr$ times. Thus the total number of vector clock operations from all the releases is $O(|S|\numthr\numlk)$. Putting it all together, the running time of \algoref{smp-upd} is $O(\trsz) + O(|S|\numthr(\numthr+\numlk))O(\numthr)$.

}
\myparagraph{Proof of \lemref{algo4-correctness}}{
Similarly to \algoref{smp-upd}, first note that the $\CUpd_t$ variable kept by \algoref{algo4}
also stores a timestamp $\s{U'}\cle \s{U}(e)$ for every event $e$ with $\s{U'}(e)(t)=\s{U}(e)(t)$ and therefore both \propref{uclock-fresh} and \propref{uclock-numentry} can be appropriately extended into this case. 
It is sufficient to prove that the $\ctimesampling^{(\tr,S)}$ timestamp for each event computed by \algoref{algo4} are the same as those computed by \algoref{algo2}. 
The proof is again established by induction on the number of events processed to date. Base case follows from the fact that sampling clocks are initialized to the same value. The inductive case follows when the processed event is not an acquire or release.  When the event is an release, the shallow copy operation changes the state of the join operation in \algoref{algo2}. When the event is an acquire, if the condition on line 12 holds, then it follows \propref{uclock-fresh} that the corresponding join operation performed by \algoref{algo2} does not alter the state. If the condition is not satisfied, the loop performs pair-wise max for the first $\lastrelu_\lk - \CUpd_t(\lastrelthr_\lk)$ entries of $\COmp_l$. The correctness follows from \propref{uclock-numentry} and the definition of the ordered list data structure. We also remark that a deepcopy of every $\COmp_t$ is created whenever $\COmp_t$ is shared among objects and needs to be updated.  

Running time:
Let us start by counting the cost incurred due to the deep copies. Now, a thread $t$ maybe forced to create a deep copy whenever an entry of $\COmp_t$ is changed. But that can happen at most $|S|$ times! Next, during join operations in the acquire handler of a thread $t$, the total number of $\COmp_l$ entries traversed(for all $l$) is at most the sum of entries of $\CUpd_t$, which is bounded by $|S|\numthr$. Thus, the total running time of \algoref{algo4} is $O(\trsz) + O(|S|\numthr)O(\numthr)$. Contrast this with $O(\trsz)+ O(|S|\numthr(\numthr + \numlk))O(\numthr)$ which is the running time of \algoref{smp-upd}. 

}
\myparagraph{Proof of \lemref{algo4-optimality}}{
Similarly to the running time proof presented for \lemref{algo4-correctness}, the lemma can be proved by evaluating the number of deep copies created and entries traversed. 
First note a deep copy is created whenever $\COmp_t$ is changed for any $t$, and by definition this is exactly $O(\s{VTWORK}(\tr))$. Also note that the sum of entries of $\s{U}_t$ for a thread $t$ is also bounded by $O(\s{VTWORK}(\tr))$, which implies that the total number of entries traversed by $t$ in acquires is at most $O(\s{VTWORK}(\tr))$. Therefore in total we conclude that the running time is $O(\trsz) + O(\s{VTWORK}(\tr)\numthr)$.

}
\newpage
\bibliographystyle{ACM-Reference-Format}
\bibliography{references}


\begin{thebibliography}{69}


\ifx \showCODEN    \undefined \def \showCODEN     #1{\unskip}     \fi
\ifx \showDOI      \undefined \def \showDOI       #1{#1}\fi
\ifx \showISBNx    \undefined \def \showISBNx     #1{\unskip}     \fi
\ifx \showISBNxiii \undefined \def \showISBNxiii  #1{\unskip}     \fi
\ifx \showISSN     \undefined \def \showISSN      #1{\unskip}     \fi
\ifx \showLCCN     \undefined \def \showLCCN      #1{\unskip}     \fi
\ifx \shownote     \undefined \def \shownote      #1{#1}          \fi
\ifx \showarticletitle \undefined \def \showarticletitle #1{#1}   \fi
\ifx \showURL      \undefined \def \showURL       {\relax}        \fi
\providecommand\bibfield[2]{#2}
\providecommand\bibinfo[2]{#2}
\providecommand\natexlab[1]{#1}
\providecommand\showeprint[2][]{arXiv:#2}

\bibitem[\protect\citeauthoryear{Abadi, Flanagan, and Freund}{Abadi
  et~al\mbox{.}}{2006}]%
        {Abadi:2006:TSL:1119479.1119480}
\bibfield{author}{\bibinfo{person}{Martin Abadi}, \bibinfo{person}{Cormac
  Flanagan}, {and} \bibinfo{person}{Stephen~N. Freund}.}
  \bibinfo{year}{2006}\natexlab{}.
\newblock \showarticletitle{{Types for Safe Locking: Static Race Detection for
  Java}}.
\newblock \bibinfo{journal}{\emph{ACM Trans. Program. Lang. Syst.}}
  \bibinfo{volume}{28}, \bibinfo{number}{2} (\bibinfo{date}{March}
  \bibinfo{year}{2006}), \bibinfo{pages}{207--255}.
\newblock


\bibitem[\protect\citeauthoryear{Abdulla, Aronis, Jonsson, and Sagonas}{Abdulla
  et~al\mbox{.}}{2014a}]%
        {abdulla2014optimal}
\bibfield{author}{\bibinfo{person}{Parosh Abdulla}, \bibinfo{person}{Stavros
  Aronis}, \bibinfo{person}{Bengt Jonsson}, {and} \bibinfo{person}{Konstantinos
  Sagonas}.} \bibinfo{year}{2014}\natexlab{a}.
\newblock \showarticletitle{Optimal dynamic partial order reduction}.
\newblock \bibinfo{journal}{\emph{ACM SIGPLAN Notices}} \bibinfo{volume}{49},
  \bibinfo{number}{1} (\bibinfo{year}{2014}), \bibinfo{pages}{373--384}.
\newblock


\bibitem[\protect\citeauthoryear{Abdulla, Aronis, Jonsson, and Sagonas}{Abdulla
  et~al\mbox{.}}{2014b}]%
        {Abdulla14}
\bibfield{author}{\bibinfo{person}{Parosh Abdulla}, \bibinfo{person}{Stavros
  Aronis}, \bibinfo{person}{Bengt Jonsson}, {and} \bibinfo{person}{Konstantinos
  Sagonas}.} \bibinfo{year}{2014}\natexlab{b}.
\newblock \showarticletitle{Optimal Dynamic Partial Order Reduction}. In
  \bibinfo{booktitle}{\emph{Proceedings of the 41st ACM SIGPLAN-SIGACT
  Symposium on Principles of Programming Languages}} (San Diego, California,
  USA) \emph{(\bibinfo{series}{POPL '14})}. \bibinfo{publisher}{ACM},
  \bibinfo{address}{New York, NY, USA}, \bibinfo{pages}{373--384}.
\newblock
\showISBNx{978-1-4503-2544-8}
\urldef\tempurl%
\url{https://doi.org/10.1145/2535838.2535845}
\showDOI{\tempurl}


\bibitem[\protect\citeauthoryear{Agarwal, Chatterjee, Pathak, Pavlogiannis, and
  Toman}{Agarwal et~al\mbox{.}}{2021}]%
        {agarwal2021stateless}
\bibfield{author}{\bibinfo{person}{Pratyush Agarwal},
  \bibinfo{person}{Krishnendu Chatterjee}, \bibinfo{person}{Shreya Pathak},
  \bibinfo{person}{Andreas Pavlogiannis}, {and} \bibinfo{person}{Viktor
  Toman}.} \bibinfo{year}{2021}\natexlab{}.
\newblock \showarticletitle{Stateless model checking under a reads-value-from
  equivalence}. In \bibinfo{booktitle}{\emph{International Conference on
  Computer Aided Verification}}. Springer, \bibinfo{pages}{341--366}.
\newblock


\bibitem[\protect\citeauthoryear{Ang and Mathur}{Ang and Mathur}{2024a}]%
        {AngPatternLanguages2024}
\bibfield{author}{\bibinfo{person}{Zhendong Ang} {and} \bibinfo{person}{Umang
  Mathur}.} \bibinfo{year}{2024}\natexlab{a}.
\newblock \showarticletitle{Predictive Monitoring against Pattern Regular
  Languages}.
\newblock \bibinfo{journal}{\emph{Proc. ACM Program. Lang.}}
  \bibinfo{volume}{8}, \bibinfo{number}{POPL}, Article \bibinfo{articleno}{73}
  (\bibinfo{date}{Jan.} \bibinfo{year}{2024}), \bibinfo{numpages}{35}~pages.
\newblock
\urldef\tempurl%
\url{https://doi.org/10.1145/3632915}
\showDOI{\tempurl}


\bibitem[\protect\citeauthoryear{Ang and Mathur}{Ang and Mathur}{2024b}]%
        {AngTracePrefixes2024}
\bibfield{author}{\bibinfo{person}{Zhendong Ang} {and} \bibinfo{person}{Umang
  Mathur}.} \bibinfo{year}{2024}\natexlab{b}.
\newblock \showarticletitle{Predictive Monitoring with Strong Trace Prefixes}.
  In \bibinfo{booktitle}{\emph{Computer Aided Verification}},
  \bibfield{editor}{\bibinfo{person}{Arie Gurfinkel} {and}
  \bibinfo{person}{Vijay Ganesh}} (Eds.). \bibinfo{publisher}{Springer Nature
  Switzerland}, \bibinfo{address}{Cham}, \bibinfo{pages}{182--204}.
\newblock
\showISBNx{978-3-031-65630-9}


\bibitem[\protect\citeauthoryear{Biswas, Cao, Zhang, Bond, and Wood}{Biswas
  et~al\mbox{.}}{2017}]%
        {racechaser}
\bibfield{author}{\bibinfo{person}{Swarnendu Biswas}, \bibinfo{person}{Man
  Cao}, \bibinfo{person}{Minjia Zhang}, \bibinfo{person}{Michael~D. Bond},
  {and} \bibinfo{person}{Benjamin~P. Wood}.} \bibinfo{year}{2017}\natexlab{}.
\newblock \showarticletitle{Lightweight Data Race Detection for Production
  Runs}. In \bibinfo{booktitle}{\emph{Proceedings of the 26th International
  Conference on Compiler Construction}} (Austin, TX, USA)
  \emph{(\bibinfo{series}{CC 2017})}. \bibinfo{publisher}{Association for
  Computing Machinery}, \bibinfo{address}{New York, NY, USA},
  \bibinfo{pages}{11–21}.
\newblock
\showISBNx{9781450352338}
\urldef\tempurl%
\url{https://doi.org/10.1145/3033019.3033020}
\showDOI{\tempurl}


\bibitem[\protect\citeauthoryear{Blackburn, Garner, Hoffmann, Khang, McKinley,
  Bentzur, Diwan, Feinberg, Frampton, Guyer, Hirzel, Hosking, Jump, Lee, Moss,
  Phansalkar, Stefanovi\'{c}, VanDrunen, von Dincklage, and
  Wiedermann}{Blackburn et~al\mbox{.}}{2006}]%
        {DaCapo2006}
\bibfield{author}{\bibinfo{person}{Stephen~M. Blackburn},
  \bibinfo{person}{Robin Garner}, \bibinfo{person}{Chris Hoffmann},
  \bibinfo{person}{Asjad~M. Khang}, \bibinfo{person}{Kathryn~S. McKinley},
  \bibinfo{person}{Rotem Bentzur}, \bibinfo{person}{Amer Diwan},
  \bibinfo{person}{Daniel Feinberg}, \bibinfo{person}{Daniel Frampton},
  \bibinfo{person}{Samuel~Z. Guyer}, \bibinfo{person}{Martin Hirzel},
  \bibinfo{person}{Antony Hosking}, \bibinfo{person}{Maria Jump},
  \bibinfo{person}{Han Lee}, \bibinfo{person}{J.~Eliot~B. Moss},
  \bibinfo{person}{Aashish Phansalkar}, \bibinfo{person}{Darko Stefanovi\'{c}},
  \bibinfo{person}{Thomas VanDrunen}, \bibinfo{person}{Daniel von Dincklage},
  {and} \bibinfo{person}{Ben Wiedermann}.} \bibinfo{year}{2006}\natexlab{}.
\newblock \showarticletitle{The DaCapo Benchmarks: Java Benchmarking
  Development and Analysis}. In \bibinfo{booktitle}{\emph{Proceedings of the
  21st Annual ACM SIGPLAN Conference on Object-oriented Programming Systems,
  Languages, and Applications}} (Portland, Oregon, USA)
  \emph{(\bibinfo{series}{OOPSLA '06})}. \bibinfo{publisher}{ACM},
  \bibinfo{address}{New York, NY, USA}, \bibinfo{pages}{169--190}.
\newblock
\showISBNx{1-59593-348-4}
\urldef\tempurl%
\url{https://doi.org/10.1145/1167473.1167488}
\showDOI{\tempurl}


\bibitem[\protect\citeauthoryear{Blackshear, Gorogiannis, O'Hearn, and
  Sergey}{Blackshear et~al\mbox{.}}{2018}]%
        {racerd2018}
\bibfield{author}{\bibinfo{person}{Sam Blackshear}, \bibinfo{person}{Nikos
  Gorogiannis}, \bibinfo{person}{Peter~W. O'Hearn}, {and} \bibinfo{person}{Ilya
  Sergey}.} \bibinfo{year}{2018}\natexlab{}.
\newblock \showarticletitle{RacerD: Compositional Static Race Detection}.
\newblock \bibinfo{journal}{\emph{Proc. ACM Program. Lang.}}
  \bibinfo{volume}{2}, \bibinfo{number}{OOPSLA}, Article
  \bibinfo{articleno}{144} (\bibinfo{date}{oct} \bibinfo{year}{2018}),
  \bibinfo{numpages}{28}~pages.
\newblock
\urldef\tempurl%
\url{https://doi.org/10.1145/3276514}
\showDOI{\tempurl}


\bibitem[\protect\citeauthoryear{Boehm}{Boehm}{2012}]%
        {evil2012}
\bibfield{author}{\bibinfo{person}{Hans-J. Boehm}.}
  \bibinfo{year}{2012}\natexlab{}.
\newblock \showarticletitle{Position Paper: Nondeterminism is Unavoidable, but
  Data Races Are Pure Evil}. In \bibinfo{booktitle}{\emph{Proceedings of the
  2012 ACM Workshop on Relaxing Synchronization for Multicore and Manycore
  Scalability}} (Tucson, Arizona, USA) \emph{(\bibinfo{series}{RACES ’12})}.
  \bibinfo{publisher}{Association for Computing Machinery},
  \bibinfo{address}{New York, NY, USA}, \bibinfo{pages}{9–14}.
\newblock
\showISBNx{9781450316323}
\urldef\tempurl%
\url{https://doi.org/10.1145/2414729.2414732}
\showDOI{\tempurl}


\bibitem[\protect\citeauthoryear{Bond, Coons, and McKinley}{Bond
  et~al\mbox{.}}{2010}]%
        {pacer-tool}
\bibfield{author}{\bibinfo{person}{Michael~D. Bond},
  \bibinfo{person}{Katherine~E. Coons}, {and} \bibinfo{person}{Kathryn~S.
  McKinley}.} \bibinfo{year}{2010}\natexlab{}.
\newblock \showarticletitle{PACER: Proportional Detection of Data Races}. In
  \bibinfo{booktitle}{\emph{Proceedings of the 31st ACM SIGPLAN Conference on
  Programming Language Design and Implementation}} (Toronto, Ontario, Canada)
  \emph{(\bibinfo{series}{PLDI '10})}. \bibinfo{publisher}{ACM},
  \bibinfo{address}{New York, NY, USA}, \bibinfo{pages}{255--268}.
\newblock
\showISBNx{978-1-4503-0019-3}
\urldef\tempurl%
\url{https://doi.org/10.1145/1806596.1806626}
\showDOI{\tempurl}


\bibitem[\protect\citeauthoryear{Bond, Kulkarni, Cao, Zhang, Fathi~Salmi,
  Biswas, Sengupta, and Huang}{Bond et~al\mbox{.}}{2013}]%
        {Bond2013}
\bibfield{author}{\bibinfo{person}{Michael~D. Bond}, \bibinfo{person}{Milind
  Kulkarni}, \bibinfo{person}{Man Cao}, \bibinfo{person}{Minjia Zhang},
  \bibinfo{person}{Meisam Fathi~Salmi}, \bibinfo{person}{Swarnendu Biswas},
  \bibinfo{person}{Aritra Sengupta}, {and} \bibinfo{person}{Jipeng Huang}.}
  \bibinfo{year}{2013}\natexlab{}.
\newblock \showarticletitle{OCTET: Capturing and Controlling Cross-thread
  Dependences Efficiently}.
\newblock \bibinfo{journal}{\emph{SIGPLAN Not.}} \bibinfo{volume}{48},
  \bibinfo{number}{10} (\bibinfo{date}{Oct.} \bibinfo{year}{2013}),
  \bibinfo{pages}{693--712}.
\newblock
\showISSN{0362-1340}
\urldef\tempurl%
\url{https://doi.org/10.1145/2544173.2509519}
\showDOI{\tempurl}


\bibitem[\protect\citeauthoryear{Burckhardt, Kothari, Musuvathi, and
  Nagarakatte}{Burckhardt et~al\mbox{.}}{2010}]%
        {PCT2010}
\bibfield{author}{\bibinfo{person}{Sebastian Burckhardt},
  \bibinfo{person}{Pravesh Kothari}, \bibinfo{person}{Madanlal Musuvathi},
  {and} \bibinfo{person}{Santosh Nagarakatte}.}
  \bibinfo{year}{2010}\natexlab{}.
\newblock \showarticletitle{A Randomized Scheduler with Probabilistic
  Guarantees of Finding Bugs}. In \bibinfo{booktitle}{\emph{Proceedings of the
  Fifteenth International Conference on Architectural Support for Programming
  Languages and Operating Systems}} (Pittsburgh, Pennsylvania, USA)
  \emph{(\bibinfo{series}{ASPLOS XV})}. \bibinfo{publisher}{Association for
  Computing Machinery}, \bibinfo{address}{New York, NY, USA},
  \bibinfo{pages}{167–178}.
\newblock
\showISBNx{9781605588391}
\urldef\tempurl%
\url{https://doi.org/10.1145/1736020.1736040}
\showDOI{\tempurl}


\bibitem[\protect\citeauthoryear{Choi, Lee, Loginov, O'Callahan, Sarkar, and
  Sridharan}{Choi et~al\mbox{.}}{2002}]%
        {Choi02}
\bibfield{author}{\bibinfo{person}{Jong-Deok Choi}, \bibinfo{person}{Keunwoo
  Lee}, \bibinfo{person}{Alexey Loginov}, \bibinfo{person}{Robert O'Callahan},
  \bibinfo{person}{Vivek Sarkar}, {and} \bibinfo{person}{Manu Sridharan}.}
  \bibinfo{year}{2002}\natexlab{}.
\newblock \showarticletitle{Efficient and Precise Datarace Detection for
  Multithreaded Object-oriented Programs}. In
  \bibinfo{booktitle}{\emph{Proceedings of the ACM SIGPLAN 2002 Conference on
  Programming Language Design and Implementation}} (Berlin, Germany)
  \emph{(\bibinfo{series}{PLDI '02})}. \bibinfo{publisher}{ACM},
  \bibinfo{address}{New York, NY, USA}, \bibinfo{pages}{258--269}.
\newblock
\showISBNx{1-58113-463-0}
\urldef\tempurl%
\url{https://doi.org/10.1145/512529.512560}
\showDOI{\tempurl}


\bibitem[\protect\citeauthoryear{Chromium Blog}{Chromium Blog}{2014}]%
        {TSanChromium}
Chromium Blog \bibinfo{year}{2014}\natexlab{}.
\newblock \bibinfo{title}{Testing Chromium: ThreadSanitizer v2, a next-gen data
  race detector}.
\newblock
  \bibinfo{howpublished}{\url{https://blog.chromium.org/2014/04/testing-chromium-threadsanitizer-v2.html}}.
\newblock
\newblock
\shownote{Accessed: 2024-07-11.}


\bibitem[\protect\citeauthoryear{Difallah, Pavlo, Curino, and
  Cudr{\'e}-Mauroux}{Difallah et~al\mbox{.}}{2013}]%
        {DifallahPCC13}
\bibfield{author}{\bibinfo{person}{Djellel~Eddine Difallah},
  \bibinfo{person}{Andrew Pavlo}, \bibinfo{person}{Carlo Curino}, {and}
  \bibinfo{person}{Philippe Cudr{\'e}-Mauroux}.}
  \bibinfo{year}{2013}\natexlab{}.
\newblock \showarticletitle{OLTP-Bench: An Extensible Testbed for Benchmarking
  Relational Databases}.
\newblock \bibinfo{journal}{\emph{PVLDB}} \bibinfo{volume}{7},
  \bibinfo{number}{4} (\bibinfo{year}{2013}), \bibinfo{pages}{277--288}.
\newblock
\urldef\tempurl%
\url{http://www.vldb.org/pvldb/vol7/p277-difallah.pdf}
\showURL{%
\tempurl}


\bibitem[\protect\citeauthoryear{Do, Elbaum, and Rothermel}{Do
  et~al\mbox{.}}{2005}]%
        {SIR2005}
\bibfield{author}{\bibinfo{person}{Hyunsook Do}, \bibinfo{person}{Sebastian
  Elbaum}, {and} \bibinfo{person}{Gregg Rothermel}.}
  \bibinfo{year}{2005}\natexlab{}.
\newblock \showarticletitle{Supporting controlled experimentation with testing
  techniques: An infrastructure and its potential impact}.
\newblock \bibinfo{journal}{\emph{Empirical Software Engineering: An
  International Journal}}  \bibinfo{volume}{10} (\bibinfo{year}{2005}),
  \bibinfo{pages}{405--435}.
\newblock


\bibitem[\protect\citeauthoryear{Elmas, Qadeer, and Tasiran}{Elmas
  et~al\mbox{.}}{2007}]%
        {elmas2007goldilocks}
\bibfield{author}{\bibinfo{person}{Tayfun Elmas}, \bibinfo{person}{Shaz
  Qadeer}, {and} \bibinfo{person}{Serdar Tasiran}.}
  \bibinfo{year}{2007}\natexlab{}.
\newblock \showarticletitle{Goldilocks: A Race and Transaction-aware Java
  Runtime}. In \bibinfo{booktitle}{\emph{Proceedings of the 28th ACM SIGPLAN
  Conference on Programming Language Design and Implementation}} (San Diego,
  California, USA) \emph{(\bibinfo{series}{PLDI '07})}.
  \bibinfo{publisher}{ACM}, \bibinfo{address}{New York, NY, USA},
  \bibinfo{pages}{245--255}.
\newblock
\showISBNx{978-1-59593-633-2}
\urldef\tempurl%
\url{https://doi.org/10.1145/1250734.1250762}
\showDOI{\tempurl}


\bibitem[\protect\citeauthoryear{Engler and Ashcraft}{Engler and
  Ashcraft}{2003}]%
        {racerx}
\bibfield{author}{\bibinfo{person}{Dawson Engler} {and} \bibinfo{person}{Ken
  Ashcraft}.} \bibinfo{year}{2003}\natexlab{}.
\newblock \showarticletitle{{RacerX: Effective, Static Detection of Race
  Conditions and Deadlocks}}.
\newblock \bibinfo{journal}{\emph{SIGOPS Oper. Syst. Rev.}}
  \bibinfo{volume}{37}, \bibinfo{number}{5} (\bibinfo{date}{Oct.}
  \bibinfo{year}{2003}), \bibinfo{pages}{237--252}.
\newblock


\bibitem[\protect\citeauthoryear{Erickson, Musuvathi, Burckhardt, and
  Olynyk}{Erickson et~al\mbox{.}}{2010}]%
        {datacollider2010}
\bibfield{author}{\bibinfo{person}{John Erickson}, \bibinfo{person}{Madanlal
  Musuvathi}, \bibinfo{person}{Sebastian Burckhardt}, {and}
  \bibinfo{person}{Kirk Olynyk}.} \bibinfo{year}{2010}\natexlab{}.
\newblock \showarticletitle{Effective Data-Race Detection for the Kernel}. In
  \bibinfo{booktitle}{\emph{Proceedings of the 9th USENIX Conference on
  Operating Systems Design and Implementation}} (Vancouver, BC, Canada)
  \emph{(\bibinfo{series}{OSDI'10})}. \bibinfo{publisher}{USENIX Association},
  \bibinfo{address}{USA}, \bibinfo{pages}{151–162}.
\newblock


\bibitem[\protect\citeauthoryear{Farchi, Nir, and Ur}{Farchi
  et~al\mbox{.}}{2003}]%
        {Farchi2003}
\bibfield{author}{\bibinfo{person}{Eitan Farchi}, \bibinfo{person}{Yarden Nir},
  {and} \bibinfo{person}{Shmuel Ur}.} \bibinfo{year}{2003}\natexlab{}.
\newblock \showarticletitle{{Concurrent Bug Patterns and How to Test Them}}. In
  \bibinfo{booktitle}{\emph{Proceedings of the 17th International Symposium on
  Parallel and Distributed Processing}} \emph{(\bibinfo{series}{IPDPS '03})}.
  \bibinfo{publisher}{IEEE Computer Society}, \bibinfo{address}{Washington, DC,
  USA}, \bibinfo{pages}{286.2--}.
\newblock


\bibitem[\protect\citeauthoryear{Farzan and Mathur}{Farzan and Mathur}{2024}]%
        {FarzanMathur2024}
\bibfield{author}{\bibinfo{person}{Azadeh Farzan} {and} \bibinfo{person}{Umang
  Mathur}.} \bibinfo{year}{2024}\natexlab{}.
\newblock \showarticletitle{Coarser Equivalences for Causal Concurrency}.
\newblock \bibinfo{journal}{\emph{Proc. ACM Program. Lang.}}
  \bibinfo{volume}{8}, \bibinfo{number}{POPL}, Article \bibinfo{articleno}{31}
  (\bibinfo{date}{Jan.} \bibinfo{year}{2024}), \bibinfo{numpages}{31}~pages.
\newblock
\urldef\tempurl%
\url{https://doi.org/10.1145/3632873}
\showDOI{\tempurl}


\bibitem[\protect\citeauthoryear{Fidge}{Fidge}{1991}]%
        {Fidge:1991:LTD:112827.112860}
\bibfield{author}{\bibinfo{person}{Colin Fidge}.}
  \bibinfo{year}{1991}\natexlab{}.
\newblock \showarticletitle{Logical Time in Distributed Computing Systems}.
\newblock \bibinfo{journal}{\emph{Computer}} \bibinfo{volume}{24},
  \bibinfo{number}{8} (\bibinfo{date}{Aug.} \bibinfo{year}{1991}),
  \bibinfo{pages}{28--33}.
\newblock
\showISSN{0018-9162}
\urldef\tempurl%
\url{https://doi.org/10.1109/2.84874}
\showDOI{\tempurl}


\bibitem[\protect\citeauthoryear{Flanagan and Freund}{Flanagan and
  Freund}{2009}]%
        {fasttrack}
\bibfield{author}{\bibinfo{person}{Cormac Flanagan} {and}
  \bibinfo{person}{Stephen~N. Freund}.} \bibinfo{year}{2009}\natexlab{}.
\newblock \showarticletitle{FastTrack: Efficient and Precise Dynamic Race
  Detection}. In \bibinfo{booktitle}{\emph{Proceedings of the 30th ACM SIGPLAN
  Conference on Programming Language Design and Implementation}} (Dublin,
  Ireland) \emph{(\bibinfo{series}{PLDI '09})}. \bibinfo{publisher}{ACM},
  \bibinfo{address}{New York, NY, USA}, \bibinfo{pages}{121--133}.
\newblock
\showISBNx{978-1-60558-392-1}
\urldef\tempurl%
\url{https://doi.org/10.1145/1542476.1542490}
\showDOI{\tempurl}


\bibitem[\protect\citeauthoryear{Flanagan and Freund}{Flanagan and
  Freund}{2013}]%
        {redcard}
\bibfield{author}{\bibinfo{person}{Cormac Flanagan} {and}
  \bibinfo{person}{Stephen~N. Freund}.} \bibinfo{year}{2013}\natexlab{}.
\newblock \showarticletitle{RedCard: Redundant Check Elimination for Dynamic
  Race Detectors}. In \bibinfo{booktitle}{\emph{Proceedings of the 27th
  European Conference on Object-Oriented Programming}} (Montpellier, France)
  \emph{(\bibinfo{series}{ECOOP'13})}. \bibinfo{publisher}{Springer-Verlag},
  \bibinfo{address}{Berlin, Heidelberg}, \bibinfo{pages}{255--280}.
\newblock
\showISBNx{978-3-642-39037-1}


\bibitem[\protect\citeauthoryear{Gorogiannis, O’Hearn, and
  Sergey}{Gorogiannis et~al\mbox{.}}{2019}]%
        {racerdx2019}
\bibfield{author}{\bibinfo{person}{Nikos Gorogiannis},
  \bibinfo{person}{Peter~W. O’Hearn}, {and} \bibinfo{person}{Ilya Sergey}.}
  \bibinfo{year}{2019}\natexlab{}.
\newblock \showarticletitle{A True Positives Theorem for a Static Race
  Detector}.
\newblock \bibinfo{journal}{\emph{Proc. ACM Program. Lang.}}
  \bibinfo{volume}{3}, \bibinfo{number}{POPL}, Article \bibinfo{articleno}{57}
  (\bibinfo{date}{Jan.} \bibinfo{year}{2019}), \bibinfo{numpages}{29}~pages.
\newblock
\urldef\tempurl%
\url{https://doi.org/10.1145/3290370}
\showDOI{\tempurl}


\bibitem[\protect\citeauthoryear{Holler, Beingessner, and Wright}{Holler
  et~al\mbox{.}}{2021}]%
        {TSanFirefox}
\bibfield{author}{\bibinfo{person}{Christian Holler}, \bibinfo{person}{Aria
  Beingessner}, {and} \bibinfo{person}{Kris Wright}.}
  \bibinfo{year}{2021}\natexlab{}.
\newblock \bibinfo{title}{Eliminating Data Races in Firefox – A Technical
  Report}.
\newblock
  \bibinfo{howpublished}{\url{https://hacks.mozilla.org/2021/04/eliminating-data-races-in-firefox-a-technical-report/}}.
\newblock
\newblock
\shownote{Accessed: 2022-07-11.}


\bibitem[\protect\citeauthoryear{Huang, Meredith, and Rosu}{Huang
  et~al\mbox{.}}{2014}]%
        {rv2014}
\bibfield{author}{\bibinfo{person}{Jeff Huang}, \bibinfo{person}{Patrick~O'Neil
  Meredith}, {and} \bibinfo{person}{Grigore Rosu}.}
  \bibinfo{year}{2014}\natexlab{}.
\newblock \showarticletitle{Maximal Sound Predictive Race Detection with
  Control Flow Abstraction}. In \bibinfo{booktitle}{\emph{Proceedings of the
  35th ACM SIGPLAN Conference on Programming Language Design and
  Implementation}} (Edinburgh, United Kingdom) \emph{(\bibinfo{series}{PLDI
  '14})}. \bibinfo{publisher}{ACM}, \bibinfo{address}{New York, NY, USA},
  \bibinfo{pages}{337--348}.
\newblock
\showISBNx{978-1-4503-2784-8}
\urldef\tempurl%
\url{https://doi.org/10.1145/2594291.2594315}
\showDOI{\tempurl}


\bibitem[\protect\citeauthoryear{Kasikci, Zamfir, and Candea}{Kasikci
  et~al\mbox{.}}{2013}]%
        {racemob}
\bibfield{author}{\bibinfo{person}{Baris Kasikci}, \bibinfo{person}{Cristian
  Zamfir}, {and} \bibinfo{person}{George Candea}.}
  \bibinfo{year}{2013}\natexlab{}.
\newblock \showarticletitle{{RaceMob: Crowdsourced Data Race Detection}}. In
  \bibinfo{booktitle}{\emph{Proceedings of the Twenty-Fourth ACM Symposium on
  Operating Systems Principles}} (Farminton, Pennsylvania)
  \emph{(\bibinfo{series}{SOSP '13})}. \bibinfo{publisher}{ACM},
  \bibinfo{address}{New York, NY, USA}, \bibinfo{pages}{406--422}.
\newblock


\bibitem[\protect\citeauthoryear{Kini, Mathur, and Viswanathan}{Kini
  et~al\mbox{.}}{2017}]%
        {wcp2017}
\bibfield{author}{\bibinfo{person}{Dileep Kini}, \bibinfo{person}{Umang
  Mathur}, {and} \bibinfo{person}{Mahesh Viswanathan}.}
  \bibinfo{year}{2017}\natexlab{}.
\newblock \showarticletitle{Dynamic Race Prediction in Linear Time}. In
  \bibinfo{booktitle}{\emph{Proceedings of the 38th ACM SIGPLAN Conference on
  Programming Language Design and Implementation}} (Barcelona, Spain)
  \emph{(\bibinfo{series}{PLDI 2017})}. \bibinfo{publisher}{ACM},
  \bibinfo{address}{New York, NY, USA}, \bibinfo{pages}{157--170}.
\newblock
\showISBNx{978-1-4503-4988-8}
\urldef\tempurl%
\url{https://doi.org/10.1145/3062341.3062374}
\showDOI{\tempurl}


\bibitem[\protect\citeauthoryear{Kokologiannakis, Marmanis, Gladstein, and
  Vafeiadis}{Kokologiannakis et~al\mbox{.}}{2022}]%
        {kokologiannakis2022truly}
\bibfield{author}{\bibinfo{person}{Michalis Kokologiannakis},
  \bibinfo{person}{Iason Marmanis}, \bibinfo{person}{Vladimir Gladstein}, {and}
  \bibinfo{person}{Viktor Vafeiadis}.} \bibinfo{year}{2022}\natexlab{}.
\newblock \showarticletitle{Truly stateless, optimal dynamic partial order
  reduction}.
\newblock \bibinfo{journal}{\emph{Proceedings of the ACM on Programming
  Languages}} \bibinfo{volume}{6}, \bibinfo{number}{POPL}
  (\bibinfo{year}{2022}), \bibinfo{pages}{1--28}.
\newblock


\bibitem[\protect\citeauthoryear{Kulkarni, Mathur, and Pavlogiannis}{Kulkarni
  et~al\mbox{.}}{2021}]%
        {kulkarniCONCUR2021}
\bibfield{author}{\bibinfo{person}{Rucha Kulkarni}, \bibinfo{person}{Umang
  Mathur}, {and} \bibinfo{person}{Andreas Pavlogiannis}.}
  \bibinfo{year}{2021}\natexlab{}.
\newblock \showarticletitle{{Dynamic Data-Race Detection Through the
  Fine-Grained Lens}}. In \bibinfo{booktitle}{\emph{32nd International
  Conference on Concurrency Theory (CONCUR 2021)}}
  \emph{(\bibinfo{series}{Leibniz International Proceedings in Informatics
  (LIPIcs)}, Vol.~\bibinfo{volume}{203})},
  \bibfield{editor}{\bibinfo{person}{Serge Haddad} {and}
  \bibinfo{person}{Daniele Varacca}} (Eds.). \bibinfo{publisher}{Schloss
  Dagstuhl -- Leibniz-Zentrum f{\"u}r Informatik}, \bibinfo{address}{Dagstuhl,
  Germany}, \bibinfo{pages}{16:1--16:23}.
\newblock
\showISBNx{978-3-95977-203-7}
\showISSN{1868-8969}
\urldef\tempurl%
\url{https://doi.org/10.4230/LIPIcs.CONCUR.2021.16}
\showDOI{\tempurl}


\bibitem[\protect\citeauthoryear{Lamport}{Lamport}{1978}]%
        {lamport1978time}
\bibfield{author}{\bibinfo{person}{Leslie Lamport}.}
  \bibinfo{year}{1978}\natexlab{}.
\newblock \showarticletitle{{Time, Clocks, and the Ordering of Events in a
  Distributed System}}.
\newblock \bibinfo{journal}{\emph{Commun. ACM}} \bibinfo{volume}{21},
  \bibinfo{number}{7} (\bibinfo{date}{July} \bibinfo{year}{1978}),
  \bibinfo{pages}{558--565}.
\newblock


\bibitem[\protect\citeauthoryear{Lattner and Adve}{Lattner and Adve}{2004}]%
        {lattner2004llvm}
\bibfield{author}{\bibinfo{person}{Chris Lattner} {and} \bibinfo{person}{Vikram
  Adve}.} \bibinfo{year}{2004}\natexlab{}.
\newblock \showarticletitle{LLVM: A compilation framework for lifelong program
  analysis \& transformation}. In \bibinfo{booktitle}{\emph{International
  symposium on code generation and optimization, 2004. CGO 2004.}} IEEE,
  \bibinfo{pages}{75--86}.
\newblock


\bibitem[\protect\citeauthoryear{Li, Lu, Musuvathi, Nath, and Padhye}{Li
  et~al\mbox{.}}{2019}]%
        {TSVD}
\bibfield{author}{\bibinfo{person}{Guangpu Li}, \bibinfo{person}{Shan Lu},
  \bibinfo{person}{Madanlal Musuvathi}, \bibinfo{person}{Suman Nath}, {and}
  \bibinfo{person}{Rohan Padhye}.} \bibinfo{year}{2019}\natexlab{}.
\newblock \showarticletitle{Efficient scalable thread-safety-violation
  detection: finding thousands of concurrency bugs during testing}. In
  \bibinfo{booktitle}{\emph{Proceedings of the 27th ACM Symposium on Operating
  Systems Principles}} (Huntsville, Ontario, Canada)
  \emph{(\bibinfo{series}{SOSP '19})}. \bibinfo{publisher}{Association for
  Computing Machinery}, \bibinfo{address}{New York, NY, USA},
  \bibinfo{pages}{162–180}.
\newblock
\showISBNx{9781450368735}
\urldef\tempurl%
\url{https://doi.org/10.1145/3341301.3359638}
\showDOI{\tempurl}


\bibitem[\protect\citeauthoryear{Marino, Musuvathi, and Narayanasamy}{Marino
  et~al\mbox{.}}{2009}]%
        {marino2009literace}
\bibfield{author}{\bibinfo{person}{Daniel Marino}, \bibinfo{person}{Madanlal
  Musuvathi}, {and} \bibinfo{person}{Satish Narayanasamy}.}
  \bibinfo{year}{2009}\natexlab{}.
\newblock \showarticletitle{LiteRace: Effective Sampling for Lightweight
  Data-race Detection}. In \bibinfo{booktitle}{\emph{Proceedings of the 30th
  ACM SIGPLAN Conference on Programming Language Design and Implementation}}
  (Dublin, Ireland) \emph{(\bibinfo{series}{PLDI '09})}.
  \bibinfo{publisher}{ACM}, \bibinfo{address}{New York, NY, USA},
  \bibinfo{pages}{134--143}.
\newblock
\showISBNx{978-1-60558-392-1}
\urldef\tempurl%
\url{https://doi.org/10.1145/1542476.1542491}
\showDOI{\tempurl}


\bibitem[\protect\citeauthoryear{Mathur}{Mathur}{2018}]%
        {rapid}
\bibfield{author}{\bibinfo{person}{Umang Mathur}.}
  \bibinfo{year}{2018}\natexlab{}.
\newblock \bibinfo{booktitle}{\emph{{{RAPID}}}}.
\newblock
\urldef\tempurl%
\url{https://github.com/umangm/rapid}
\showURL{%
\tempurl}
\newblock
\shownote{Accessed: 2024-07-01.}


\bibitem[\protect\citeauthoryear{Mathur, Kini, and Viswanathan}{Mathur
  et~al\mbox{.}}{2018}]%
        {shb2018}
\bibfield{author}{\bibinfo{person}{Umang Mathur}, \bibinfo{person}{Dileep
  Kini}, {and} \bibinfo{person}{Mahesh Viswanathan}.}
  \bibinfo{year}{2018}\natexlab{}.
\newblock \showarticletitle{What Happens-after the First Race? Enhancing the
  Predictive Power of Happens-before Based Dynamic Race Detection}.
\newblock \bibinfo{journal}{\emph{Proc. ACM Program. Lang.}}
  \bibinfo{volume}{2}, \bibinfo{number}{OOPSLA}, Article
  \bibinfo{articleno}{145} (\bibinfo{date}{Oct.} \bibinfo{year}{2018}),
  \bibinfo{numpages}{29}~pages.
\newblock
\showISSN{2475-1421}
\urldef\tempurl%
\url{https://doi.org/10.1145/3276515}
\showDOI{\tempurl}


\bibitem[\protect\citeauthoryear{Mathur, Pavlogiannis, Tun\c{c}, and
  Viswanathan}{Mathur et~al\mbox{.}}{2022}]%
        {MathurTreeClocks2022}
\bibfield{author}{\bibinfo{person}{Umang Mathur}, \bibinfo{person}{Andreas
  Pavlogiannis}, \bibinfo{person}{H\"{u}nkar~Can Tun\c{c}}, {and}
  \bibinfo{person}{Mahesh Viswanathan}.} \bibinfo{year}{2022}\natexlab{}.
\newblock \showarticletitle{A Tree Clock Data Structure for Causal Orderings in
  Concurrent Executions}. In \bibinfo{booktitle}{\emph{Proceedings of the 27th
  ACM International Conference on Architectural Support for Programming
  Languages and Operating Systems}} (Lausanne, Switzerland)
  \emph{(\bibinfo{series}{ASPLOS 2022})}. \bibinfo{publisher}{Association for
  Computing Machinery}, \bibinfo{address}{New York, NY, USA},
  \bibinfo{pages}{710–725}.
\newblock
\showISBNx{9781450392051}
\urldef\tempurl%
\url{https://doi.org/10.1145/3503222.3507734}
\showDOI{\tempurl}


\bibitem[\protect\citeauthoryear{Mathur, Pavlogiannis, and Viswanathan}{Mathur
  et~al\mbox{.}}{2020}]%
        {Mathur20}
\bibfield{author}{\bibinfo{person}{Umang Mathur}, \bibinfo{person}{Andreas
  Pavlogiannis}, {and} \bibinfo{person}{Mahesh Viswanathan}.}
  \bibinfo{year}{2020}\natexlab{}.
\newblock \showarticletitle{The Complexity of Dynamic Data Race Prediction}. In
  \bibinfo{booktitle}{\emph{Proceedings of the 35th Annual ACM/IEEE Symposium
  on Logic in Computer Science}} (Saarbr\"{u}cken, Germany)
  \emph{(\bibinfo{series}{LICS ’20})}. \bibinfo{publisher}{Association for
  Computing Machinery}, \bibinfo{address}{New York, NY, USA},
  \bibinfo{pages}{713–727}.
\newblock
\showISBNx{9781450371049}
\urldef\tempurl%
\url{https://doi.org/10.1145/3373718.3394783}
\showDOI{\tempurl}


\bibitem[\protect\citeauthoryear{Mathur, Pavlogiannis, and Viswanathan}{Mathur
  et~al\mbox{.}}{2021}]%
        {SyncP2021}
\bibfield{author}{\bibinfo{person}{Umang Mathur}, \bibinfo{person}{Andreas
  Pavlogiannis}, {and} \bibinfo{person}{Mahesh Viswanathan}.}
  \bibinfo{year}{2021}\natexlab{}.
\newblock \showarticletitle{Optimal Prediction of Synchronization-Preserving
  Races}.
\newblock \bibinfo{journal}{\emph{Proc. ACM Program. Lang.}}
  \bibinfo{volume}{5}, \bibinfo{number}{POPL}, Article \bibinfo{articleno}{36}
  (\bibinfo{date}{jan} \bibinfo{year}{2021}), \bibinfo{numpages}{29}~pages.
\newblock
\urldef\tempurl%
\url{https://doi.org/10.1145/3434317}
\showDOI{\tempurl}


\bibitem[\protect\citeauthoryear{Mathur and Viswanathan}{Mathur and
  Viswanathan}{2020}]%
        {MathurAtomicity20}
\bibfield{author}{\bibinfo{person}{Umang Mathur} {and} \bibinfo{person}{Mahesh
  Viswanathan}.} \bibinfo{year}{2020}\natexlab{}.
\newblock \showarticletitle{Atomicity Checking in Linear Time Using Vector
  Clocks}. In \bibinfo{booktitle}{\emph{Proceedings of the Twenty-Fifth
  International Conference on Architectural Support for Programming Languages
  and Operating Systems}} (Lausanne, Switzerland)
  \emph{(\bibinfo{series}{ASPLOS ’20})}. \bibinfo{publisher}{Association for
  Computing Machinery}, \bibinfo{address}{New York, NY, USA},
  \bibinfo{pages}{183–199}.
\newblock
\showISBNx{9781450371025}
\urldef\tempurl%
\url{https://doi.org/10.1145/3373376.3378475}
\showDOI{\tempurl}


\bibitem[\protect\citeauthoryear{Mattern}{Mattern}{1988}]%
        {Mattern1988}
\bibfield{author}{\bibinfo{person}{Friedemann Mattern}.}
  \bibinfo{year}{1988}\natexlab{}.
\newblock \showarticletitle{{Virtual Time and Global States of Distributed
  Systems}}. In \bibinfo{booktitle}{\emph{Parallel and Distributed
  Algorithms}}. \bibinfo{publisher}{North-Holland}, \bibinfo{pages}{215--226}.
\newblock


\bibitem[\protect\citeauthoryear{M\"{u}ehlenfeld and Wotawa}{M\"{u}ehlenfeld
  and Wotawa}{2007}]%
        {Helgrind}
\bibfield{author}{\bibinfo{person}{Arndt M\"{u}ehlenfeld} {and}
  \bibinfo{person}{Franz Wotawa}.} \bibinfo{year}{2007}\natexlab{}.
\newblock \showarticletitle{Fault Detection in Multi-threaded C++ Server
  Applications}. In \bibinfo{booktitle}{\emph{Proceedings of the 12th ACM
  SIGPLAN Symposium on Principles and Practice of Parallel Programming}} (San
  Jose, California, USA) \emph{(\bibinfo{series}{PPoPP '07})}.
  \bibinfo{publisher}{ACM}, \bibinfo{address}{New York, NY, USA},
  \bibinfo{pages}{142--143}.
\newblock
\showISBNx{978-1-59593-602-8}
\urldef\tempurl%
\url{https://doi.org/10.1145/1229428.1229457}
\showDOI{\tempurl}


\bibitem[\protect\citeauthoryear{Musuvathi, Qadeer, Ball, Basler, Nainar, and
  Neamtiu}{Musuvathi et~al\mbox{.}}{2008}]%
        {heisenbugs}
\bibfield{author}{\bibinfo{person}{Madanlal Musuvathi}, \bibinfo{person}{Shaz
  Qadeer}, \bibinfo{person}{Thomas Ball}, \bibinfo{person}{Gerard Basler},
  \bibinfo{person}{Piramanayagam~Arumuga Nainar}, {and} \bibinfo{person}{Iulian
  Neamtiu}.} \bibinfo{year}{2008}\natexlab{}.
\newblock \showarticletitle{{Finding and Reproducing Heisenbugs in Concurrent
  Programs}}. In \bibinfo{booktitle}{\emph{Proceedings of the 8th USENIX
  Conference on Operating Systems Design and Implementation}} (San Diego,
  California) \emph{(\bibinfo{series}{OSDI'08})}. \bibinfo{publisher}{USENIX
  Association}, \bibinfo{address}{Berkeley, CA, USA},
  \bibinfo{pages}{267--280}.
\newblock


\bibitem[\protect\citeauthoryear{Nethercote and Seward}{Nethercote and
  Seward}{2007}]%
        {Valgrind}
\bibfield{author}{\bibinfo{person}{Nicholas Nethercote} {and}
  \bibinfo{person}{Julian Seward}.} \bibinfo{year}{2007}\natexlab{}.
\newblock \showarticletitle{Valgrind: a framework for heavyweight dynamic
  binary instrumentation}. In \bibinfo{booktitle}{\emph{Proceedings of the 28th
  ACM SIGPLAN Conference on Programming Language Design and Implementation}}
  (San Diego, California, USA) \emph{(\bibinfo{series}{PLDI '07})}.
  \bibinfo{publisher}{Association for Computing Machinery},
  \bibinfo{address}{New York, NY, USA}, \bibinfo{pages}{89–100}.
\newblock
\showISBNx{9781595936332}
\urldef\tempurl%
\url{https://doi.org/10.1145/1250734.1250746}
\showDOI{\tempurl}


\bibitem[\protect\citeauthoryear{Pavlogiannis}{Pavlogiannis}{2019}]%
        {PavlogiannisPOPL20}
\bibfield{author}{\bibinfo{person}{Andreas Pavlogiannis}.}
  \bibinfo{year}{2019}\natexlab{}.
\newblock \showarticletitle{Fast, Sound, and Effectively Complete Dynamic Race
  Prediction}.
\newblock \bibinfo{journal}{\emph{Proc. ACM Program. Lang.}}
  \bibinfo{volume}{4}, \bibinfo{number}{POPL}, Article \bibinfo{articleno}{17}
  (\bibinfo{date}{Dec.} \bibinfo{year}{2019}), \bibinfo{numpages}{29}~pages.
\newblock
\urldef\tempurl%
\url{https://doi.org/10.1145/3371085}
\showDOI{\tempurl}


\bibitem[\protect\citeauthoryear{Pozniansky and Schuster}{Pozniansky and
  Schuster}{2003}]%
        {Pozniansky:2003:EOD:966049.781529}
\bibfield{author}{\bibinfo{person}{Eli Pozniansky} {and} \bibinfo{person}{Assaf
  Schuster}.} \bibinfo{year}{2003}\natexlab{}.
\newblock \showarticletitle{Efficient On-the-fly Data Race Detection in
  Multithreaded C++ Programs}. In \bibinfo{booktitle}{\emph{Proceedings of the
  Ninth ACM SIGPLAN Symposium on Principles and Practice of Parallel
  Programming}} (San Diego, California, USA) \emph{(\bibinfo{series}{PPoPP
  '03})}. \bibinfo{publisher}{ACM}, \bibinfo{address}{New York, NY, USA},
  \bibinfo{pages}{179--190}.
\newblock
\showISBNx{1-58113-588-2}
\urldef\tempurl%
\url{https://doi.org/10.1145/781498.781529}
\showDOI{\tempurl}


\bibitem[\protect\citeauthoryear{Rhodes, Flanagan, and Freund}{Rhodes
  et~al\mbox{.}}{2017}]%
        {bigfoot}
\bibfield{author}{\bibinfo{person}{Dustin Rhodes}, \bibinfo{person}{Cormac
  Flanagan}, {and} \bibinfo{person}{Stephen~N. Freund}.}
  \bibinfo{year}{2017}\natexlab{}.
\newblock \showarticletitle{BigFoot: Static Check Placement for Dynamic Race
  Detection}. In \bibinfo{booktitle}{\emph{Proceedings of the 38th ACM SIGPLAN
  Conference on Programming Language Design and Implementation}} (Barcelona,
  Spain) \emph{(\bibinfo{series}{PLDI 2017})}. \bibinfo{publisher}{ACM},
  \bibinfo{address}{New York, NY, USA}, \bibinfo{pages}{141--156}.
\newblock
\showISBNx{978-1-4503-4988-8}
\urldef\tempurl%
\url{https://doi.org/10.1145/3062341.3062350}
\showDOI{\tempurl}


\bibitem[\protect\citeauthoryear{Roemer, Gen\c{c}, and Bond}{Roemer
  et~al\mbox{.}}{2018}]%
        {Roemer18}
\bibfield{author}{\bibinfo{person}{Jake Roemer}, \bibinfo{person}{Kaan
  Gen\c{c}}, {and} \bibinfo{person}{Michael~D. Bond}.}
  \bibinfo{year}{2018}\natexlab{}.
\newblock \showarticletitle{High-coverage, Unbounded Sound Predictive Race
  Detection}. In \bibinfo{booktitle}{\emph{Proceedings of the 39th ACM SIGPLAN
  Conference on Programming Language Design and Implementation}} (Philadelphia,
  PA, USA) \emph{(\bibinfo{series}{PLDI 2018})}. \bibinfo{publisher}{ACM},
  \bibinfo{address}{New York, NY, USA}, \bibinfo{pages}{374--389}.
\newblock
\showISBNx{978-1-4503-5698-5}
\urldef\tempurl%
\url{https://doi.org/10.1145/3192366.3192385}
\showDOI{\tempurl}


\bibitem[\protect\citeauthoryear{Roemer, Gen\c{c}, and Bond}{Roemer
  et~al\mbox{.}}{2020}]%
        {Roemer20}
\bibfield{author}{\bibinfo{person}{Jake Roemer}, \bibinfo{person}{Kaan
  Gen\c{c}}, {and} \bibinfo{person}{Michael~D. Bond}.}
  \bibinfo{year}{2020}\natexlab{}.
\newblock \showarticletitle{SmartTrack: Efficient Predictive Race Detection}.
  In \bibinfo{booktitle}{\emph{Proceedings of the 41st ACM SIGPLAN Conference
  on Programming Language Design and Implementation}} (London, UK)
  \emph{(\bibinfo{series}{PLDI 2020})}. \bibinfo{publisher}{Association for
  Computing Machinery}, \bibinfo{address}{New York, NY, USA},
  \bibinfo{pages}{747–762}.
\newblock
\showISBNx{9781450376136}
\urldef\tempurl%
\url{https://doi.org/10.1145/3385412.3385993}
\showDOI{\tempurl}


\bibitem[\protect\citeauthoryear{Rosu}{Rosu}{2018}]%
        {rvpredict}
\bibfield{author}{\bibinfo{person}{Grigore Rosu}.}
  \bibinfo{year}{2018}\natexlab{}.
\newblock \bibinfo{booktitle}{\emph{{{RV-Predict, Runtime Verification}}}}.
\newblock
\newblock
\shownote{Accessed: 2018-04-01.}


\bibitem[\protect\citeauthoryear{Said, Wang, Yang, and Sakallah}{Said
  et~al\mbox{.}}{2011}]%
        {Said2011}
\bibfield{author}{\bibinfo{person}{Mahmoud Said}, \bibinfo{person}{Chao Wang},
  \bibinfo{person}{Zijiang Yang}, {and} \bibinfo{person}{Karem Sakallah}.}
  \bibinfo{year}{2011}\natexlab{}.
\newblock \showarticletitle{{Generating Data Race Witnesses by an SMT-based
  Analysis}}. In \bibinfo{booktitle}{\emph{Proceedings of the Third
  International Conference on NASA Formal Methods}} (Pasadena, CA)
  \emph{(\bibinfo{series}{NFM'11})}. \bibinfo{publisher}{Springer-Verlag},
  \bibinfo{address}{Berlin, Heidelberg}, \bibinfo{pages}{313--327}.
\newblock


\bibitem[\protect\citeauthoryear{Savage, Burrows, Nelson, Sobalvarro, and
  Anderson}{Savage et~al\mbox{.}}{1997}]%
        {savage1997eraser}
\bibfield{author}{\bibinfo{person}{Stefan Savage}, \bibinfo{person}{Michael
  Burrows}, \bibinfo{person}{Greg Nelson}, \bibinfo{person}{Patrick
  Sobalvarro}, {and} \bibinfo{person}{Thomas Anderson}.}
  \bibinfo{year}{1997}\natexlab{}.
\newblock \showarticletitle{Eraser: A Dynamic Data Race Detector for
  Multithreaded Programs}.
\newblock \bibinfo{journal}{\emph{ACM Trans. Comput. Syst.}}
  \bibinfo{volume}{15}, \bibinfo{number}{4} (\bibinfo{date}{Nov.}
  \bibinfo{year}{1997}), \bibinfo{pages}{391–411}.
\newblock
\showISSN{0734-2071}
\urldef\tempurl%
\url{https://doi.org/10.1145/265924.265927}
\showDOI{\tempurl}


\bibitem[\protect\citeauthoryear{Sen, Ro\c{s}u, and Agha}{Sen
  et~al\mbox{.}}{2005}]%
        {sen2005detecting}
\bibfield{author}{\bibinfo{person}{Koushik Sen}, \bibinfo{person}{Grigore
  Ro\c{s}u}, {and} \bibinfo{person}{Gul Agha}.}
  \bibinfo{year}{2005}\natexlab{}.
\newblock \showarticletitle{{Detecting Errors in Multithreaded Programs by
  Generalized Predictive Analysis of Executions}}. In
  \bibinfo{booktitle}{\emph{Proceedings of the 7th IFIP WG 6.1 International
  Conference on Formal Methods for Open Object-Based Distributed Systems}}
  (Athens, Greece) \emph{(\bibinfo{series}{FMOODS'05})}.
  \bibinfo{publisher}{Springer-Verlag}, \bibinfo{address}{Berlin, Heidelberg},
  \bibinfo{pages}{211--226}.
\newblock


\bibitem[\protect\citeauthoryear{Serebryany and Iskhodzhanov}{Serebryany and
  Iskhodzhanov}{2009}]%
        {threadsanitizer}
\bibfield{author}{\bibinfo{person}{Konstantin Serebryany} {and}
  \bibinfo{person}{Timur Iskhodzhanov}.} \bibinfo{year}{2009}\natexlab{}.
\newblock \showarticletitle{{ThreadSanitizer: Data Race Detection in
  Practice}}. In \bibinfo{booktitle}{\emph{Proceedings of the Workshop on
  Binary Instrumentation and Applications}} (New York, New York, USA)
  \emph{(\bibinfo{series}{WBIA '09})}. \bibinfo{publisher}{ACM},
  \bibinfo{address}{New York, NY, USA}, \bibinfo{pages}{62--71}.
\newblock


\bibitem[\protect\citeauthoryear{Serebryany, Kennelly, Phillips, Denton, Elver,
  Potapenko, Morehouse, Tsyrklevich, Holler, Lettner, Kilzer, and
  Brandt}{Serebryany et~al\mbox{.}}{2024}]%
        {GWPASan2024}
\bibfield{author}{\bibinfo{person}{Kostya Serebryany}, \bibinfo{person}{Chris
  Kennelly}, \bibinfo{person}{Mitch Phillips}, \bibinfo{person}{Matt Denton},
  \bibinfo{person}{Marco Elver}, \bibinfo{person}{Alexander Potapenko},
  \bibinfo{person}{Matt Morehouse}, \bibinfo{person}{Vlad Tsyrklevich},
  \bibinfo{person}{Christian Holler}, \bibinfo{person}{Julian Lettner},
  \bibinfo{person}{David Kilzer}, {and} \bibinfo{person}{Lander Brandt}.}
  \bibinfo{year}{2024}\natexlab{}.
\newblock \showarticletitle{GWP-ASan: Sampling-Based Detection of Memory-Safety
  Bugs in Production}. In \bibinfo{booktitle}{\emph{Proceedings of the 46th
  International Conference on Software Engineering: Software Engineering in
  Practice}} (Lisbon, Portugal) \emph{(\bibinfo{series}{ICSE-SEIP '24})}.
  \bibinfo{publisher}{Association for Computing Machinery},
  \bibinfo{address}{New York, NY, USA}, \bibinfo{pages}{168–177}.
\newblock
\showISBNx{9798400705014}
\urldef\tempurl%
\url{https://doi.org/10.1145/3639477.3640328}
\showDOI{\tempurl}


\bibitem[\protect\citeauthoryear{Shi, Mathur, and Pavlogiannis}{Shi
  et~al\mbox{.}}{2024}]%
        {Shi2024}
\bibfield{author}{\bibinfo{person}{Zheng Shi}, \bibinfo{person}{Umang Mathur},
  {and} \bibinfo{person}{Andreas Pavlogiannis}.}
  \bibinfo{year}{2024}\natexlab{}.
\newblock \showarticletitle{Optimistic Prediction of Synchronization-Reversal
  Data Races}. In \bibinfo{booktitle}{\emph{Proceedings of the IEEE/ACM 46th
  International Conference on Software Engineering}} (Lisbon, Portugal)
  \emph{(\bibinfo{series}{ICSE '24})}. \bibinfo{publisher}{Association for
  Computing Machinery}, \bibinfo{address}{New York, NY, USA}, Article
  \bibinfo{articleno}{134}, \bibinfo{numpages}{13}~pages.
\newblock
\showISBNx{9798400702174}
\urldef\tempurl%
\url{https://doi.org/10.1145/3597503.3639099}
\showDOI{\tempurl}


\bibitem[\protect\citeauthoryear{Smaragdakis, Evans, Sadowski, Yi, and
  Flanagan}{Smaragdakis et~al\mbox{.}}{2012}]%
        {cp2012}
\bibfield{author}{\bibinfo{person}{Yannis Smaragdakis}, \bibinfo{person}{Jacob
  Evans}, \bibinfo{person}{Caitlin Sadowski}, \bibinfo{person}{Jaeheon Yi},
  {and} \bibinfo{person}{Cormac Flanagan}.} \bibinfo{year}{2012}\natexlab{}.
\newblock \showarticletitle{Sound Predictive Race Detection in Polynomial
  Time}. In \bibinfo{booktitle}{\emph{Proceedings of the 39th Annual ACM
  SIGPLAN-SIGACT Symposium on Principles of Programming Languages}}
  (Philadelphia, PA, USA) \emph{(\bibinfo{series}{POPL '12})}.
  \bibinfo{publisher}{ACM}, \bibinfo{address}{New York, NY, USA},
  \bibinfo{pages}{387--400}.
\newblock
\showISBNx{978-1-4503-1083-3}
\urldef\tempurl%
\url{https://doi.org/10.1145/2103656.2103702}
\showDOI{\tempurl}


\bibitem[\protect\citeauthoryear{Thokair, Zhang, Mathur, and
  Viswanathan}{Thokair et~al\mbox{.}}{2023}]%
        {RPT2023}
\bibfield{author}{\bibinfo{person}{Mosaad~Al Thokair}, \bibinfo{person}{Minjian
  Zhang}, \bibinfo{person}{Umang Mathur}, {and} \bibinfo{person}{Mahesh
  Viswanathan}.} \bibinfo{year}{2023}\natexlab{}.
\newblock \showarticletitle{Dynamic Race Detection with O(1) Samples}.
\newblock \bibinfo{journal}{\emph{Proc. ACM Program. Lang.}}
  \bibinfo{volume}{7}, \bibinfo{number}{POPL}, Article \bibinfo{articleno}{45}
  (\bibinfo{date}{jan} \bibinfo{year}{2023}), \bibinfo{numpages}{30}~pages.
\newblock
\urldef\tempurl%
\url{https://doi.org/10.1145/3571238}
\showDOI{\tempurl}


\bibitem[\protect\citeauthoryear{Tun\c{c}, Mathur, Pavlogiannis, and
  Viswanathan}{Tun\c{c} et~al\mbox{.}}{2023}]%
        {syncpdeadlocks2023}
\bibfield{author}{\bibinfo{person}{H\"{u}nkar~Can Tun\c{c}},
  \bibinfo{person}{Umang Mathur}, \bibinfo{person}{Andreas Pavlogiannis}, {and}
  \bibinfo{person}{Mahesh Viswanathan}.} \bibinfo{year}{2023}\natexlab{}.
\newblock \showarticletitle{Sound Dynamic Deadlock Prediction in Linear Time}.
\newblock \bibinfo{journal}{\emph{Proc. ACM Program. Lang.}}
  \bibinfo{volume}{7}, \bibinfo{number}{PLDI}, Article \bibinfo{articleno}{177}
  (\bibinfo{date}{June} \bibinfo{year}{2023}), \bibinfo{numpages}{26}~pages.
\newblock
\urldef\tempurl%
\url{https://doi.org/10.1145/3591291}
\showDOI{\tempurl}


\bibitem[\protect\citeauthoryear{Wilcox, Finch, Flanagan, and Freund}{Wilcox
  et~al\mbox{.}}{2015}]%
        {Wilcox2015}
\bibfield{author}{\bibinfo{person}{James~R. Wilcox}, \bibinfo{person}{Parker
  Finch}, \bibinfo{person}{Cormac Flanagan}, {and} \bibinfo{person}{Stephen~N.
  Freund}.} \bibinfo{year}{2015}\natexlab{}.
\newblock \showarticletitle{Array Shadow State Compression for Precise Dynamic
  Race Detection (T)}. In \bibinfo{booktitle}{\emph{Proceedings of the 2015
  30th IEEE/ACM International Conference on Automated Software Engineering
  (ASE)}} \emph{(\bibinfo{series}{ASE '15})}. \bibinfo{publisher}{IEEE Computer
  Society}, \bibinfo{address}{Washington, DC, USA}, \bibinfo{pages}{155--165}.
\newblock
\showISBNx{978-1-5090-0025-8}
\urldef\tempurl%
\url{https://doi.org/10.1109/ASE.2015.19}
\showDOI{\tempurl}


\bibitem[\protect\citeauthoryear{Wolff, Shi, Duck, Mathur, and
  Roychoudhury}{Wolff et~al\mbox{.}}{2024}]%
        {rff2024}
\bibfield{author}{\bibinfo{person}{Dylan Wolff}, \bibinfo{person}{Zheng Shi},
  \bibinfo{person}{Gregory~J. Duck}, \bibinfo{person}{Umang Mathur}, {and}
  \bibinfo{person}{Abhik Roychoudhury}.} \bibinfo{year}{2024}\natexlab{}.
\newblock \showarticletitle{Greybox Fuzzing for Concurrency Testing}. In
  \bibinfo{booktitle}{\emph{Proceedings of the 29th ACM International
  Conference on Architectural Support for Programming Languages and Operating
  Systems, Volume 2}} (La Jolla, CA, USA) \emph{(\bibinfo{series}{ASPLOS
  '24})}. \bibinfo{publisher}{Association for Computing Machinery},
  \bibinfo{address}{New York, NY, USA}, \bibinfo{pages}{482–498}.
\newblock
\showISBNx{9798400703850}
\urldef\tempurl%
\url{https://doi.org/10.1145/3620665.3640389}
\showDOI{\tempurl}


\bibitem[\protect\citeauthoryear{Wood, Cao, Bond, and Grossman}{Wood
  et~al\mbox{.}}{2017}]%
        {Wood2017}
\bibfield{author}{\bibinfo{person}{Benjamin~P. Wood}, \bibinfo{person}{Man
  Cao}, \bibinfo{person}{Michael~D. Bond}, {and} \bibinfo{person}{Dan
  Grossman}.} \bibinfo{year}{2017}\natexlab{}.
\newblock \showarticletitle{Instrumentation Bias for Dynamic Data Race
  Detection}.
\newblock \bibinfo{journal}{\emph{Proc. ACM Program. Lang.}}
  \bibinfo{volume}{1}, \bibinfo{number}{OOPSLA}, Article
  \bibinfo{articleno}{69} (\bibinfo{date}{Oct.} \bibinfo{year}{2017}),
  \bibinfo{numpages}{31}~pages.
\newblock
\showISSN{2475-1421}
\urldef\tempurl%
\url{https://doi.org/10.1145/3133893}
\showDOI{\tempurl}


\bibitem[\protect\citeauthoryear{Yuan, Yang, and Gu}{Yuan
  et~al\mbox{.}}{2018}]%
        {pos}
\bibfield{author}{\bibinfo{person}{Xinhao Yuan}, \bibinfo{person}{Junfeng
  Yang}, {and} \bibinfo{person}{Ronghui Gu}.} \bibinfo{year}{2018}\natexlab{}.
\newblock \showarticletitle{Partial order aware concurrency sampling}. In
  \bibinfo{booktitle}{\emph{Computer Aided Verification: 30th International
  Conference, CAV 2018, Held as Part of the Federated Logic Conference, FloC
  2018, Oxford, UK, July 14-17, 2018, Proceedings, Part II 30}}. Springer,
  \bibinfo{pages}{317--335}.
\newblock


\bibitem[\protect\citeauthoryear{Zhang, Jung, and Lee}{Zhang
  et~al\mbox{.}}{2017}]%
        {prorace}
\bibfield{author}{\bibinfo{person}{Tong Zhang}, \bibinfo{person}{Changhee
  Jung}, {and} \bibinfo{person}{Dongyoon Lee}.}
  \bibinfo{year}{2017}\natexlab{}.
\newblock \showarticletitle{ProRace: Practical Data Race Detection for
  Production Use}.
\newblock \bibinfo{journal}{\emph{SIGPLAN Not.}} \bibinfo{volume}{52},
  \bibinfo{number}{4} (\bibinfo{date}{April} \bibinfo{year}{2017}),
  \bibinfo{pages}{149–162}.
\newblock
\showISSN{0362-1340}
\urldef\tempurl%
\url{https://doi.org/10.1145/3093336.3037708}
\showDOI{\tempurl}


\bibitem[\protect\citeauthoryear{Zhao, Wolff, Mathur, and Roychoudhury}{Zhao
  et~al\mbox{.}}{2025}]%
        {SURW2025}
\bibfield{author}{\bibinfo{person}{Huan Zhao}, \bibinfo{person}{Dylan Wolff},
  \bibinfo{person}{Umang Mathur}, {and} \bibinfo{person}{Abhik Roychoudhury}.}
  \bibinfo{year}{2025}\natexlab{}.
\newblock \showarticletitle{Selectively Uniform Concurrency Testing}. In
  \bibinfo{booktitle}{\emph{Proceedings of the 30th ACM International
  Conference on Architectural Support for Programming Languages and Operating
  Systems, Volume 1}} (Rotterdam, Netherlands) \emph{(\bibinfo{series}{ASPLOS
  '25})}. \bibinfo{publisher}{Association for Computing Machinery},
  \bibinfo{address}{New York, NY, USA}, \bibinfo{pages}{1003–1019}.
\newblock
\showISBNx{9798400706981}
\urldef\tempurl%
\url{https://doi.org/10.1145/3669940.3707214}
\showDOI{\tempurl}


\bibitem[\protect\citeauthoryear{Zhao, P\^{\i}rlea, Ang, Mathur, and
  Sergey}{Zhao et~al\mbox{.}}{2024}]%
        {treeClocksCPP2024}
\bibfield{author}{\bibinfo{person}{Qiyuan Zhao}, \bibinfo{person}{George
  P\^{\i}rlea}, \bibinfo{person}{Zhendong Ang}, \bibinfo{person}{Umang Mathur},
  {and} \bibinfo{person}{Ilya Sergey}.} \bibinfo{year}{2024}\natexlab{}.
\newblock \showarticletitle{Rooting for Efficiency: Mechanised Reasoning about
  Array-Based Trees in Separation Logic}. In
  \bibinfo{booktitle}{\emph{Proceedings of the 13th ACM SIGPLAN International
  Conference on Certified Programs and Proofs}} (London, UK)
  \emph{(\bibinfo{series}{CPP 2024})}. \bibinfo{publisher}{Association for
  Computing Machinery}, \bibinfo{address}{New York, NY, USA},
  \bibinfo{pages}{45–59}.
\newblock
\showISBNx{9798400704888}
\urldef\tempurl%
\url{https://doi.org/10.1145/3636501.3636944}
\showDOI{\tempurl}


\bibitem[\protect\citeauthoryear{Zheng, Ravi, Qin, and Agrawal}{Zheng
  et~al\mbox{.}}{2014}]%
        {GMrace}
\bibfield{author}{\bibinfo{person}{Mai Zheng}, \bibinfo{person}{Vignesh~T.
  Ravi}, \bibinfo{person}{Feng Qin}, {and} \bibinfo{person}{Gagan Agrawal}.}
  \bibinfo{year}{2014}\natexlab{}.
\newblock \showarticletitle{GMRace: Detecting Data Races in GPU Programs via a
  Low-Overhead Scheme}.
\newblock \bibinfo{journal}{\emph{IEEE Transactions on Parallel and Distributed
  Systems}} \bibinfo{volume}{25}, \bibinfo{number}{1} (\bibinfo{year}{2014}),
  \bibinfo{pages}{104--115}.
\newblock
\urldef\tempurl%
\url{https://doi.org/10.1109/TPDS.2013.44}
\showDOI{\tempurl}


\end{thebibliography}

\end{document}